 \documentclass[final,5p,times,twocolumn,authoryear]{elsarticle}

\usepackage{amssymb}
\usepackage{lipsum}
\usepackage{url}
\usepackage{amsmath}
\usepackage{xcolor}
\usepackage{hyperref}

\journal{New Astronomy}

\begin{document}

\begin{frontmatter}



\title{Analysis and simulations of binary black hole merger spins\\ --- the question of spin-axis tossing at black hole formation}


\author[label1,label3]{Hans C.~G. Larsen}
\fntext[label3]{These authors contributed equally to this work.}
\author[label1,label3]{Casper C. Pedersen}
\author[label1]{Thomas M. Tauris\corref{cor1}}
\cortext[cor1]{Corresponding author. Email: tauris@mp.aau.dk}
\author[label1,label2]{Ali Sepas}
\author[label1]{Claudia Larsen}
\author[label2]{Christophe A.~N. Biscio}
\affiliation[label1]{organization={Department of Materials and Production, Aalborg University},
             addressline={Fibigerstr{\ae}de 16},
             city={Aalborg},
             postcode={9220},
             state={},
             country={Denmark}}
\affiliation[label2]{organization={Department of Mathematical Sciences, Aalborg University},
             addressline={Fredrik Bajers Vej 7K},
             city={Aalborg},
             postcode={9220},
             state={},
             country={Denmark}}

\begin{abstract}
    The origin of binary black hole (BH) mergers remains a topic of active debate, with effective spins ($\chi_{\rm eff}$) measured by the LIGO-Virgo-KAGRA (LVK) Collaboration providing crucial insights.
    In this study, our objective is to investigate the empirical $\chi_{\rm eff}$ distribution (and constrain individual spin components) of binary BH mergers and compare them with extensive simulations, assuming that they originate purely from isolated binaries or a mixture of formation channels. We explore scenarios using BH kicks with and without the effect of spin-axis tossing during BH formation.
    We employ simple yet robust Monte Carlo simulations of the final core collapse forming the second-born BH, using minimal assumptions to ensure transparency and reproducibility. The synthetic $\chi_{\rm eff}$ distribution is compared to the empirical data from LVK science runs O1--O3 using functional data analysis, kernel density estimations, and three different statistical tests, accounting for data uncertainties.
    We find strong indications for spin-axis tossing during BH formation if LVK sources are dominated by the isolated binary channel. Simulations with spin-axis tossing achieve high p-values (up to 0.882) using Kolmogorov-Smirnov, Cramer-von Mises, and Anderson-Darling tests, while without tossing, all p-values drop below 0.001 for isolated binaries.
    A statistically acceptable solution without tossing, however, emerges if $\sim 72\pm 8\%$ of detected binary BH mergers result from dynamical interactions causing random BH spin directions. Finally, for an isolated binary origin, we find a preference for mass reversal in $\sim 30\,\%$ of the progenitor binaries.
    Predictions from this study can be tested with LVK O4+O5 data as well as the 3G detectors, Einstein Telescope and Cosmic Explorer, enabling improved constraints on formation channel ratios and the critical question of BH spin-axis tossing.
\end{abstract}




\end{frontmatter}




\section{Introduction}\label{sec:intro}
Over the last decade, gravitational wave (GW) astronomy has richly revealed the existence of binary black hole (BH+BH) mergers \citep{aaa+16a}. So far, 83 such BH+BH merger events have been reported in LIGO-Virgo-KAGRA (LVK) science runs O1--O3, and this number is expected to more than double after the ongoing O4a+b (\url{https://gracedb.ligo.org/}). 
The origin of BH+BH mergers remains controversial and has sparked a plethora of new research on massive binary star evolution to understand the formation process of such compact object mergers \citep[e.g.][]{mlp+16,rcr16,md16b,ktl+16,zsr+19,mrf+19,bfq+20,bkf+20,lss+20,knic20,vtr+21,slm+22,fl22,mbs+22,bbs+22,sl22,spl23}; for reviews, see \citet{mf22,tv23,mb24}.

\begin{figure}
\centering
\vspace*{+0.2cm}\hspace*{-0.0cm}
\includegraphics[width=0.45\textwidth]{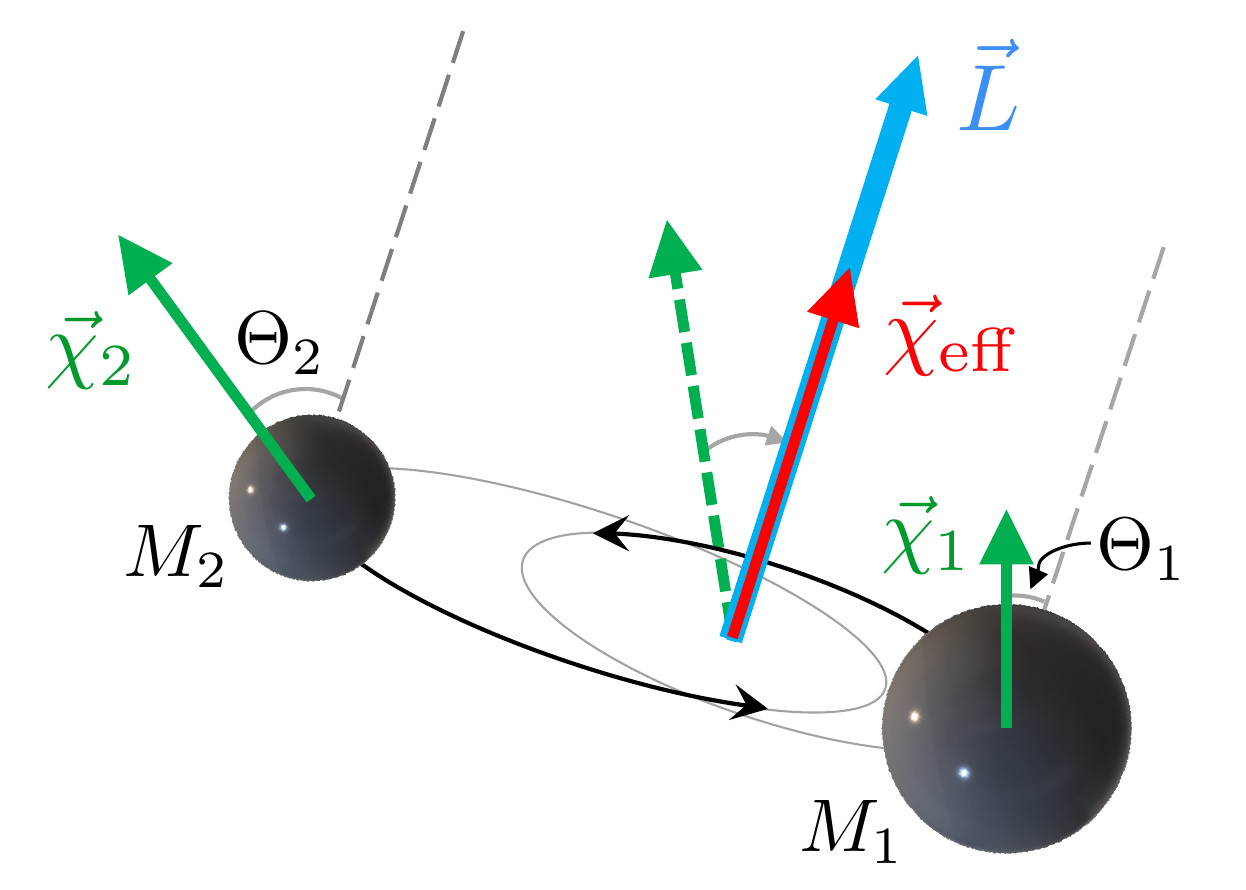} 
\caption{Illustration of the effective spin of a BH+BH merger --- see text.}
\label{fig:illustration_chi_eff}
\end{figure}

\begin{figure}
\centering
\vspace*{-0.0cm}\hspace*{-0.5cm}
\includegraphics[width=0.45\textwidth]{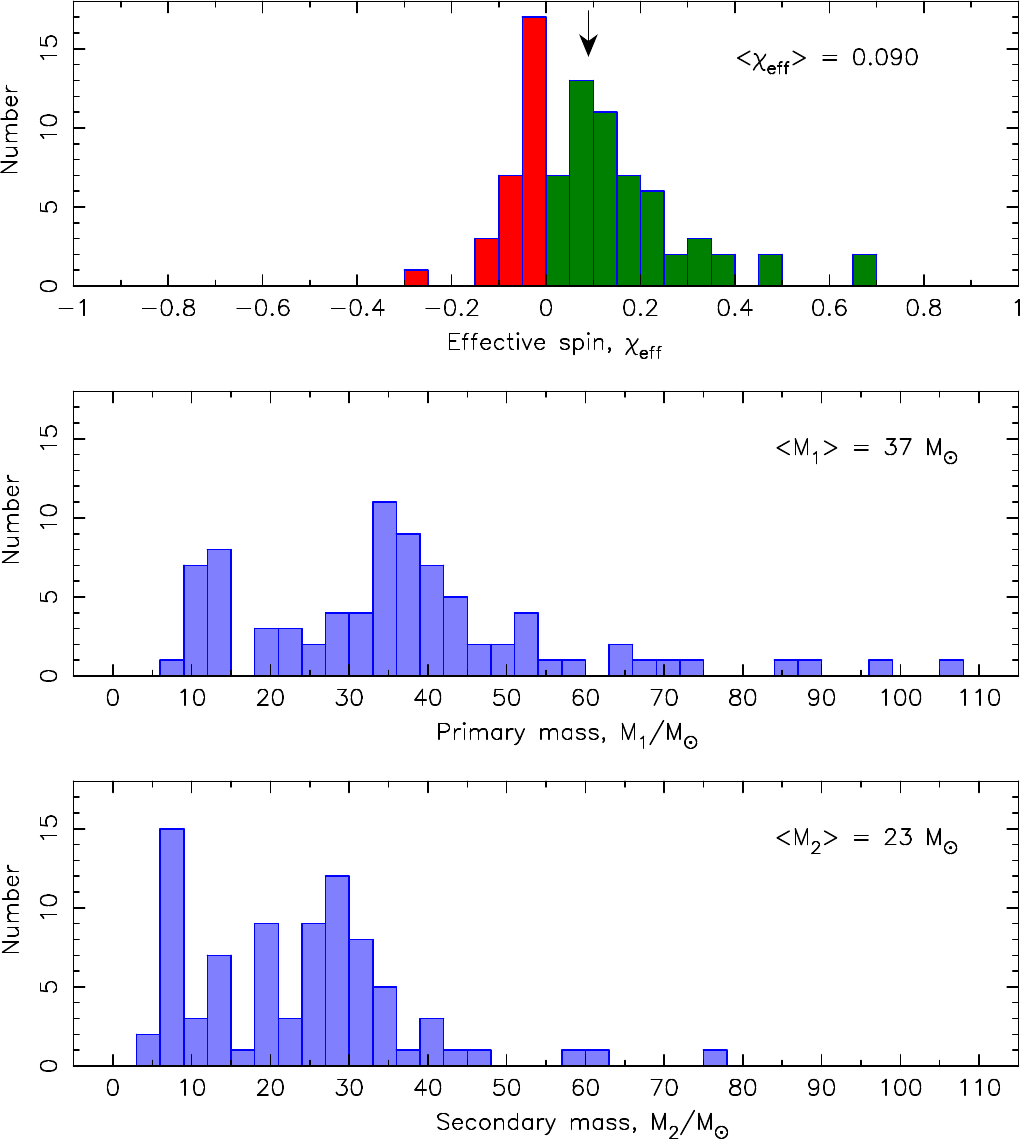}
\caption{Top: Distribution of individually measured mean values of effective inspiral spins of 83 observed BH+BH merger events from O1--O3. Green and red colors indicate $\chi_{\rm eff}>0$ and $\chi_{\rm eff}<0$, respectively, and the arrow marks the average value, $\langle \chi_{\rm eff} \rangle = 0.090$. Central and bottom: Inferred component masses of primary and secondary BHs ($M_2>3.0\;M_\odot$) with average values of $37\;M_\odot$ and $23\;M_\odot$, respectively.
Data is taken from the LVK catalogue: \url{https://gwosc.org/eventapi/html/GWTC/}.} 
\label{fig:GWTC3_BH+BH}
\end{figure}

The individual masses of the BH components derived from the detected BH+BH mergers fall in the range of a few solar masses ($M_\odot$) to more than $100\;M_\odot$ \citep{aaa+23b}. 
The other observable fundamental parameter that describes an astrophysical BH is spin. The dimensionless spin magnitude, with a value between 0 and 1, is given by $\chi=cJ/GM^2$, where $J$ is the spin angular momentum of the BH, $M$ is its mass and $c$ and $G$ are the speed of light in vacuum and the gravitational constant, respectively. 
It is challenging to directly measure the magnitude of individual BH spins ($\chi_1$ and $\chi_2$), and their misalignment (tilt) angles ($\Theta_1$ and $\Theta_2$) with respect to the orbital angular momentum vector of the binary ($\vec{L}$) during inspiral --- see Fig.~\ref{fig:illustration_chi_eff}. 
This is because it is difficult to disentangle the contributions of each individual BH’s spin from the GW signal of a BH+BH merger, which primarily reflects the system's overall angular momentum. Spin-orbit couplings are complex, especially when the BH spins are misaligned, and there can be degeneracies in the parameters that influence the GW signal \citep{GWTC-1}.
However, the sum of the weighted projected BH spins along the orbital angular momentum is measurable, i.e. the effective spin ($-1\le \chi_{\rm eff}\le1$):
\begin{eqnarray}\label{eq:chi_eff2}
                 \chi _{\rm eff} & \equiv & \frac{\left(M_1\vec{\chi}_1+M_2\vec{\chi}_2\right)}{M_T} \cdot \frac{\vec{L}}{|\vec{L}|} \nonumber \\
                 & = & \frac{\chi_1 \cos \Theta_1 + q\,\chi_2\cos\Theta_2}{1+q}\,,
\end{eqnarray}
where $M_T = M_1 + M_2$ is the total mass of the two BHs and $q\equiv M_2/M_1\le 1$ is their mass ratio.
A value of $\chi_{\rm eff}<0$ indicates that $\vec{\chi}_{\rm eff}$ is anti-aligned with $\vec{L}$.
Whereas misaligned BH spins are subject to precession, $\chi_{\rm eff}$ is constant during the long (often several Gyr) phase of inspiral prior to merger \citep{gks+15} and therefore it is an important diagnostic tool for the origin of BH+BH systems. 

Figure~\ref{fig:GWTC3_BH+BH} (top panel) shows the distribution of the {\em mean} values of $\chi_{\rm eff}$ of all BH+BH mergers from LVK science runs O1--O3. The distribution is somewhat of a puzzle because it is both asymmetric \citep{aaa+20e,zbb+21,rco+21,aaa+23b} and has a significant number of systems with $\chi _{\rm eff}<0$ (although caution should be taken due to the precision of measured $\chi_{\rm eff}$ values that typically have uncertainties between 0.1 and 0.4 for each event). 
A perfect symmetric distribution of $\chi_{\rm eff}$ centered on $\chi_{\rm eff}=0$ would be expected if all events originated from binaries that assemble pairs of BHs with random individual spin directions via dynamical interactions, for example, in globular clusters \citep{mo10,rzp+16}. 

An asymmetric $\chi_{\rm eff}$ distribution, on the other hand, provides evidence for a substantial contribution from BH+BH systems produced in isolated binaries. In tight binaries, torques exerted by the inflowing material --- such as Lense-Thirring precession, disk viscosity, and back-reaction torques --- prior to the collapse of the progenitor of the second-born BH, tend to align the spin of the accreting first-born BH with the orbital angular momentum vector \citep[e.g.][]{bp75}. Since the first-born BH is expected to align its spin via accretion --- and often is the more massive of the two BHs (but see Section~\ref{subsec:mass-reversal}) --- the contribution from $\chi_1$ is in most cases positive.

Besides BH masses and spins, orbital eccentricities could also be used as discriminators between different BH+BH formation models. However, with current GW detectors, the residual eccentricities are too small to be detected once the BH+BH systems enter the sensitivity window of GW detection \citep{frg+24}.
For further discussions on interpretation of the empirical $\chi_{\rm eff}$ data, see e.g. \citet{sbm17,qfm+18,bfq+20,ob21,cmcf22,mf22,tau22,cf24,mkcc24}.

Negative effective spins (requiring at least one misalignment angle component, $\Theta_i>90^{\circ}$) can therefore only result in isolated binaries if either: 
\begin{enumerate}[i)]
    \item very large asymmetric momentum kicks are imparted onto BHs in their formation process following the collapse of a massive stellar core \citep{wgo+18,cfr21,flr21} such that the orbit is flipped, resulting in retrograde motion; or 
   \item the formation of a BH is accompanied by tossing of its spin axis \citep{tau22}.
\end{enumerate}
However, large BH kicks of several $100\;{\rm km\,s}^{-1}$ are incompatible with the kinematics of the observed Galactic BH binary systems \citep[e.g.][]{man16,mir17,ne25},
and in most cases kicks exceeding even $500\;{\rm km\,s}^{-1}$ are needed to produce a post-SN retrograde orbit.

In this work, we will focus on the second possibility: spin-axis tossing during core collapse. 
\cite{tau22} performed an investigation comparing LVK data with simulations of BH+BH mergers and found qualitative evidence for spin-axis tossing during BH formation.
In addition, \citet[][and references therein]{tau22} presented unambiguous observational evidence of SN spin-axis tossing during the formation of radio pulsars in double neutron star systems.  
It is therefore reasonable to expect a similar effect when BHs form; see also \citet{Janka_2022} who investigated this.
Furthermore, there is recent evidence in favor of spin-axis tossing in BHs in some X-ray binaires, e.g. MAXI~J1820+070 and Cygnus~X-3, see \citet{pvb+22,vpb+24,dzm+24}; but see also \cite{zbs+25} for discussions on GX~399--4.
Here, in this study, we present a statistical quantitative investigation of this important topic and, as we shall see, we find significant evidence in favor of spin-axis tossing among LVK BH+BH mergers, supporting the conclusion of the early work by \citet{tau22}.
However, the robustness of this result depends on the mixing ratio between different BH+BH formation channels. If dynamical formation channels (in globular cluster or AGN disk environments) dominate, then the empirical LVK data is better explained without spin-axis tossing at BH formation.

Section~\ref{sec:stat} summarizes the test statistics applied in this work (further details are given in \ref{Appendix:A}), and in Section~\ref{sec:funboxplot}, we present our analysis of effective spin data from LVK. In Section~\ref{sec:MCMC}, we introduce and discuss our Monte Carlo simulations of synthetic populations of BH+BH mergers under various assumptions, including trials with and without spin-axis tossing, and testing for mixed populations of progenitor channels. We continuously interleave the discussion with results from our statistical analysis.
Our conclusions are summarized in Section~\ref{sec:conclusions}.


\section{Data statistics and Methodology}\label{sec:stat}
We begin this investigation with a description of our applied test statistics. Then we analyze the empirical LVK data in Section~\ref{sec:funboxplot} before simulating the underlying spin distribution through Monte Carlo simulations of the formation of BH+BH binaries in Section~\ref{sec:MCMC}.
The applied methodology is a logical stepwise process, following well-established techniques in population synthesis: we first constrain BH spin magnitudes using the empirical $\chi_{\rm eff}$ and mass distributions, carefully accounting for measurement errors. Later, in Section~\ref{subsec:final-sim}, we explicitly test the robustness of our results by re-running simulations using independent spin distributions retrieved by the LVK Collaboration.
This ensures that our results are not sensitive to initial spin assumptions, i.e. our results are not biased by separate spin magnitude fitting because we perform statistical tests (Kolmogorov-Smirnov, Cramér-von Mises, and Anderson-Darling tests) on the final outputs, ensuring independence.

It is not possible to fit all BH+BH parameters simultaneously in a reliable manner. While Bayesian hierarchical fitting, which could attempt this, can be useful in certain contexts, it is well known that such methods can produce misleading posteriors when applied to poorly constrained spin parameters \citep[e.g.][]{fsm+17,zbc+20}. 
Bayesian methodologies often suffer from severe limitations: strong prior dependence, overconfidence in poorly constrained parameters, sensitivity to likelihood assumptions, and difficulty in interpreting posterior probabilities. We believe that our chosen approach in this work provides a more transparent, straightforward and assumption-minimizing alternative.
We note that the posterior distributions of $\chi_{\rm eff}$ provided by the LVK Collaboration are obtained through Bayesian parameter estimation and thus depend on the choice of spin priors. Specifically, LVK analyses typically assume uniform priors on the spin magnitudes and isotropic spin directions. As such, the posterior samples we use throughout this work (e.g. in our statistical comparisons) are conditioned on these assumptions. While alternative, astrophysically motivated priors could shift the shape of the $\chi_{\rm eff}$ posteriors, our approach is consistent with the publicly released data products and thus valid within the adopted LVK framework.

We emphasize that BH spin tilts ($\Theta_1$ and $\Theta_2$) cannot be measured directly with current technology and sensitivity of GW detectors, nor can they be disentangled from the empirical distributions of $\chi_{\rm eff}$ and BH masses. For this reason, our study focuses on what can actually be measured and compared with simulations. (However, see Section~\ref{subsec:relative-BH-spins} for a discussion of the individual BH contributions to $\chi_{\rm eff}$.)

In \ref{Appendix:A}, we outline the statistical theory applied on the LVK data, and for later comparison between this empirical data and our simulations. Estimation of the optimal function that best describes the distribution of data points is of central relevance when comparing the simulated results with the observed values. 
Kernel density estimation (KDE) is the general framework that can be used for finding the optimal distribution of data that describes histograms.  
For a more meticulous analysis of band width estimation, we describe and compare the least-squares cross-validation (LSCV) method and the bandwidth selector method \citep{scikit-learn,Garcia-Portugues2023}. 
There are many other methods, but these two are the most frequent methods that are used. Further details are provided in \ref{Appendix:A} where we also discuss functional boxplots, which we apply in our investigation. 

\subsection{Test Statistics}
When comparing our Monte Carlo simulations of $\chi_{\rm eff}$ with the observational LVK data, it is important to provide a statistical measure of how well the simulated results fit the observed data. This can be achieved by different statistical hypothesis tests. Each method has its benefits and disadvantages. We consider three different tests --- two-sided Kolmogorov–Smirnov, Cramér–von Mises, and Anderson-Darling --- that we describe in \ref{Appendix:A} and apply in Sections~\ref{sec:funboxplot} and \ref{sec:MCMC}.


\section{Analysis of effective spin data from LVK}\label{sec:funboxplot}
The statistics briefly outlined in the previous section (see \ref{Appendix:A} for further details) will be applied here in Section~\ref{sec:funboxplot} to analyze and interpret the LVK BH+BH merger data. This will be done by first analyzing the measured effective spins, $\chi_{\rm eff}$, and then in Section~\ref{sec:MCMC}, exploring the $\chi_{\rm eff}$ distribution through Monte Carlo simulations of the formation of BH+BH binaries by examining the effect of the second supernova (SN) that produces the last-formed BH. 

\begin{figure*}
\hspace{1.0cm}
 \includegraphics[width=0.85\columnwidth]{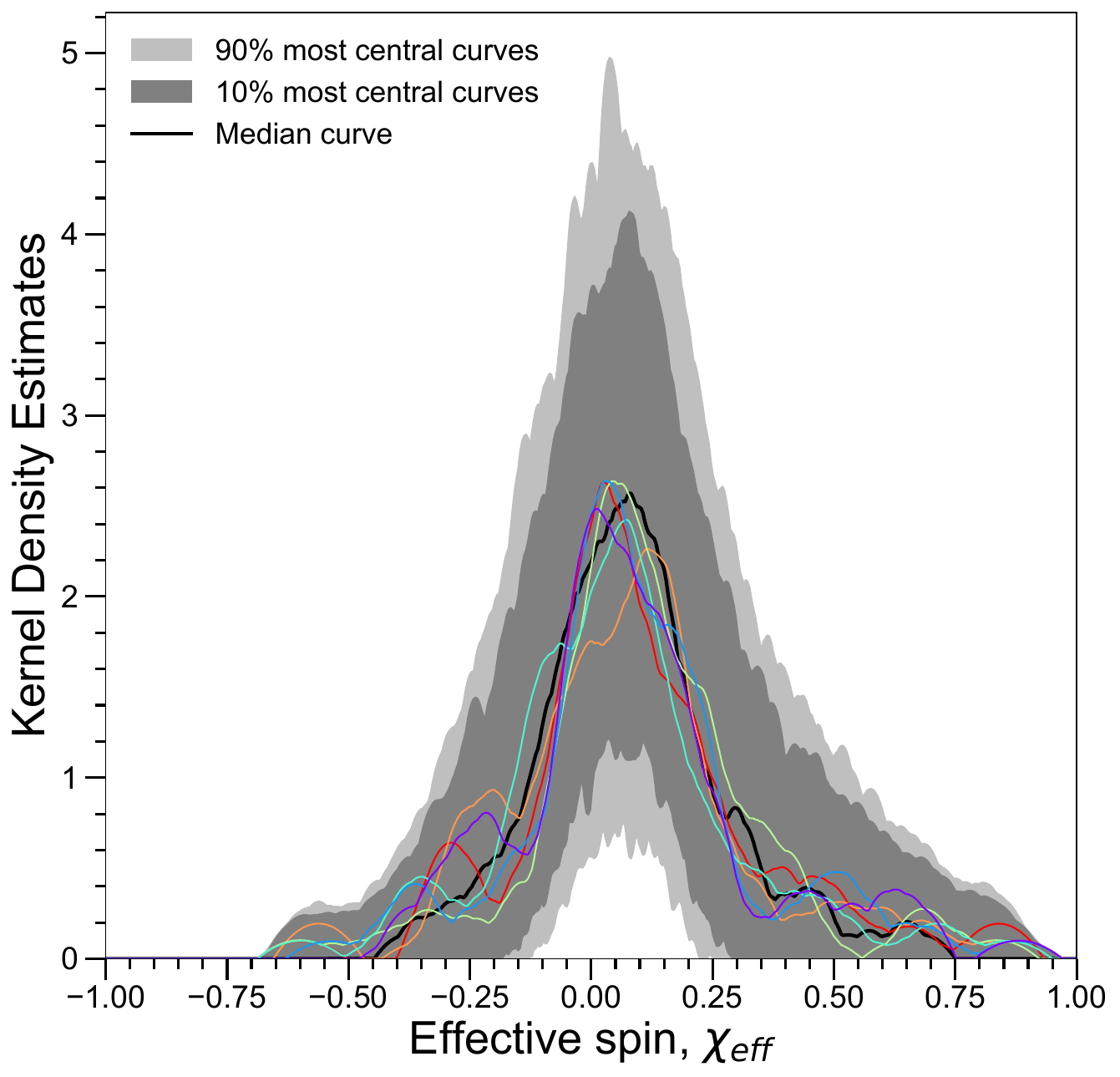}
\hspace{1.0cm}
 \includegraphics[width=0.85\columnwidth]{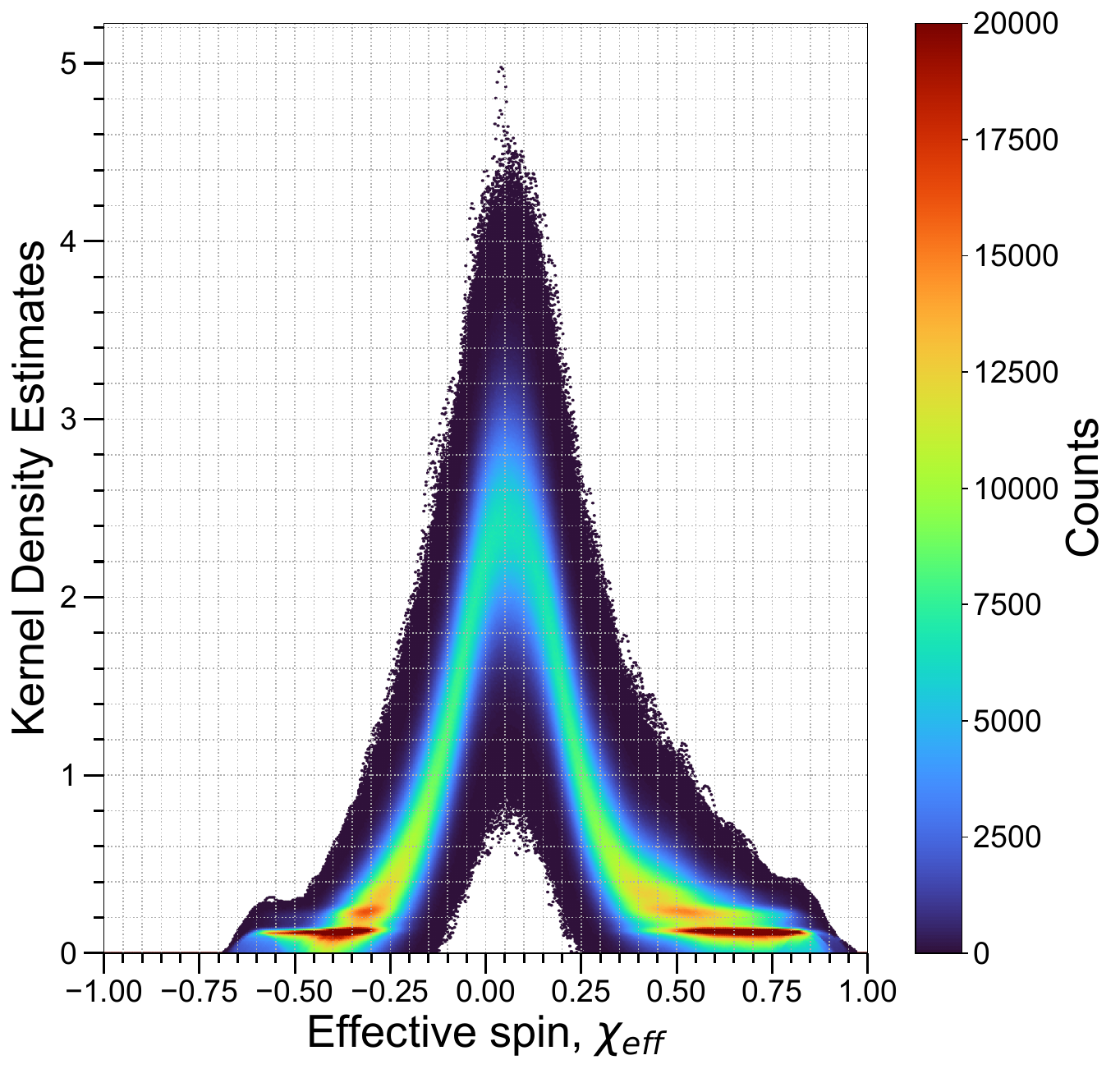}
\caption{(Left) Functional boxplot of $10^6$ curves, where each curve is a KDE of a permutation of the $\chi_{\rm eff}$ data, based on an Epanechnikov kernel. The bandwidth was calculated with the \texttt{scikit-learn} bandwidth estimator with $n=12$. 
Each data point was permuted within its 90\% confidence interval using a fitted extended skewed-normal distribution, as no universally optimal waveform model for estimating $\chi_{\rm eff}$ exists \citep{aaa+23}.
The depth of the curves was calculated using a MBD; the black curve is the one with the highest depth and is denoted the median curve. The dark gray region is the filled area between the 10\% most central curves, and the light gray region is the filled area between the 90\% most central curves. The colored curves are examples of outliers, e.g. near $\chi_{\rm eff}\simeq 1$. (Right) A binned scatterplot of the KDE's data points. Due to high densities in the tails, some pixels have more than 20\,000 counts.}
\label{fig:FuncBox_ESN}
\end{figure*}

\subsection{Effective spin data}
Parameterizing the $\chi_{\rm eff}$ distribution with a Gaussian fit (see \ref{Appendix:A}), calculating the mean, median, and standard deviation of the 83 LVK data points shown in Fig.~\ref{fig:GWTC3_BH+BH} yields: $\mu = 0.09$, $m= 0.06$ and $\sigma = 0.17$, respectively. 
The difference in mean and median implies that the data are somewhat right-skewed, as the mean lies at a higher value than the median. 
This is also supported by the skewness parameter, calculated using the following relation \citep{oa12}:
\begin{equation}
    \alpha = \sum_{i=0}^n \frac{(X_i-\mu)^3}{n\,\sigma^3}\,.
\end{equation}
Here, $n$ is the sample size and $X_i$ are the data points of the sample. 
This yields a skewness of $\alpha = +1.18$ for the LVK data, meaning that the observed effective spin distribution is skewed to the right (as verified from a visual inspection of Fig.~\ref{fig:GWTC3_BH+BH}). 

Originally, after LVK observing runs O1, O2 and O3a, in \citet{Abbott_2021}, it was found that the minimum value of the $\chi_{\rm eff}$ distribution is less than 0 at 99\% credibility. However, in \citet{Roulet+2021} it was argued that there is no evidence for negative $\chi_{\rm eff}$ values. 
In principle, it is correct that we cannot definitely point to an individual BH+BH merger with $\chi_{\rm eff}<0$, since all observed mergers have confidence intervals with at least some part in the positive $\chi_{\rm eff}$ value interval. As an example, the observing run O3b candidate with the most significant support for $\chi_{\rm eff}<0$ is GW191109\_010717 with 90\% probability \citep{aaa+23}. 
However, the probability that {\it all} 83 measured effective spins have $\chi_{\rm eff}\ge0$, assuming, for simplicity, a uniform distribution over the credibility intervals and independent probabilities for individual events, results in a combined probability of only $3.01\times 10^{-19}$. Thus, the hypothesis that $\chi_{\rm eff}\ge0$ for all BH+BH merger events is difficult to reconcile with the data under these assumptions.

This is, of course, only a first-order estimate, as the true underlying probability distributions for the LVK $\chi_{\rm eff}$ values are not uniform. However, even using the true distributions, we do not expect the probability to be substantially higher.
The low probability arises simply because, as soon as part of the credibility interval for a given data point is $<0$, the probability of $\chi_{\rm eff}\ge 0$ for that event drops below 100\%. In fact, this is the case for 66 merger events out of 83, with 28 of these events having a median value of $\chi_{\rm eff}<0$.
Therefore, we expect an extremely low probability that all data points would have $\chi_{\rm eff}\ge 0$, even when using the ``true'' (best LVK posterior) distribution. 
Note that the posterior PDF depends on the waveform model used in the parameter estimation. This is a well-known and important limitation, as no universally optimal waveform model exists \citep{aaa+23}. This limitation motivated us to permute each data point $10^6$ times within its 90\% confidence interval using a fitted extended skewed-normal distribution (Fig.~\ref{fig:FuncBox_ESN}).

The $\chi_{\rm eff}$ posterior samples used in this work are based on the underlying waveform models adopted by the LVK Collaboration. For GWTC-3 \citep[see][for a detailed description]{aaa+23}, these are the \texttt{IMRPhenomXPHM} and \texttt{SEOBNRv4PHM} waveform models, both of which incorporate precession and inspiral–merger–ringdown consistency. These waveform choices are documented in the corresponding LVK data releases. While systematic differences between waveform models can affect spin measurements, these differences are generally small for the bulk of the LVK BH+BH population \citep[e.g.][]{ccg+19}. Therefore, we do not expect such effects to significantly bias our comparative statistical analysis.

The question of whether or not the distribution of effective spin is symmetric around $\chi_{\rm eff}=0$ is somewhat more straightforward to answer. \citet{Abbott_2021} argue for a non-symmetric distribution. Although solutions where the mean value of all BH+BH mergers is close to $\chi_{\rm eff}=0$ do exist, the conclusion in \citet{Roulet+2021} is that the distribution is not symmetric around $\chi_{\rm eff}=0$ at 95\% credibility because the number of systems with more-or-less aligned spins dominate over the ones with anti-aligned spins. This argument of more-or-less aligned spins, and thus $\chi_{\rm eff}>0$ for a given merger, closely follows expectations from isolated binary star evolution \citep[see e.g.][]{kal00,fsm+17}.
Similarly, \citet{Abbott_2021} argued that the distribution of $\chi_{\rm eff}$ is not symmetric because it represents a mix of sources that have i) a symmetric distribution around 0, i.e. as expected from a dynamical BH+BH formation channel in a dense stellar environment \citep[e.g. via dynamical interactions in globular clusters,][]{mo10,rzp+16}; and ii) a population of sources having only positive values of $\chi_{\rm eff}$. The latter is the case for isolated binary star evolution with small to moderate BH kicks and without the possibility of BH spin-axis tossing at birth \citep{tau22}. Evidence for the latter effect, however, will be investigated in Section~\ref{sec:MCMC}.

Based on both the current statistics after LVK observing runs O1--O3, where we find a skewness parameter of $\alpha = +1.18$, and the above evolutionary arguments of a mix of progenitor populations, it is expected that the distribution of $\chi_{\rm eff}$ is not symmetric. 
This preliminary finding comes with the caveat, of course, that we have not yet considered the error for each data point in these calculations, and that should still be taken into account when choosing a KDE for the $\chi_{\rm eff}$ data. To account for the error of each data point, we now apply functional data analysis to get an overview of the effect it has on the shape of the KDEs. 

\subsection{Errors of effective spins and fits}\label{subsec:ErrorChiEff}
To demonstrate the effect that errors have on our KDE curves, functional boxplots of $10^6$ trials have been made based on the $\chi_{\rm eff}$ data. Each data point has been selected within its credibility interval, either with a uniform distribution or an extended skewed normal (ESN) distribution. An Epanechnikov kernel \citep[][see also \ref{appendix:KDE}]{epa69} has been used for both cases to avoid oversmoothing of data.

Figure~\ref{fig:FuncBox_ESN} (left panel), shows a functional boxplot of $10^6$ curves, where each curve is a KDE of a permutation of $\chi_{\rm eff}$ data using fitted ESN distributions to the credibility intervals of each data point. 
The depth was calculated using modified band depth (MBD, see \ref{subsec:functionalboxplot}), as it is less sensitive to shape outliers, which can easily arise due to the variation of data points within their respective credibility intervals.
A region made up of the area (light gray) between the 90\% most central curves and the area with the 10\% most central curves (dark gray) is shown in the plot. 
(Due to the large number of curves, traditionally chosen areas of the 50\% and 90\% most central curves, are almost identical. However, it is still important to have many curves as this will average out some of the variations that arise from the different permutations, such that we can determine the confidence bounds of our fits for the $\chi_{\rm eff}$ data.)

Furthermore, in the left panel of Fig.~\ref{fig:FuncBox_ESN}, we also see some very wide bounds created by the 90\% most central curves. However, this was to be expected when considering the tendency for somewhat large confidence intervals of the data points. 
Many different fits can be made within the confidence bound. 
For example, we can construct a symmetric distribution around 0, although the likelihood of this is very low, particularly when considering higher probabilities concentrated near the median of the confidence intervals for each data point. 
However, we cannot make a fit that excludes negative $\chi_{\rm eff}$ values. 
The median curve (black) peaks at $\chi_{\rm eff}>0$, which is in good agreement with our findings from a simple Gaussian fit and the related skewness. We have plotted examples of outlier curves in colors, some of which appear to have multiple modes.

The right panel of Fig.~\ref{fig:FuncBox_ESN} shows a binned scatterplot, included to highlight regions of high curve density not easily seen in the functional boxplot on the left. High densities appear in the tails, as expected, since most curves begin and end in similar regions. This indicates that the outer bounds of the functional boxplot are shaped by curve extremities.

For the functional boxplot seen in Fig.~\ref{fig:FuncBox_ESN}, we fitted ESN distributions to the confidence intervals of each data point in order to gain a better approximation of the underlying density functions. Some of these distributions are illustrated in fig.~7 of \citet{aaa+23}. The formula for the ESN distribution is given by \citep{ESN}:
\begin{align}
    f(x; \alpha, \tau)= \phi(x) \;\frac{\Phi(\tau \sqrt{1+\alpha^2}+\alpha x)}{\Phi(\tau)}\,,
\end{align}
where $\phi$ is the standard normal probability distribution function (PDF), $\Phi$ is the standard normal cumulative distribution function (CDF), and $\alpha$ and $\tau$ are two parameters that can be varied to optimize the fit. To be more specific, we can transform $x\rightarrow (x -\xi )/\omega$ and adjust $\xi$ to move the distribution to the corresponding confidence interval, and alter $\omega$ to gain a reasonable width of the distribution within the confidence bounds. Here $\xi$ and $\omega$ can be seen as analogues to the mean and the standard deviation, respectively, of a normal distribution.
The value $\xi$ was adjusted such that the median of the distribution is at the true location in accordance with the LVK data, and afterwards the parameters $\tau$ and $\omega$ were adjusted such that the area of the region under the lower bound and over the upper bound is 0.05. 
This process was then repeated until the area under the PDF from the lower bound to the median is equal to the area under the PDF from the median to the upper bound. 
To speed-up the computation time, some error was allowed in the algorithm such that the area under the different regions were within $\pm 1.5\%$ of the desired values. 
These distributions were then used as weights when generating values within the 90\% confidence intervals for the data points for the permutations.
It is clear that the distributions used for the confidence intervals affect our confidence bounds, and random effects of the different permutations could play a role in this boxplot too. 

For the subsequent analysis in this work, we chose the confidence bounds created by the 10\% most central curves in the functional boxplot where ESN distributions were used. 
The 10\% most central curves region was selected because it already covers a quite broad area, and we do not expect the true distribution to be as extreme as the confidence bound created by the 90\% most central curves allowed (but rather somewhere in between those bounds and the median curve.)

\subsubsection{Detection bias and $\chi_{\rm eff}$ sensitivity}\label{subsubsec:ChiEff-detection}
Systems with positive $\chi_{\rm eff}$ tend to have higher peak GW amplitudes and longer-duration signals compared to those with negative $\chi_{\rm eff}$. This is because positive $\chi_{\rm eff}$ indicates spins aligned with the orbital angular momentum, which reduces the rate of inspiral and results in a stronger and longer-lasting GW signal. Conversely, negative $\chi_{\rm eff}$ corresponds to anti-aligned spins, leading to a faster inspiral and a weaker signal. However, current observational data have not conclusively confirmed that this leads to any significant selection bias \citep[e.g.][]{vbt22}. For this reason, we have not accounted for this bias in our analysis presented here.


\section{Monte Carlo simulations}\label{sec:MCMC}
\begin{figure}
\centering
\includegraphics[width=0.44\textwidth]{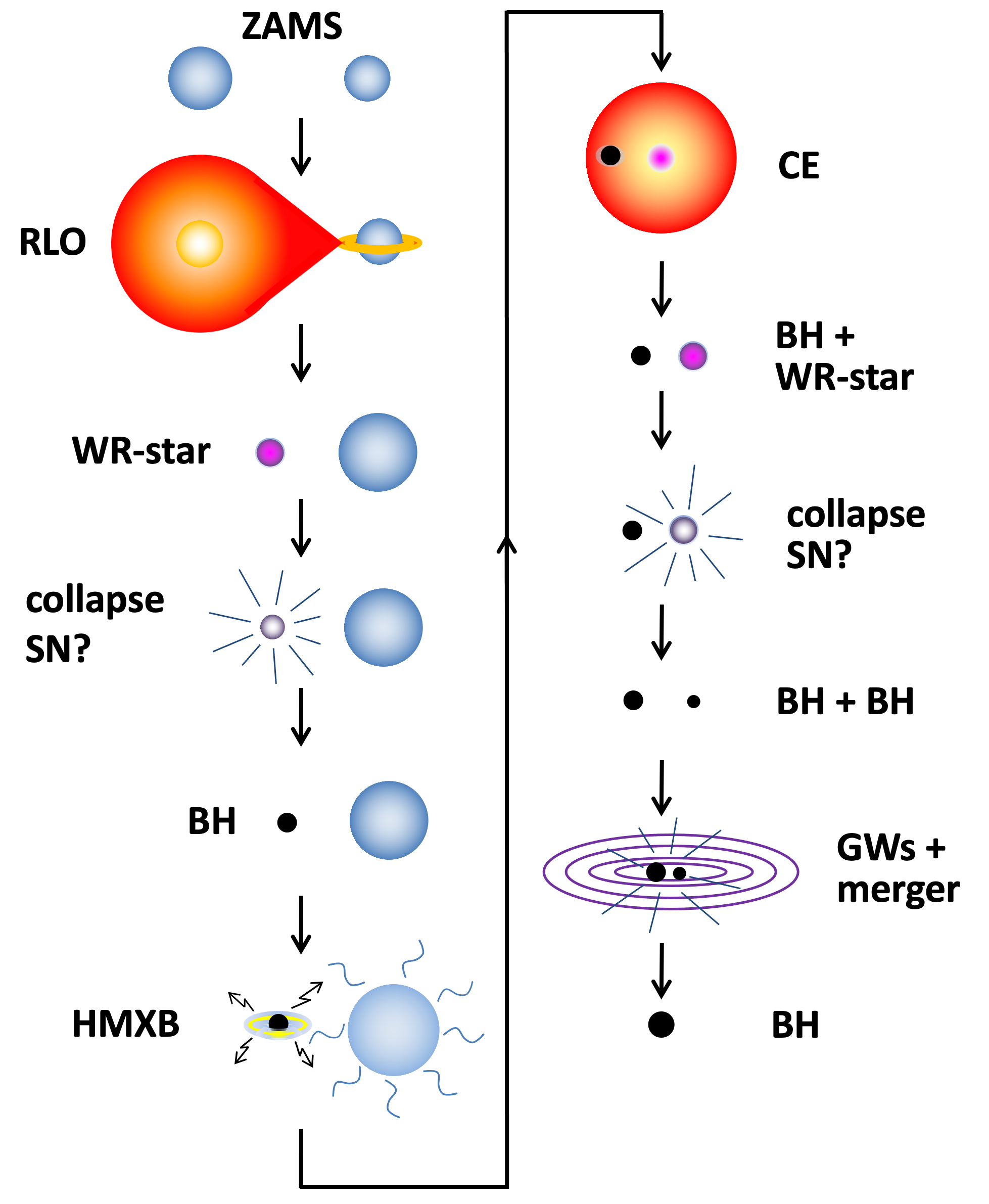}
\caption{Formation model of a double BH merger as a ﬁnal product of close massive binary star evolution. Acronyms --- ZAMS: zero-age main sequence; RLO: Roche-lobe overflow (mass transfer); WR: Wolf-Rayet star (He-star); SN: supernova; BH: black hole; HMXB: high-mass X-ray binary; CE: common envelope; GWs: gravitational waves. After \citet{tv23}.
\label{fig:vdH-BH+BH}}
\end{figure}

The standard formation scenario of a BH+BH system produced in an isolated binary system with common envelope spiral-in is illustrated in Fig.~\ref{fig:vdH-BH+BH}.  
To investigate the underlying distribution of the recorded LVK data on BH+BH merger systems, Monte Carlo simulations were performed on the second SN explosion in a synthetic population of progenitor binaries, following the recipe outlined in \citet{tkf+17,tau22,tv23}. The simulations presented here cover a 9-dimensional phase space with the following parameters: the masses of the two BHs, $M_{\mathrm{BH,1}}$ and $M_{\mathrm{BH,2}}$, their spins, $\chi_1$ and $\chi_2$, the pre-SN orbital separation, $a_i$, the magnitude of the kick velocity, $w$, and two kick angles $\phi$ and $\theta$. 
The last parameter is the tossing angle of the spin axis of the second-born BH, $\Phi_2$. The BH masses and spins are discussed in Sections~\ref{subsec:BH-masses} and \ref{subsec:BH-spins}. The values of $a_i$ and $w$ are either fixed or assumed to follow a uniform PDF within a given interval (see the discussions in Section~\ref{subsec:kick-separation}).
The kick angles $\theta$ and $\phi$ are chosen so that the kick directions are isotropic (i.e. uniformly distributed on a sphere). 

For isolated BH+BH merger systems, the tilt angles of the two BH spins by the time the system merges ($\Theta_1$ and $\Theta_2$, see Eq.~\ref{eq:chi_eff2} and Fig.~\ref{fig:illustration_chi_eff}) are derived as follows \citep[see][for further arguments]{tau22}. 
The tilt angle $\Theta_1$ of the spin axis of the {\bf first-born BH} ($\vec{\chi}_1$) is equal to the misalignment angle, $\delta$ caused by a momentum kick imparted by the second SN in the system, i.e. $\Theta_1=\delta$. The reason is that at the time of the stellar collapse of the second helium star, the first-born BH will possess prograde spin with its spin vector ($\vec{\chi}_1$) parallel to the pre-SN orbital angular momentum vector, $\vec{L}_0$ as a result of accretion \citep[e.g.][]{bp75,klop05}. 
For the spin tilt angle of the {\bf second-born BH}, we consider two cases: 
i) in the case of {\em no tossing}: $\Theta_2 = \Theta_1=\delta$ (this follows from the assumption that tidal torques will align the spin axis of the collapsing helium star with that of $\vec{L}_0$); ii) if the secondary BH is subject to {\em tossing} of its spin axis during its formation, then $\Theta_2$ may take a value such that the direction of $\vec{\chi}_2$ deviates significantly from that of $\vec{\chi}_1$.
The PDF for the degree of tossing is assumed to follow an isotropic distribution of post-SN spin-axis directions: $P(\Theta_2)=\frac{1}{2} \sin (\Theta_2)$, where $\Theta_2 \in [0,180^\circ]$. 

In Section~\ref{subsec:spin-axis-tossing}, we explore isolated binaries with either non-isotropic BH spin-axis tossing or without any spin-axis tossing at all (the default option in the current literature) and demonstrate that they cannot reproduce empirical data (but see Section~\ref{subsubsec:dynamical-results}). 
The question of mass reversal in isolated binaries is investigated in Section~\ref{subsec:mass-reversal}; and in Section~\ref{subsec:final-sim} we present our best simulations of isolated systems when applying improved BH spin distributions. Finally, in Section~\ref{subsec:dynamical}, we explore the possibility that the empirical LVK sample is composed of a mixture of binaries that originate from either isolated binaries or were produced via dynamical interactions (e.g. in globular clusters and AGN disks).

The formation of BH+BH merger candidates requires that the systems are not disrupted in the SN, and that the post-SN systems merge within a Hubble time ($\sim 13.7\;{\rm Gyr}$). For a given bound post-SN system, we calculate the merger time following \citet{pet64}.
We simulated in each setup a number of systems $\mathcal{O}(10^6)$ until we have $N=100\,000$ post-SN systems that successfully become GW merger sources within a Hubble time. 

The simulation methodology and setup applied here is explained in detail in Sections~\ref{subsec:BH-masses}--\ref{subsec:dynamical} and similar to those presented in \citet{tau22}, except for the following:
\begin{itemize}
    \item BH masses are selected here based on LVK data.
    \item BH spins $(\chi_1,\,\chi_2)$ are chosen from $\chi_{\rm eff}$-fits to LVK data\\(or following LVK spin components, cf. Sections~\ref{subsec:final-sim}--\ref{subsec:dynamical}).
    \item Mass reversal is explored more thoroughly.
    \item Both isolated and dynamical origins are considered.
\end{itemize}
Mass reversal refers to the situation where the most massive BH is born second as a consequence of mass transfer in the progenitor binary, originally with two ZAMS stellar components with somewhat similar masses \citep[see e.g.][for the lower mass analogue case of $\beta$-Lyrae]{dgl94}.

\subsection{Black hole masses}\label{subsec:BH-masses}
\begin{figure*}
\hspace{1.0cm}
  \includegraphics[width=0.90\columnwidth]{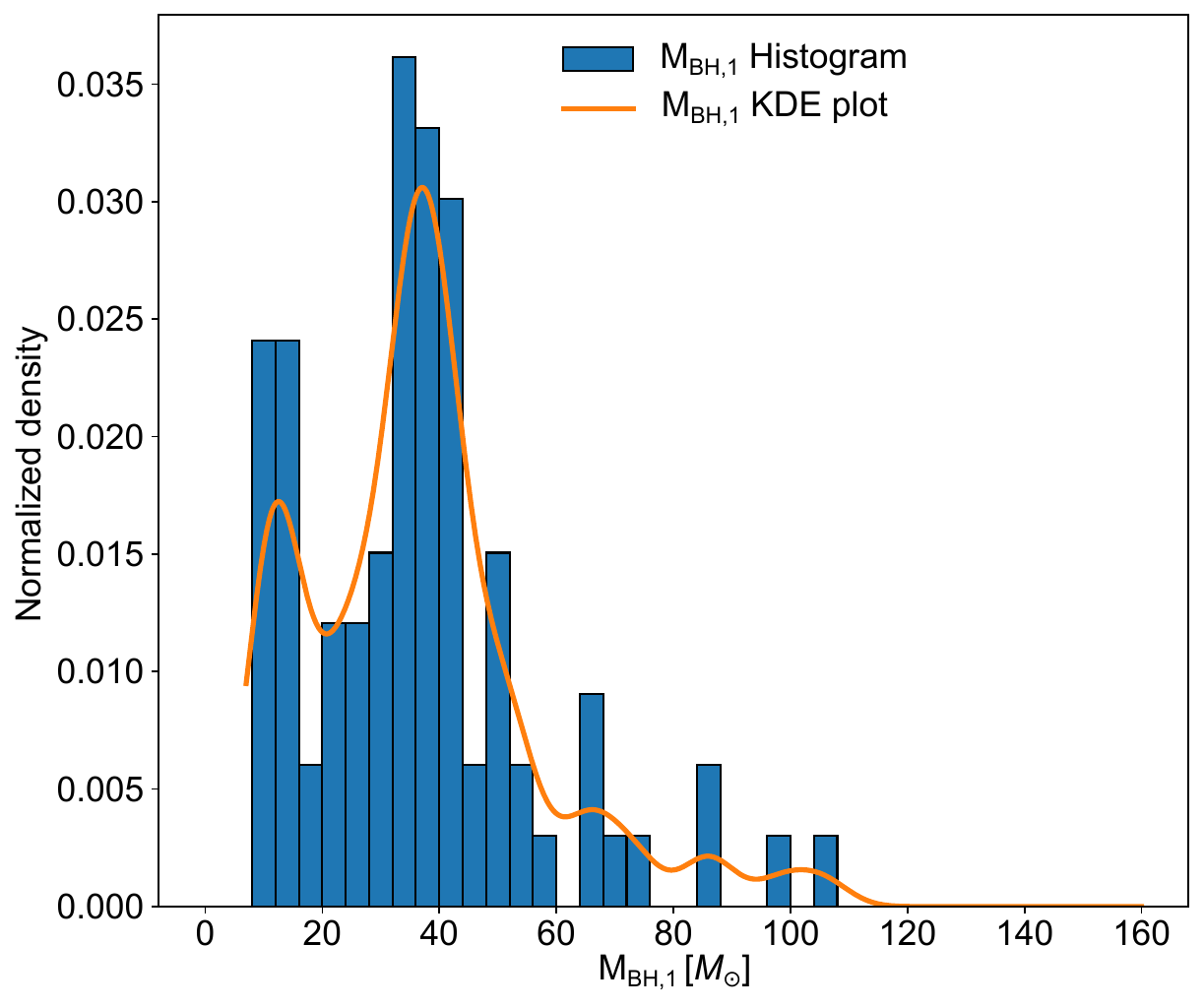}
\hspace{1.0cm}
  \includegraphics[width=0.85\columnwidth]{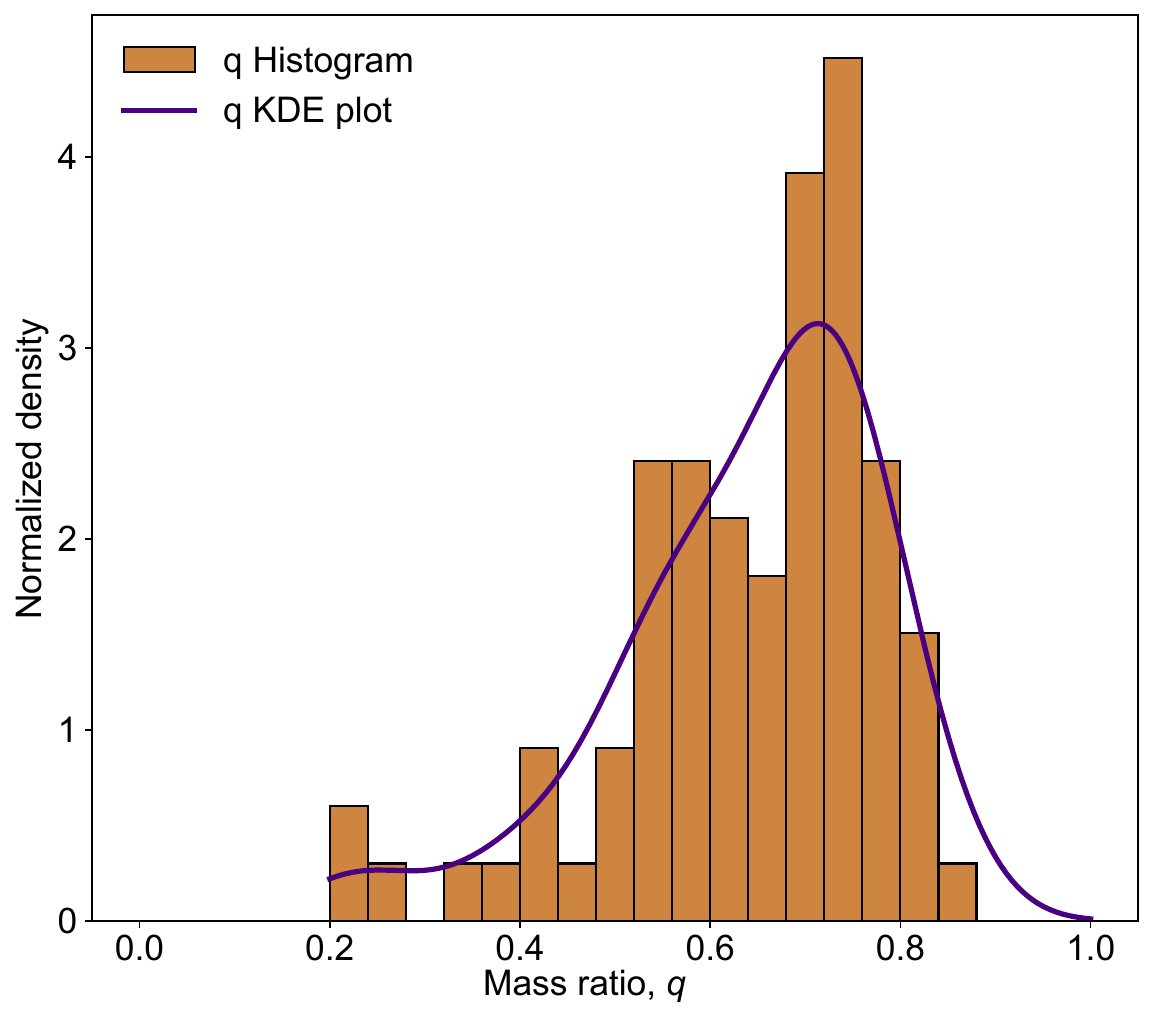}
\caption{Distributions of $M_{\mathrm{BH,1}}$ (left) and mass ratio $q$ (right) based on empirical LVK data. The plots show smooth kernel density estimates (KDEs; solid curves) over the ranges $M_{\rm BH,1} \in [7, 160];M_\odot$ and $q \in [0.2, 1]$, which are used in our simulations for selecting BH masses (see Figs.~\ref{fig:Bimodal2D} and \ref{fig:BimodalMBH1MBH2}) to mimic the empirical masses.}
\label{fig:KDEMBH1andq}
\end{figure*}
\begin{figure}
\centering
 \includegraphics[width=0.78\columnwidth]{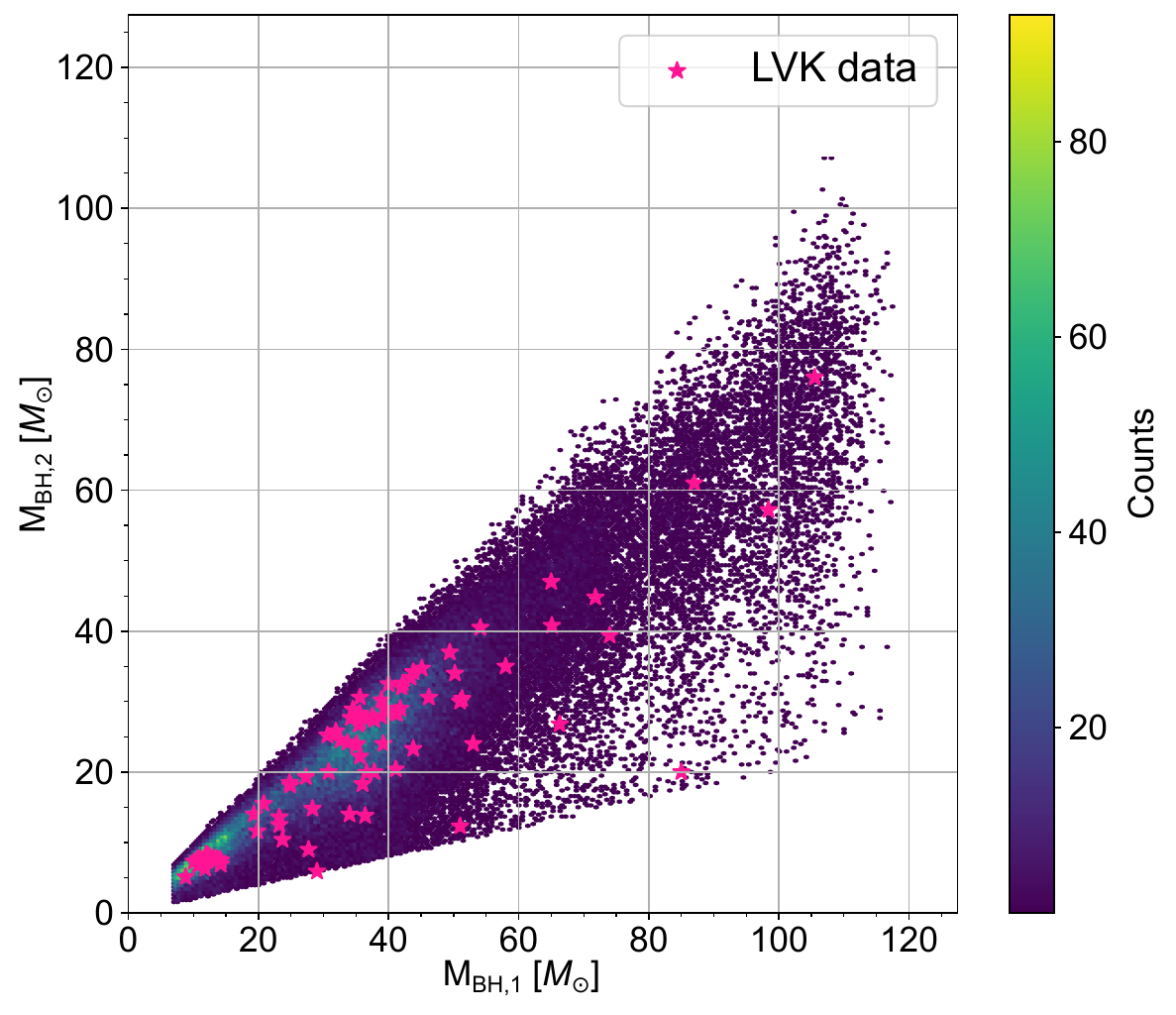}
 \caption{Binned scatterplot of the selected distribution of BH+BH mass pairs using $KDE(\text{LVK data})$, see Fig.~\ref{fig:KDEMBH1andq}. The magenta stars show the empirical median-value masses from the individual GW mergers in the LVK data set. The number of simulated systems plotted here is $N = 10^5$.}
 \label{fig:Bimodal2D}
\end{figure}
\begin{figure*}
\hspace{1.0cm}
 \includegraphics[width=0.85\columnwidth]{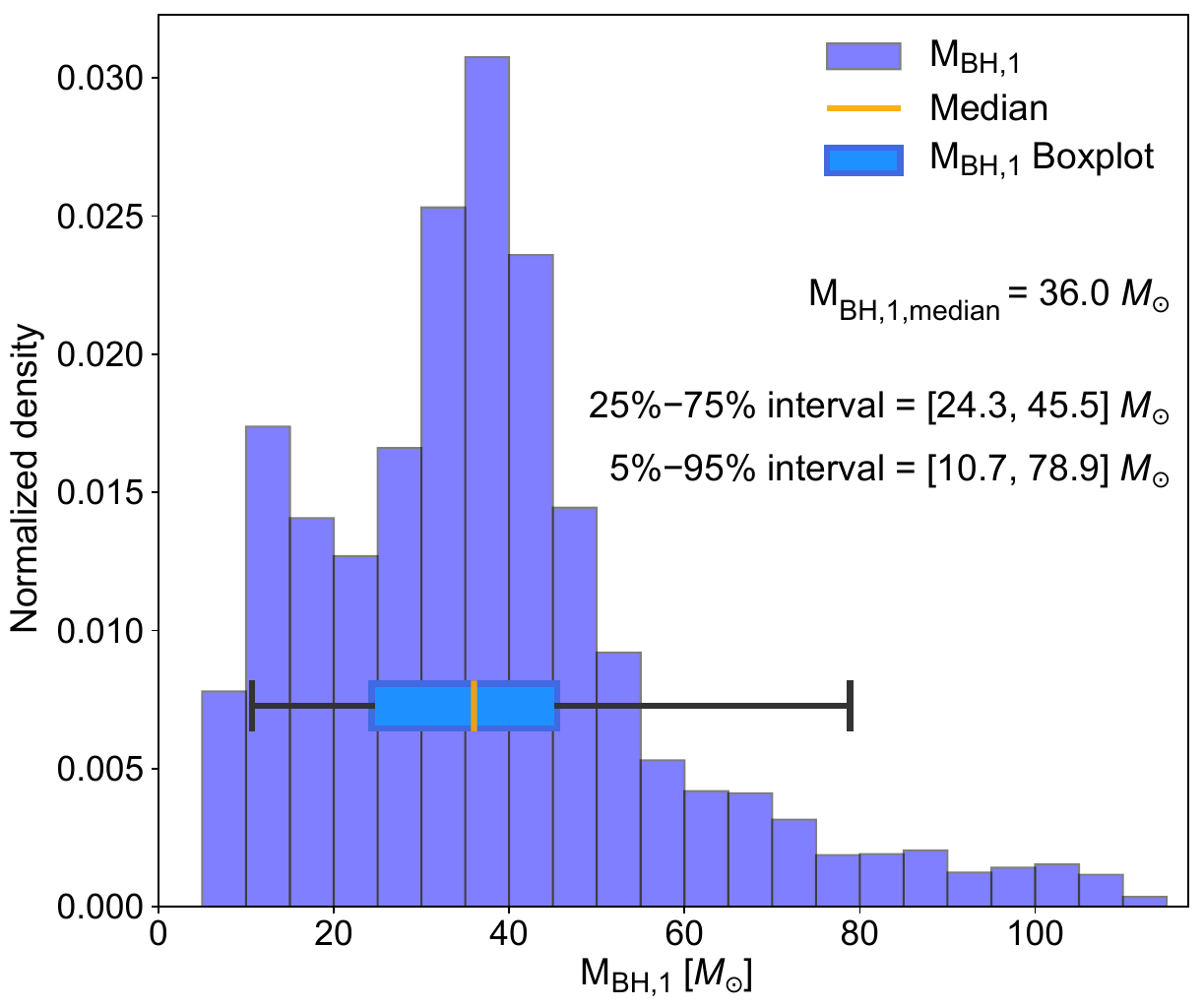}
\hspace{1.0cm}
 \includegraphics[width=0.85\columnwidth]{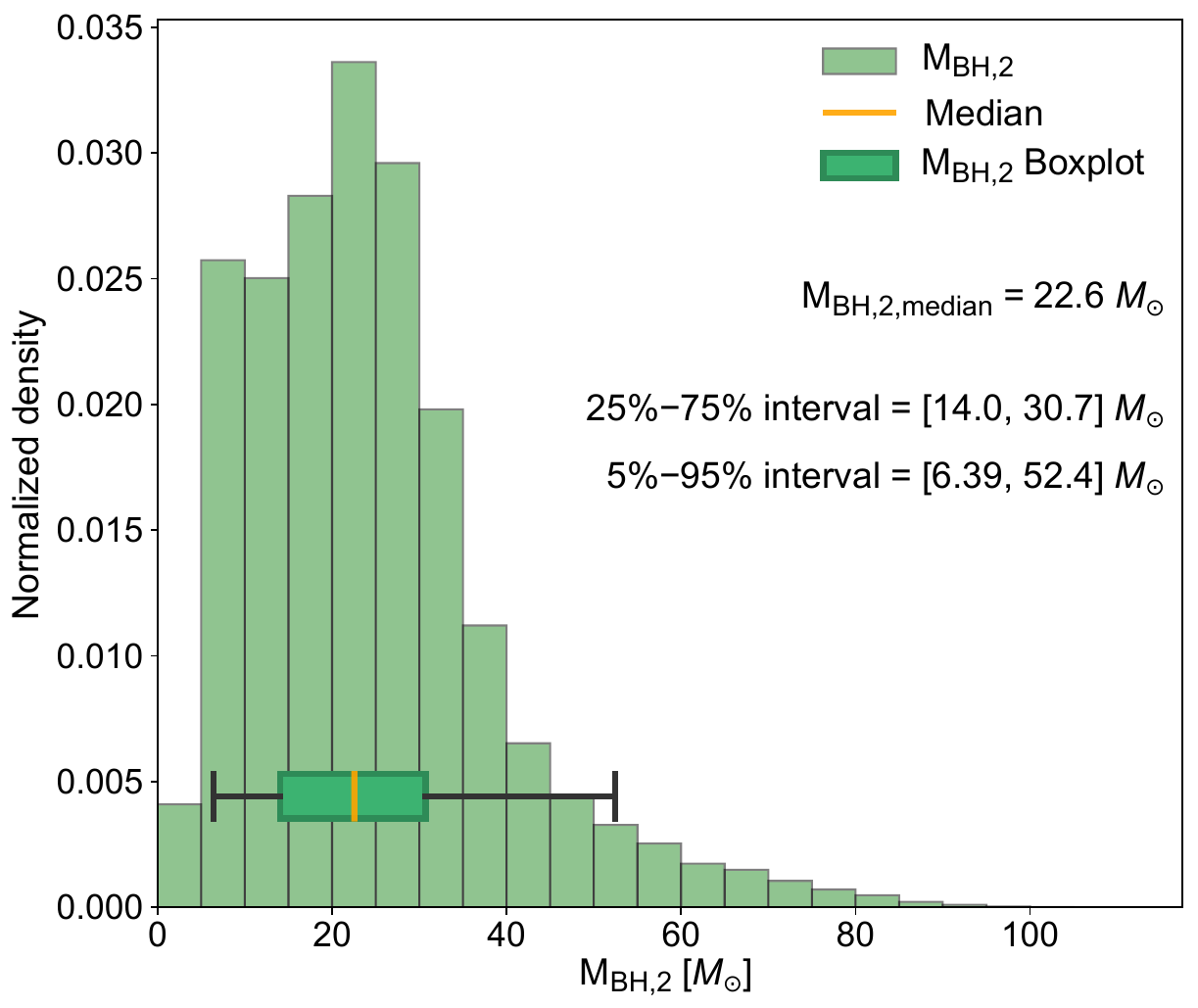}
\caption{Distribution of BH component masses $M_{\rm BH,1}$ and $M_{\rm BH,2}$ (left and right panels, respectively) selected for our simulations using empirical LVK data, based on $KDE(\text{LVK data})$ shown in Fig.~\ref{fig:KDEMBH1andq}. The number of shown selected BH mass pairs is $N=10^5$.}
\label{fig:BimodalMBH1MBH2}
\end{figure*}

The BH masses for the simulations are not trivial to choose and some considerations must be made due to both selection effects and because the underlying distribution is poorly known. BH masses can be chosen from either a binary evolution point of view \citep[e.g.][]{spl23}, BH formation physics \citep{wsj20,bwv24}, or directly from empirical BH data. In the latter case, masses can be distributed in pairs according to the LVK data, for example, using an empirical probability distribution function, $ePDF(\text{LVK data},p)$ with an allowed perturbation that follows a uniform distribution over a percentage, $p$ of individual BH mass credibility intervals. Alternatively, for a smoother mass distribution, we can estimate the PDF of both $M_{\rm BH,1}$ and $q = M_{\rm BH,2}/M_{\rm BH,1}$ using KDE applied to the LVK data, as explained in \ref{appendix:KDE}. The resulting distribution is thus both smoothed out and exhibits a bimodal pattern, while complying with the underlying mass ratios from the LVK data. 
The Gaussian KDE of $M_{\mathrm{BH,1}}$ and $q$ are shown in Fig.~\ref{fig:KDEMBH1andq}, including the histograms of the LVK data. We notice a bimodal mass distribution of $M_{\rm BH,1}$ with peaks near $\sim 10\;M_\odot$ (minor) and $\sim 36\;M_\odot$ (major).
The mass ranges were chosen such that $M_{\rm BH,1}$ is between $7-160\;M_{\odot}$, thereby including all primary BH masses in the LVK data set. Furthermore, the $q$-value ranges between $0.2 - 1$, where a lower bound of $0.2$ was chosen as this is the lowest ratio from the median values of the empirical LVK data. This way of distributing the masses yields the distribution of BH+BH pairs shown in Fig.~\ref{fig:Bimodal2D}. The individual mass distributions can be seen in Fig.~\ref{fig:BimodalMBH1MBH2}.
Notice, this prescription may result in a BH mass of less than $3.6\;M_\odot$ ($3.0\;M_\odot$) in 0.54\% (0.26\%) of the cases.

We assume that 20\,\% of the mass of the collapsing He-star is lost during its collapse to a BH (i.e., $M_{\rm BH,2} = 0.8 \times M_{\rm He}$), primarily due to the release of gravitational binding energy. While this fraction may be somewhat high --- partly depending on fallback and how efficiently neutrinos escape during the collapse \citep[see e.g.][for discussions]{mhlc16,wsj20} --- it provides a reasonable estimate. Importantly, as our simulated $\chi_{\rm eff}$ distributions are only weakly sensitive to the exact BH mass distribution \citep{tau22}, using a smaller mass-loss fraction, such as 5\,\%, leads to very similar results.

\subsubsection{Selections effects and parameter correlations}\label{subsubsec:selection}
We note that selection effects are at work but mitigated in our results: While LVK-detected masses include selection effects (e.g. heavier BHs are overrepresented) and do not reflect the intrinsic astrophysical BH+BH population, this is precisely why we use empirical data directly in our simulations. To account for large uncertainties and minimize statistical bias, we apply functional data analysis (Section~\ref{sec:funboxplot}). This ensures that our simulated systems (and consequently the $\chi_{\rm eff}$ distribution) match observations rather than relying on population synthesis models with significant uncertainties. Thus, rather than introducing selection bias, our approach explicitly avoids it.

Evidence suggesting an anti-correlation between effective spin ($\chi_{\rm eff}$) and BH mass ratio ($q$) has been presented by e.g. \citet{chn+21,alt23,bo24}; but see also \citet{hvb+24,hmv25}, who find less conclusive significance using flexible hierarchical models. For a brief discussions and the rationale for neglecting any potential correlation in our analysis, see \ref{Appendix:B}.

\subsection{Black hole spins}\label{subsec:BH-spins}
Before analyzing the simulation results, we first describe 
the optimization algorithm applied to find the best fitting parameters for the distributions of the individual BH spins $\chi_1$ and $\chi_2$.

\subsubsection{Optimization algorithm}\label{subsubsec:optimization}
Our applied optimization algorithm to fit the empirical distribution of individual BH spins, $\chi_1$ and $\chi_2$, is based on the findings of Section~\ref{sec:funboxplot} for the distribution of $\chi_{\rm eff}$. We now use these results to find a curve that is the best fit within the confidence bounds. This is done by assuming that the probability distributions of individual spin parameters follow a beta distribution, 
$P(\chi_i)\sim \beta(\alpha_i,\beta_i)$, where $i=\{1,2\}$ for the two BHs, and where the formula for the beta distribution is given by: 
\begin{equation}
    f(x;\alpha,\beta)=\frac{\Gamma(\alpha+\beta)}{\Gamma(\alpha)\Gamma(\beta)}\,x^{\alpha-1}(1-x)^{\beta-1}, \quad \alpha,\beta \in \mathbb{R}^+
\end{equation} 
where $\alpha$ and $\beta$ are shape parameters \citep{oa12}, and where $\Gamma$ is the gamma function:
\begin{equation}
    \Gamma(u)=\int_0^{\,\infty} t^{u-1} e^{-1} \,dt\,.
\end{equation}
Beta distributions were applied to account for skewed distributions in the optimization process. 
The mean value of a beta distribution is simply given by:
\begin{equation}
    E[X] = \frac{\alpha}{\alpha + \beta}\,.
\end{equation}
The calculations were done using the \texttt{SciPy} library in \texttt{Python}.
This popular method was also used in \citet{vlo19,aaa+19,aaa+23b}, since it is defined in the interval $0\leq x\leq 1$, just as the magnitude of the individual spin parameters are, and it is very flexible when it comes to the shape of the distribution.
Thus by using the predetermined variables from the Monte Carlo simulations, we are able to find a series of best fit solutions for the effective spin of the BH+BH systems by only varying $\alpha_{1,2}$ and $\beta_{1,2}$ in the beta distribution.

The best fit solutions are found by minimizing the root-mean-square error (RMSE) to a predetermined normalized curve that fits within the confidence bounds from the functional box plot. Six types of curves were investigated: "LSCV", "median", "peak", "low", "left", and "right". The LSCV and median curves were gathered directly from Section~\ref{sec:funboxplot}, while the rest (allowing to explore more extreme $\chi$ distributions) were constructed using the functions defining upper and lower bounds, together with a normal distribution in the following manner:
\begin{equation}
    H(x;\sigma,\mu) = \frac{U(x)-L(x)}{K}\frac{1}{\sigma\sqrt{2\pi}}\exp{\left[-\frac{1}{2}\left(\frac{x-\mu}{\sigma}\right)^2\right]}\,,
\end{equation}
where $K$ is a normalization constant, and $U(x)$ and $L(x)$ are the interpolated functions for the upper and lower bound, respectively. The result is a distribution which follows the shape of the confidence bounds, however with some flexibility concerning the shape and placement of the peak. 
By assuming a given mass distribution, it is then possible to gain information concerning the spin parameters of the individual BHs. Here we continue with our results obtained from the LSCV curve.

\subsubsection{Spin parameter distributions}\label{subsubsec:spin-param-dist}
Using the KDE distributed BH masses derived in Section~\ref{subsec:BH-masses} (Fig.~\ref{fig:Bimodal2D}), we are able to simulate different distributions of $\chi_{\rm eff}$. In order to minimize the degrees of freedom, we begin with the following assumptions for the simulations:
\begin{itemize}
    \item Pre-SN He-star mass of $M_{\rm He}=M_{\rm BH,2}/0.8.$
    \item Mass reversal is disregarded.
     \item Uniform pre-SN orbital separation of $a_i \in [4,40]\;R_\odot$.
    \item Constant SN kick of $w=50\;{\rm km\,s}^{-1}$ (isotropic direction).
    \item Spin-axis tossing of second-born BH (isotropic direction).
    \end{itemize}
These assumptions (e.g. spin-axis tossing) will be tested and adjusted later. We adopted the range of $a_i$ values following the arguments of \citet{tau22}: the lower limit is set to fit the size of the collapsing helium star within its Roche lobe, and the upper limit is chosen such that most of the surviving post-SN systems will remain in orbits tight enough ensure a BH+BH merger within a Hubble time. \citet{tau22} demonstrated that the value of $a_i$ is not important for the simulated  distribution of $\chi_{\rm eff}$ (unless BH kicks are extremely large {\em and} pre-SN orbits are very wide). Only the number of BH+BH mergers within a Hubble time decreases with increasing $a_i$.
Our new simulations confirm this conclusion (see Section~\ref{subsec:kick-separation}, where we include tests with $a_i \in [4,400]\;R_\odot$; however, note the caveat discussed there).

\begin{figure}
\centering
\includegraphics[width=0.38\textwidth]{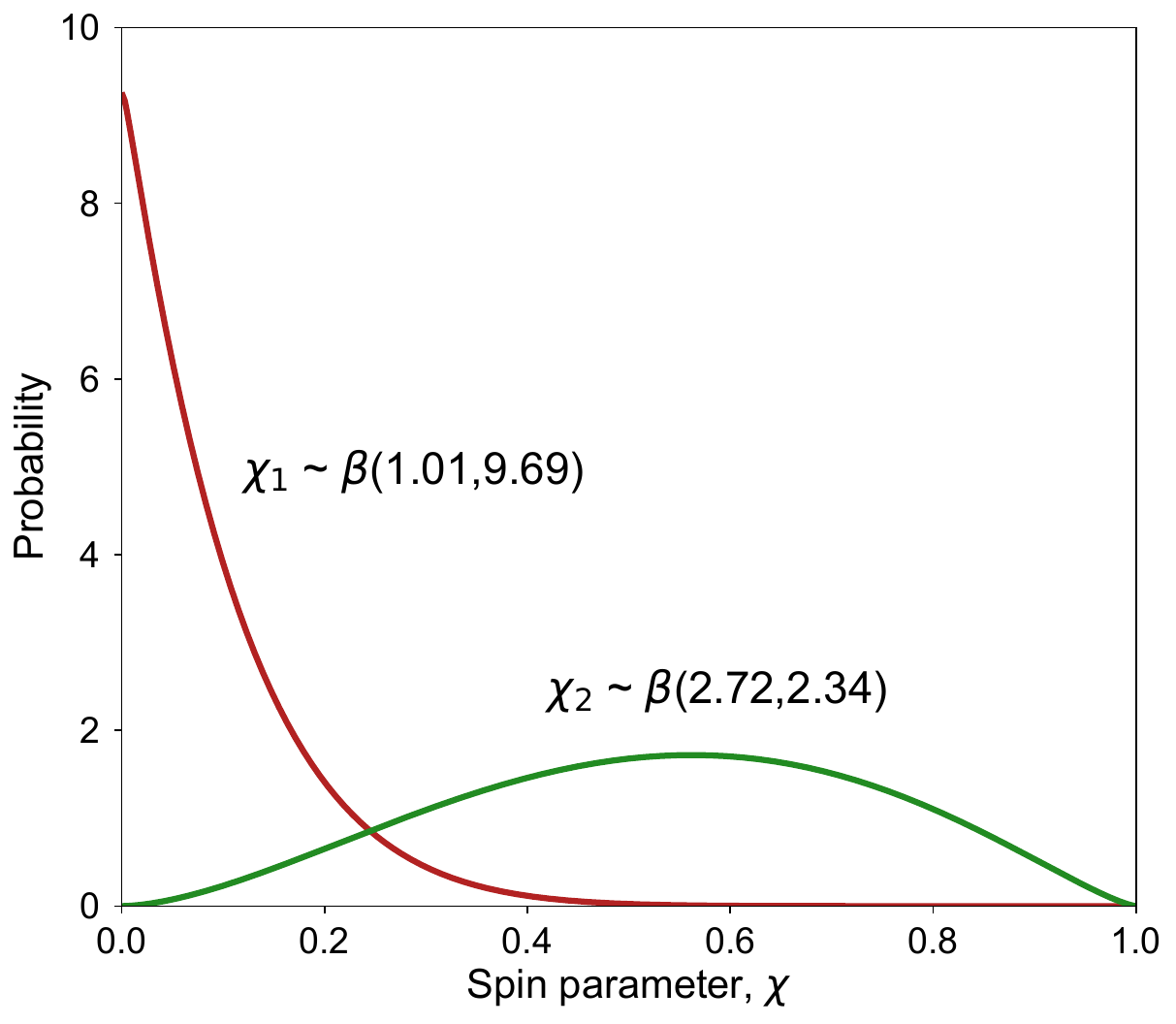}
\caption{Resulting $\chi_{1,2}$ distributions that yield the best-fit solutions, shown in panel~A of Fig.~\ref{fig:BimodalVarDistPlots}, between simulated values of $\chi_{\rm eff}$ and empirical data. See examples of other spin component distributions investigated in Fig.~\ref{fig:LVKchidits}.}
\label{fig:chi12-dist}
\end{figure}

Figure~\ref{fig:chi12-dist} shows the best fits for the $\chi_1$ and $\chi_2$ distributions that follow the beta distributions introduced in Section~\ref{subsubsec:optimization}. These results were obtained via iteration by optimizing the fit between the resulting simulated data and the empirical LVK $\chi_{\rm eff}$ data. These $\chi_{1,2}$ fits will be optimized further in Section~\ref{subsec:final-sim}.

\begin{figure*}
\hspace*{0.2cm}
\includegraphics[width=0.45\textwidth]{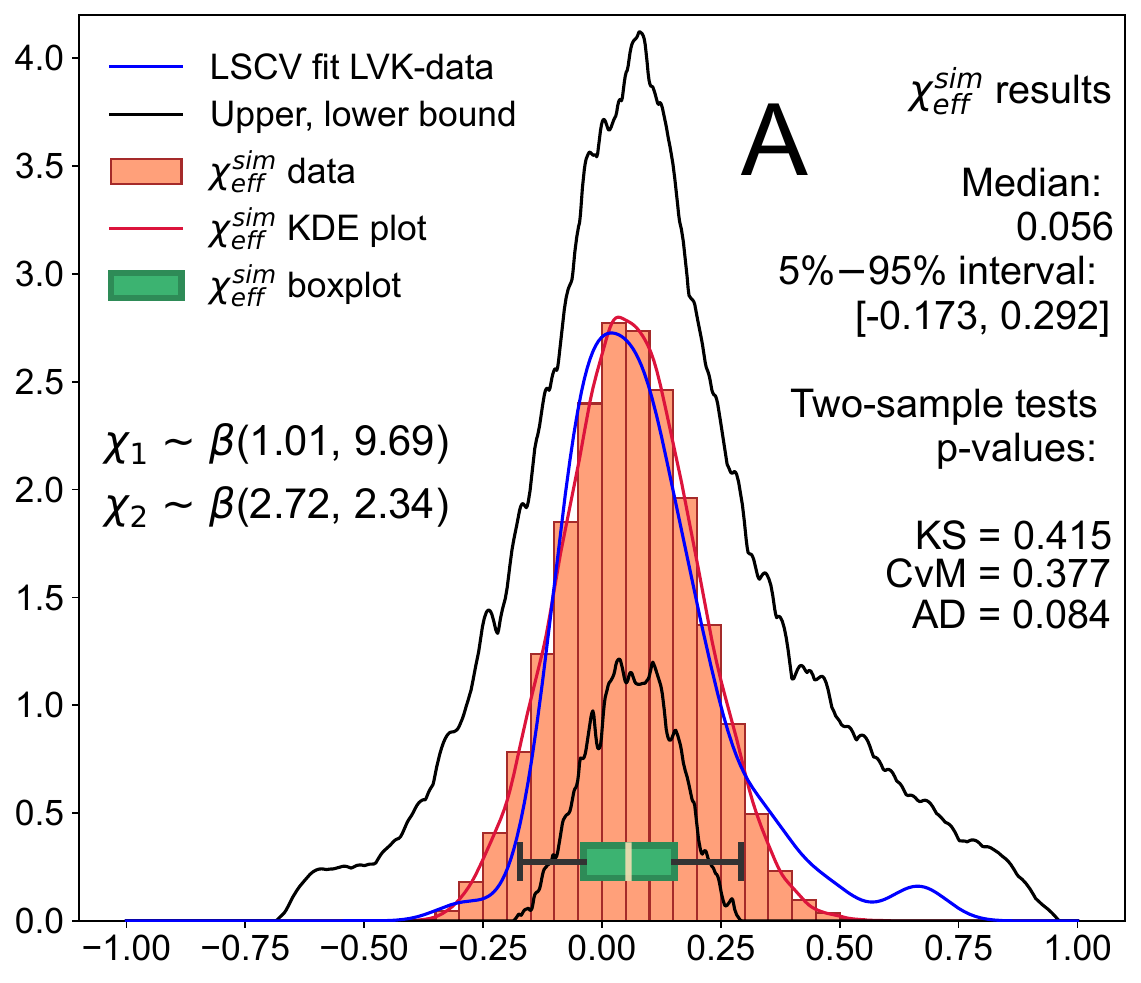}
\hspace*{1.0cm}
\includegraphics[width=0.45\textwidth]{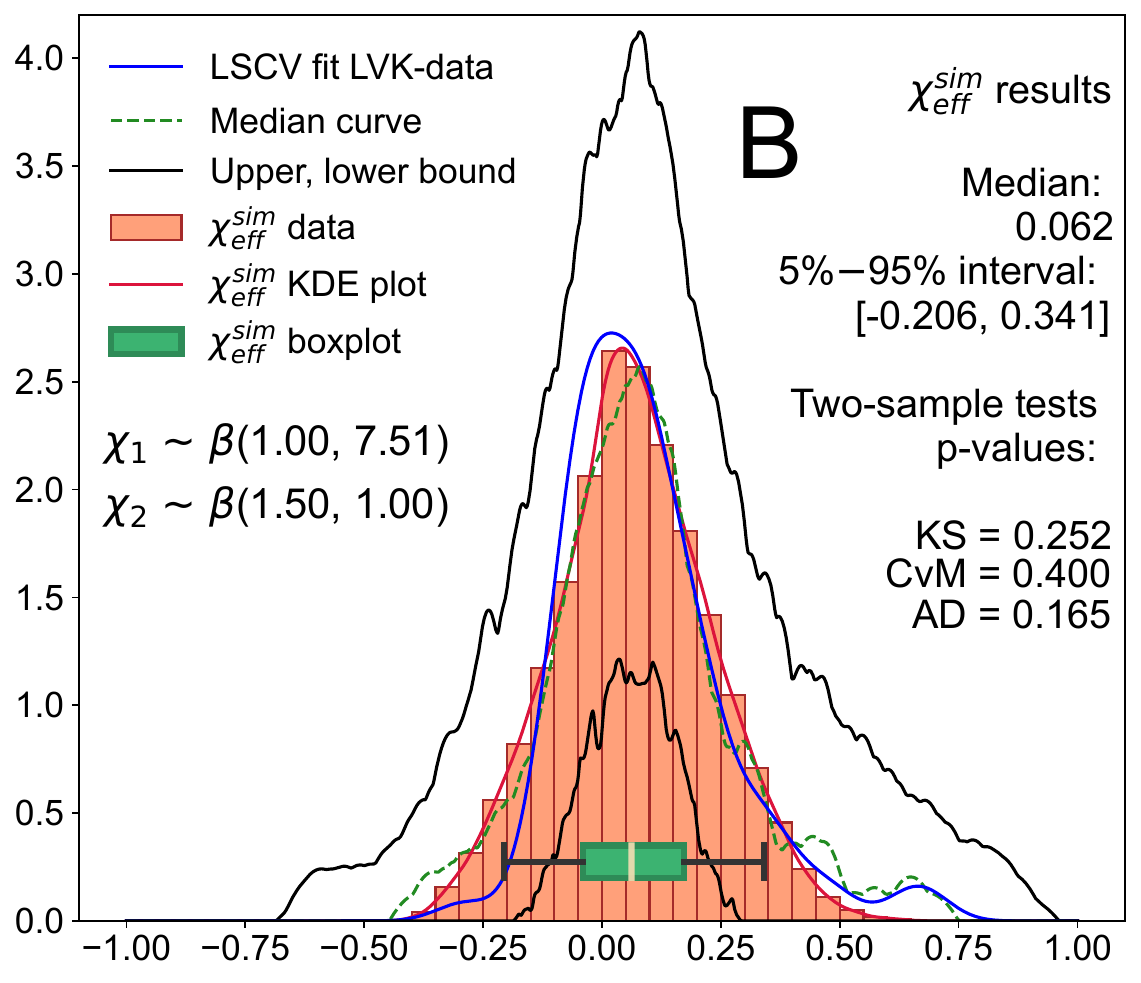}
\caption{Histograms of the distribution of $\chi_{\rm eff}$ (x-axis) when fitted to two different curves within the 90\% range (black lines) of the functional boxplot using the empirical LVK data of BH+BH mergers. Panel~A displays the optimal fit to the LSCV curve, while panel~B exhibits the optimal fit to the median curve. Additional fits ("peak", "low", "left", and "right") testing extreme cases from diverse non-empirical distributions, $H(\chi_{\rm eff})$ were not plotted due to their low statistical probability. The simulations were conducted assuming spin-axis tossing, a constant kick of $w=50\;{\rm km\,s}^{-1}$, and a pre-SN orbital separation of $a_i\in[4,40]\;R_\odot$. The total count of simulated BH+BH merger candidates is $N^{\rm sim}=10^5$, with $N^{\rm LVK}=83$ observed BH+BH mergers utilized. The y-axis represents the normalized density.}
\label{fig:BimodalVarDistPlots}
\end{figure*}

It has been argued that BHs formed from single stars, as well as first-born BHs in binaries, are expected to have very low spin values \citep{hws05,fm19} --- a prediction consistent with $\chi_1$ and the BH+BH GW observations. In contrast, spin estimates of BHs in HMXBs have traditionally been very high, with reported values of $0.84$–$0.99$ \citep{mbo+21}. However, \citet{zmv+25} recently pointed out that current spin measurement techniques in X-ray binaries may be affected by substantial systematic uncertainties, and they suggest that the true BH spins in HMXBs might be low --- consistent with the spins inferred from BH+BH mergers.

Figure~\ref{fig:BimodalVarDistPlots} shows the best-fitting $\chi_{\rm eff}$ solution to the LSCV curve (panel~A) and to the median curve (panel~B). Since these two curves best represent the empirical LVK data, the corresponding $\chi_{1,2}$ distributions for the individual BH spins are assumed to most closely reflect the true distributions, given their resulting high p-values from the two-sample tests. The black lines in Fig.~\ref{fig:BimodalVarDistPlots} indicate the upper and lower bounds of the 90\% confidence interval of $\chi_{\rm eff}$ from the empirical LVK data.

\subsubsection{Negative effective spins and spin-axis tossing}
It is generally difficult to obtain large negative effective spin parameters --- especially if BH formation is considered without spin-axis tossing. This is due to the relatively small misalignment angles, $\delta$ (i.e. tilt of the orbital plane due to the SN) that result from our Monte Carlo simulations. 
The maximum misalignment angle is achieved \citep[see eq.~5 and fig.~3 in][]{tau22} for SN kick angles $\theta=180^\circ$, or $\phi=0\pm 180^\circ$ if $\theta > \theta_{\rm retro} =\cos ^{-1}(-v_{\rm rel}/w)$, i.e. the critical angle for post-SN retrograde orbits. Obtaining $\chi_{\rm eff}<0$ based only on a SN kick requires at least (but not necessarily sufficient, cf. Eq.~\ref{eq:chi_eff2}) $\Theta_1=\delta > 90^\circ$, resulting in retrograde spin of the first-born BH).
Since $\delta > 90^\circ$ is impossible for kick magnitudes $w<v_{\rm rel}$, 
the only way to produce $\chi_{\rm eff}<0$, for kicks typically up to $400-500\;{\rm km\,s}^{-1}$, is therefore to allow for tossing of the BH spin axis; relevant here when the second BH forms (Section~\ref{subsec:spin-axis-tossing}).

\subsection{Kick and orbital separation dependencies}\label{subsec:kick-separation}
In the previous section, the kick magnitude was set to a constant value of $50\;{\rm km\,s}^{-1}$ and a uniform distribution of pre-SN orbital separations limited to $[4,40]\;R_\odot$ was applied. In this section, we investigate simulations of more extreme kicks and pre-SN orbital separation ranges to test these dependencies. We adapt the best fits for the $\chi_{1,2}$ distributions obtained in Section~\ref{subsubsec:spin-param-dist} (Fig.~\ref{fig:chi12-dist}). Therefore, the following assumptions are used in our next set of Monte Carlo simulations:

\begin{itemize}
    \item Pre-SN He-star mass of $M_{\rm He}=M_{\rm BH,2}/0.8.$
    \item Mass reversal is disregarded.
    \item Uniform pre-SN orbital separation of $a_i \in [4,40]\;R_\odot$,\\$a_i \in [4,400]\;R_\odot$, or an extreme fixed value.
    \item Uniform kick magnitude, $w \in [0,350]\;{\rm km\,s}^{-1}$ or an\\ extreme fixed value. Isotropic kick direction.
    \item Spin-axis tossing of second-born BH (isotropic direction).
    \item $\chi_1$ distributed as $\beta(1.01,9.69)$.
    \item $\chi_2$ distributed as $\beta(2.72,2.34)$. 
\end{itemize}

\begin{figure*}
\vspace{-0.2cm}
\hspace*{0.4cm}
\includegraphics[width=0.43\textwidth]{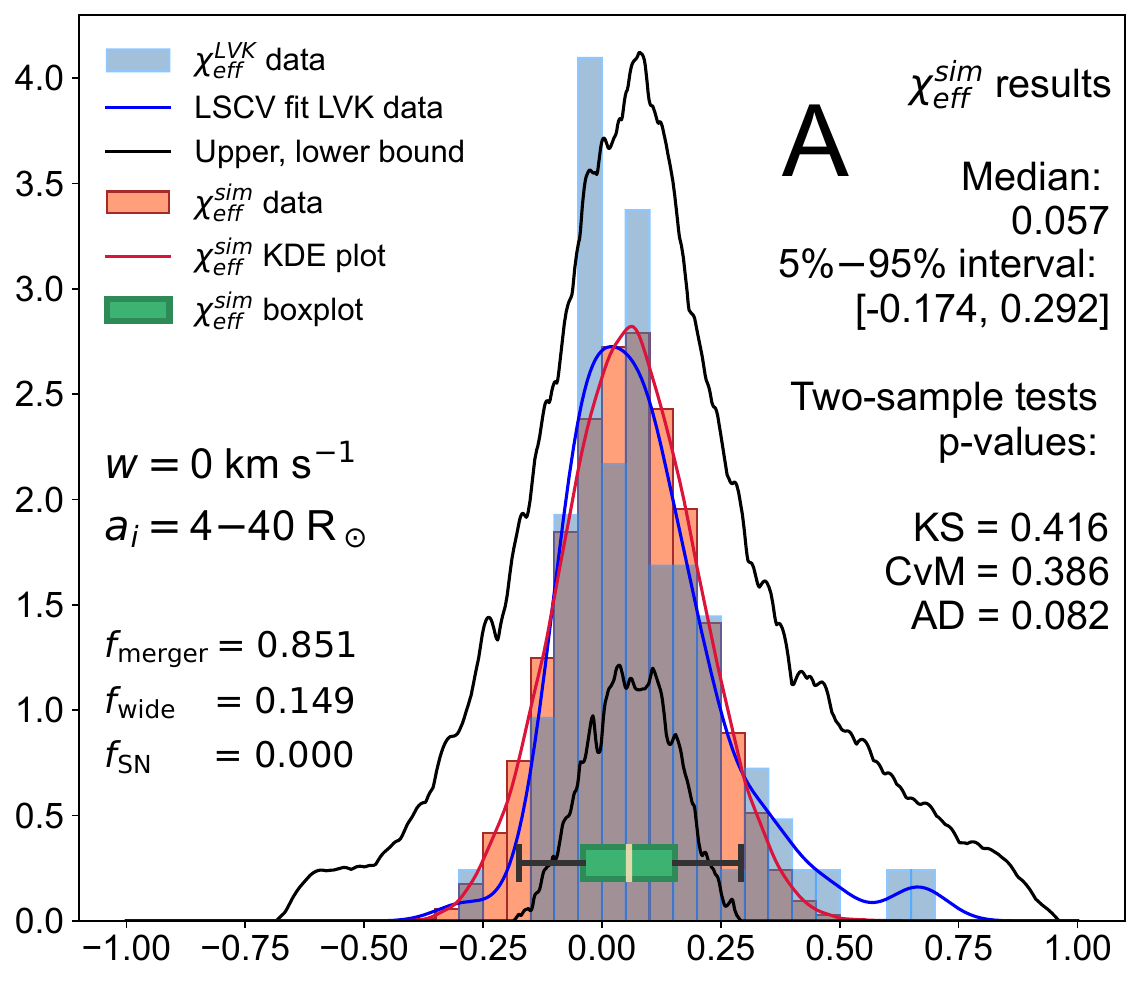}
\hspace*{1.0cm}
\includegraphics[width=0.43\textwidth]{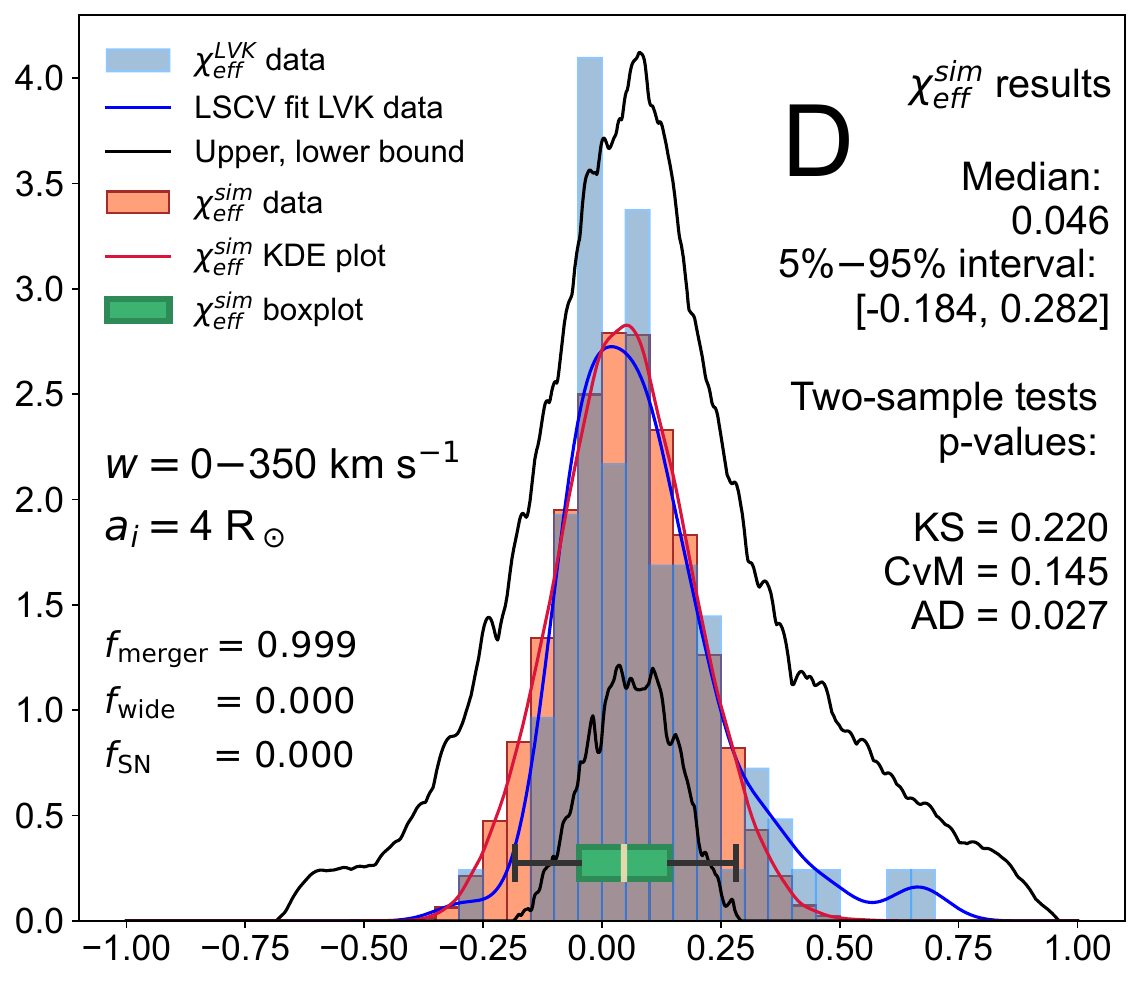}

\vspace*{0.3cm}
\hspace*{0.4cm}
\includegraphics[width=0.43\textwidth]{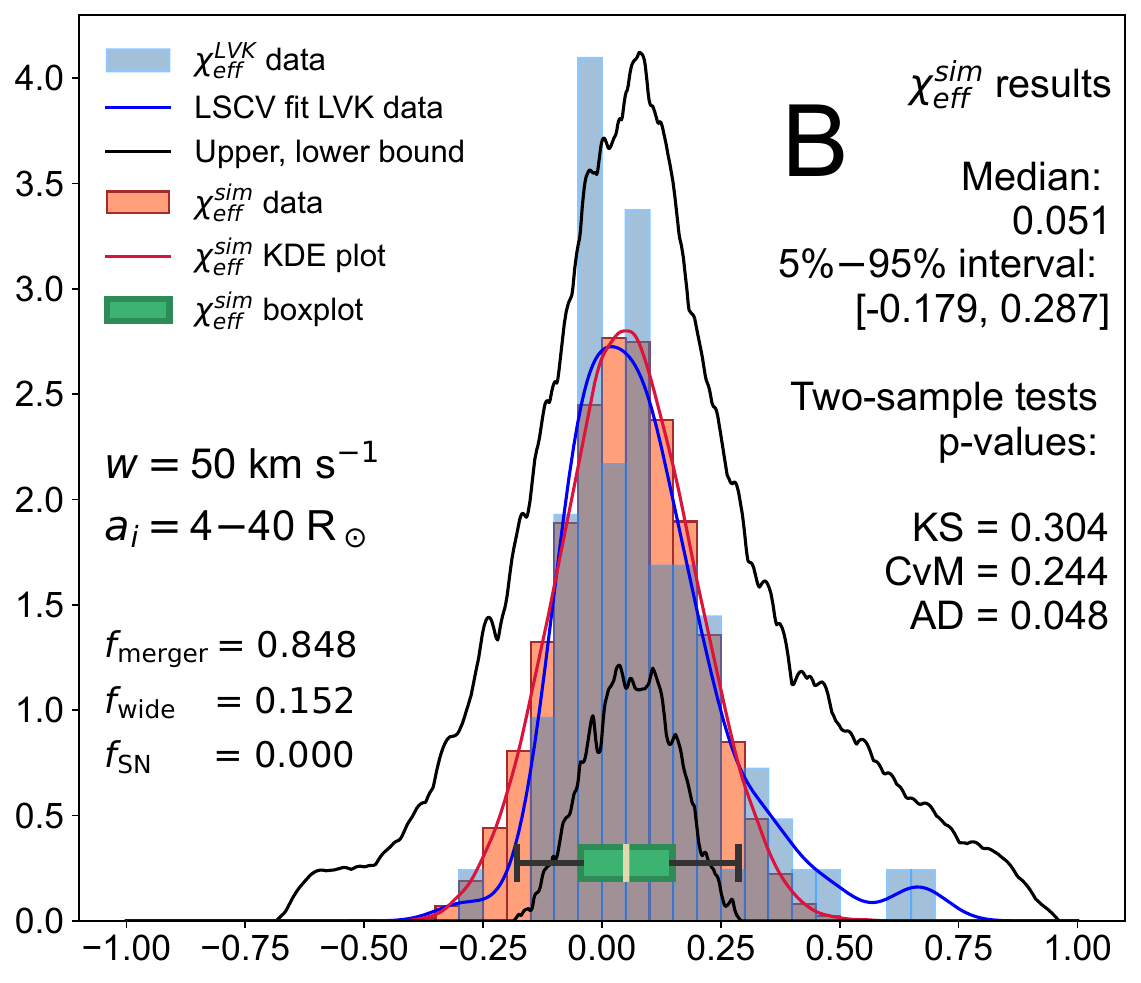}
\hspace*{1.0cm}
\includegraphics[width=0.43\textwidth]{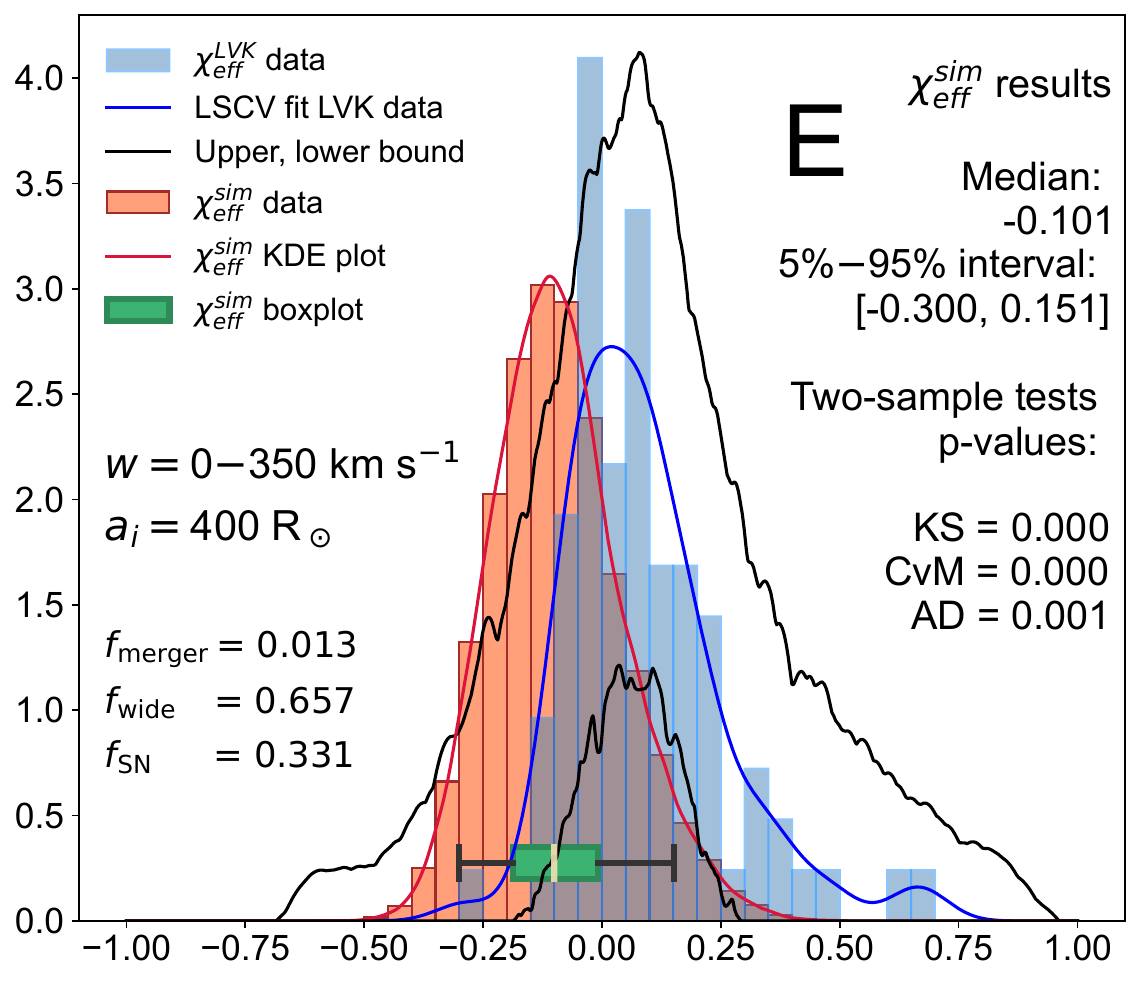}

\vspace*{0.3cm}
\hspace*{0.4cm}
    \includegraphics[width=0.43\textwidth]{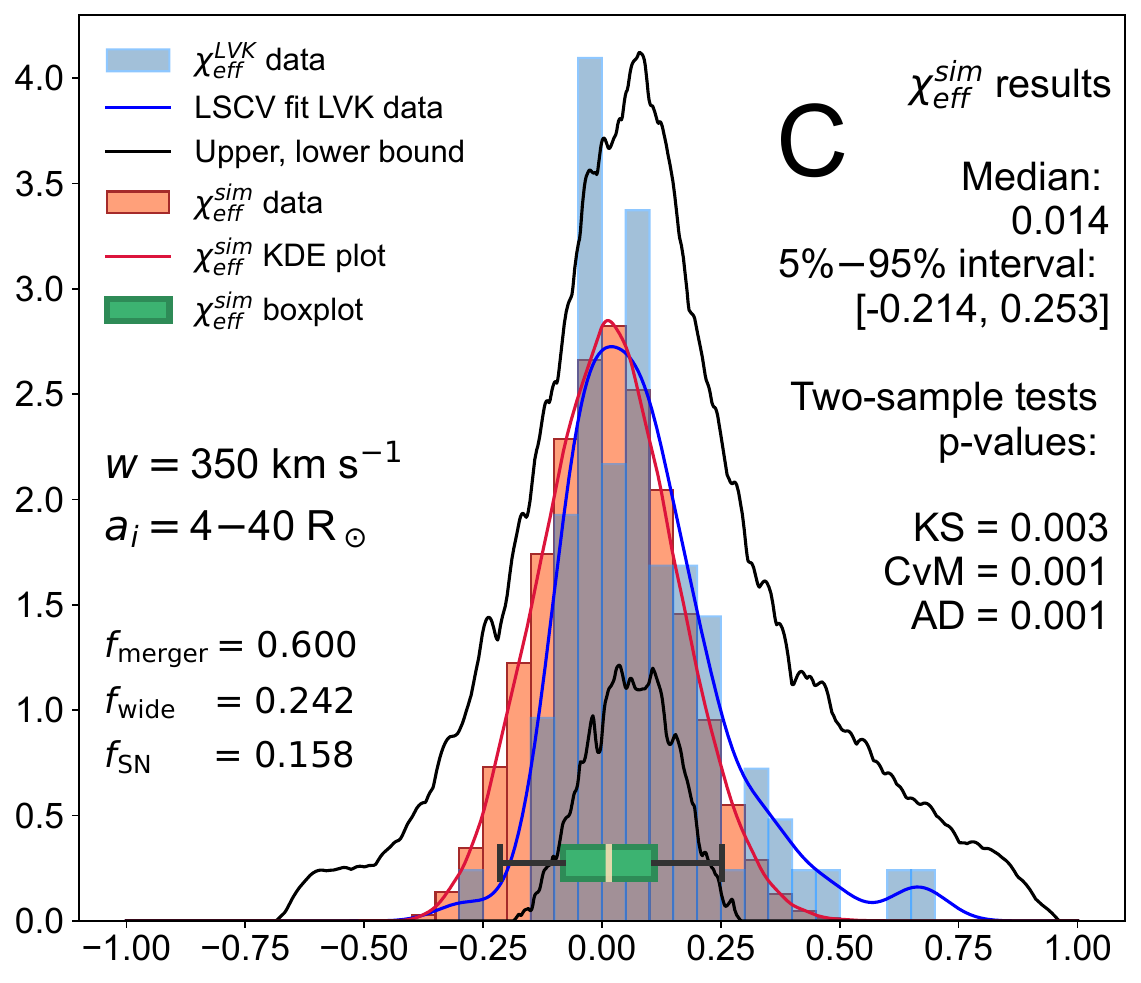}
\hspace*{1.0cm}
\includegraphics[width=0.43\textwidth]{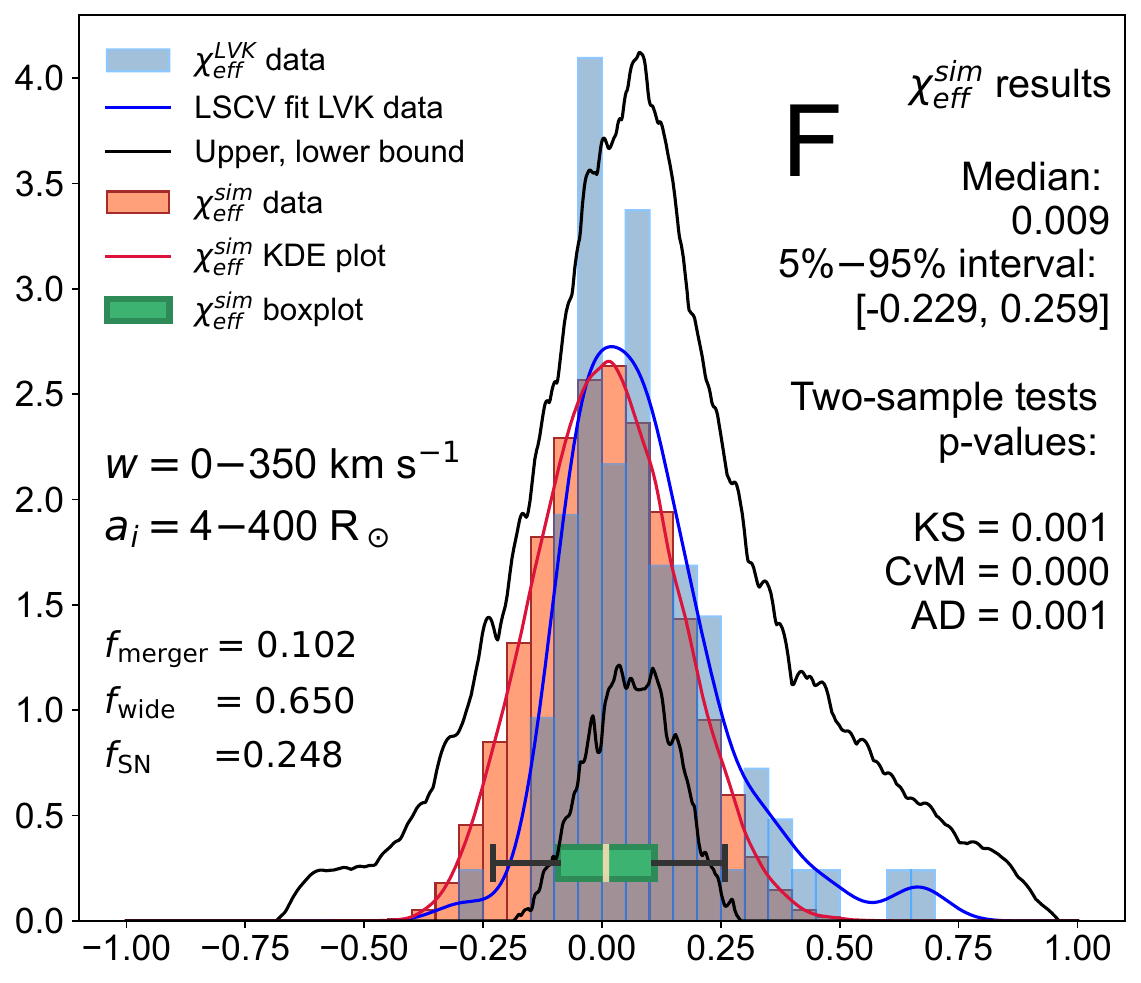}
\caption{Distribution of $\chi_{\rm eff}$ (x-axis) for $N=10^5$ BH+BH systems that merge within 13.7~Gyr, based on simulations testing dependencies on natal kicks and orbital separations. All simulations assume {\bf {\em full} spin-axis tossing}. The y-axis shows the normalized density of systems. Panels A, B, and C use fixed kick magnitudes of $w=0$, $w=50\;{\rm km\,s}^{-1}$, and $w=350\;{\rm km\,s}^{-1}$, respectively, with pre-SN orbital separations $a_i$ drawn uniformly from $[4,40]\;R_\odot$. Panels D, E, and F sample $w$ uniformly from $[0,350]\;{\rm km\,s}^{-1}$, while fixing $a_i$ at $4\;R_\odot$, $400\;R_\odot$, or varying $a_i$ uniformly over $[4,400]\;R_\odot$, respectively. The best agreement with empirical data (highest p-values) is found for relatively modest values of both $w$ and $a_i$. The fractions of systems resulting in mergers ($f_{\rm merger}$), wide-orbit binaries that do not merge within a Hubble time ($f_{\rm wide}$), or SN-disrupted systems ($f_{\rm SN}$) are listed in each panel. For comparison to simulations {\em without} spin-axis tossing, see Fig.~\ref{fig:rangeTester_no-tossing}.}
\label{fig:rangeTester}
\end{figure*}

These simulations yield the $\chi_{\rm eff}$ distributions shown in Fig.~\ref{fig:rangeTester}. As is evident from the plots, there are no major changes in the distribution when these variables are changed, other than slight changes in the median of the distributions. The only exception is panel~E that assumes an extremely wide fixed pre-SN orbit of $a_i=400\;R_\odot$. Producing bound post-SN BH+BH systems that will merge within a Hubble time in this setup requires large backward kicks (i.e. large values of the kick angle, $\theta$) which therefore results in retrograde spin of the first-born BH, resulting in a distribution shifted toward negative values of $\chi_{\rm eff}$.
For the setup in panel~E, only $\sim 1\%$ of the trial systems ($f_{\rm merger}=0.013$) will result in BH+BH systems that merge within a Hubble time. The rest of the systems either result in post-SN binaries too wide to merge ($f_{\rm wide}=0.657$) or become disrupted in the SN ($f_{\rm SN}=0.331$).

Although we confirm that varying $a_{\rm i}$ over wide bounds has little impact on the resulting $\chi_{\rm eff}$ distribution, we note that in very wide binaries, tidal interactions might not be sufficiently strong to enable efficient tidal locking prior to core collapse, thereby questioning in these cases the use of a large spin magnitude for the second-born BH (cf. Sections~\ref{subsubsec:spin-param-dist} and \ref{subsec:final-sim}).

\begin{figure}
\centering
\includegraphics[width=0.38\textwidth]{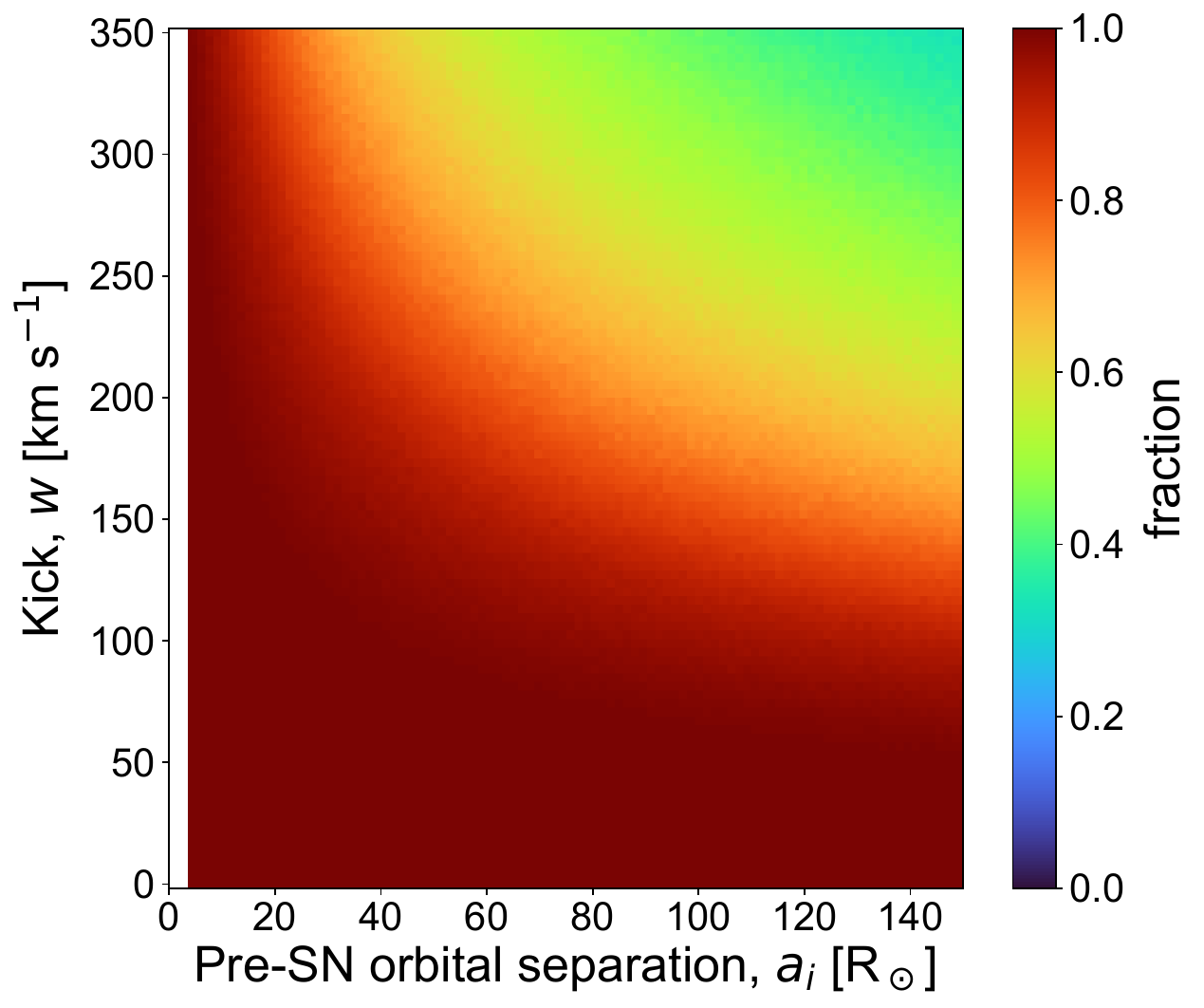}
\caption{Fraction of trial systems that survive the SN explosion as a function of pre-SN orbital separation, $a_i$ and kick velocity, $w$. As expected, wide-orbit pre-SN systems that experience large kicks are more likely to disrupt.}
\label{fig:f_bound}
\end{figure}

\begin{figure}
\centering
\includegraphics[width=0.38\textwidth]{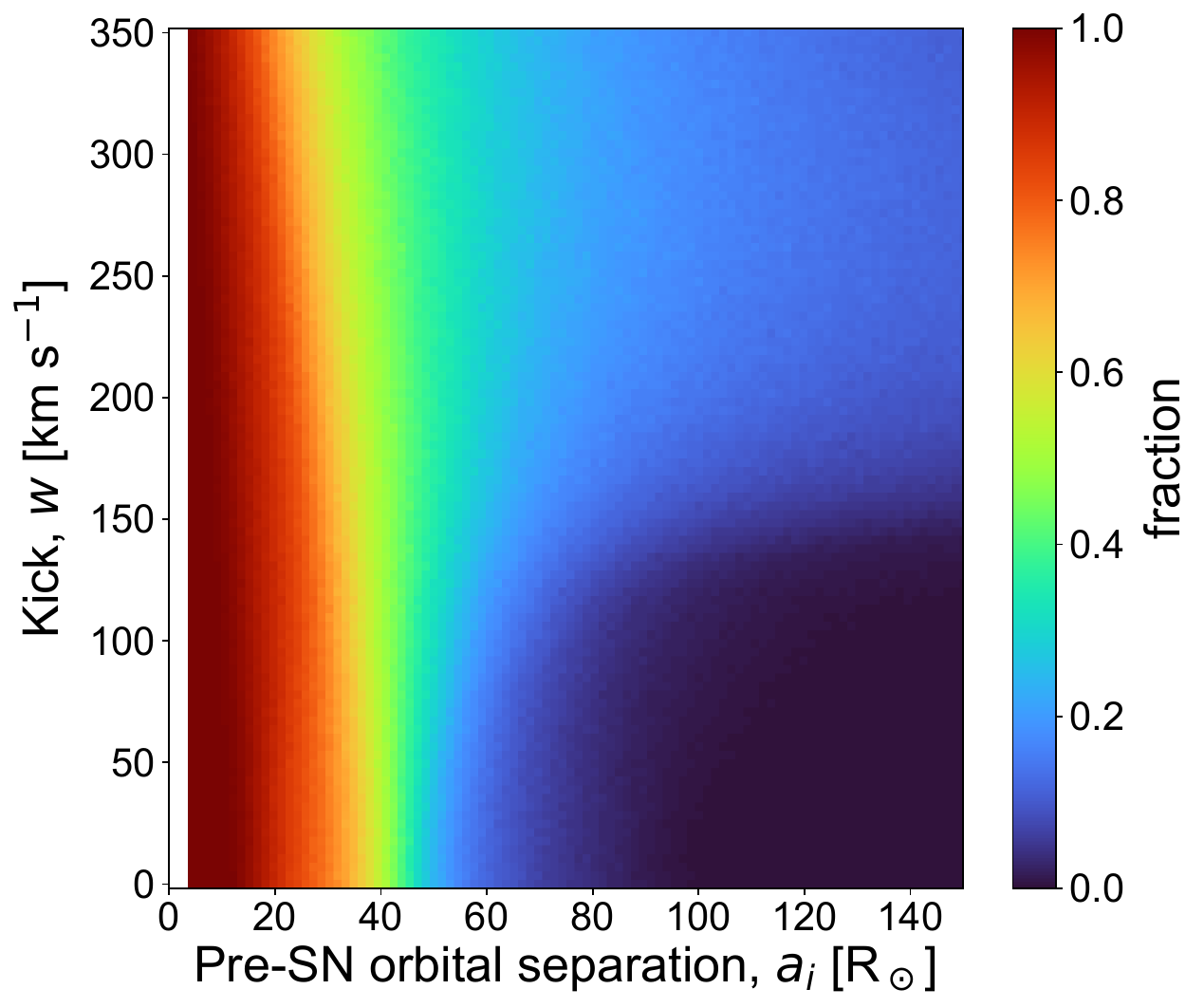}
\caption{Fraction of trial systems that become detectable BH+BH mergers within a Hubble time as a function of pre-SN orbital separation, $a_i$ and kick velocity, $w$. It is clear that only few BH+BH mergers result if $a_i>40\;R_\odot$.}
\label{fig:f_detect}
\end{figure}

The fraction of trial systems that survived the SN ($1-f_{\rm SN}$) as a function of pre-SN orbital separation and kick velocity is plotted in Fig.~\ref{fig:f_bound}. This plot can be convolved with the fraction of systems that merge within a Hubble time to produce the fraction of trial systems ($f_{\rm merger}$) that become detectable as BH+BH mergers (Fig.~\ref{fig:f_detect}). It is evident that the far majority of these BH+BH mergers produced from isolated binary evolutions had pre-SN orbital separations of $a_i \in [4,40]\;R_\odot$.

\begin{figure}
\centering
\includegraphics[width=0.38\textwidth]{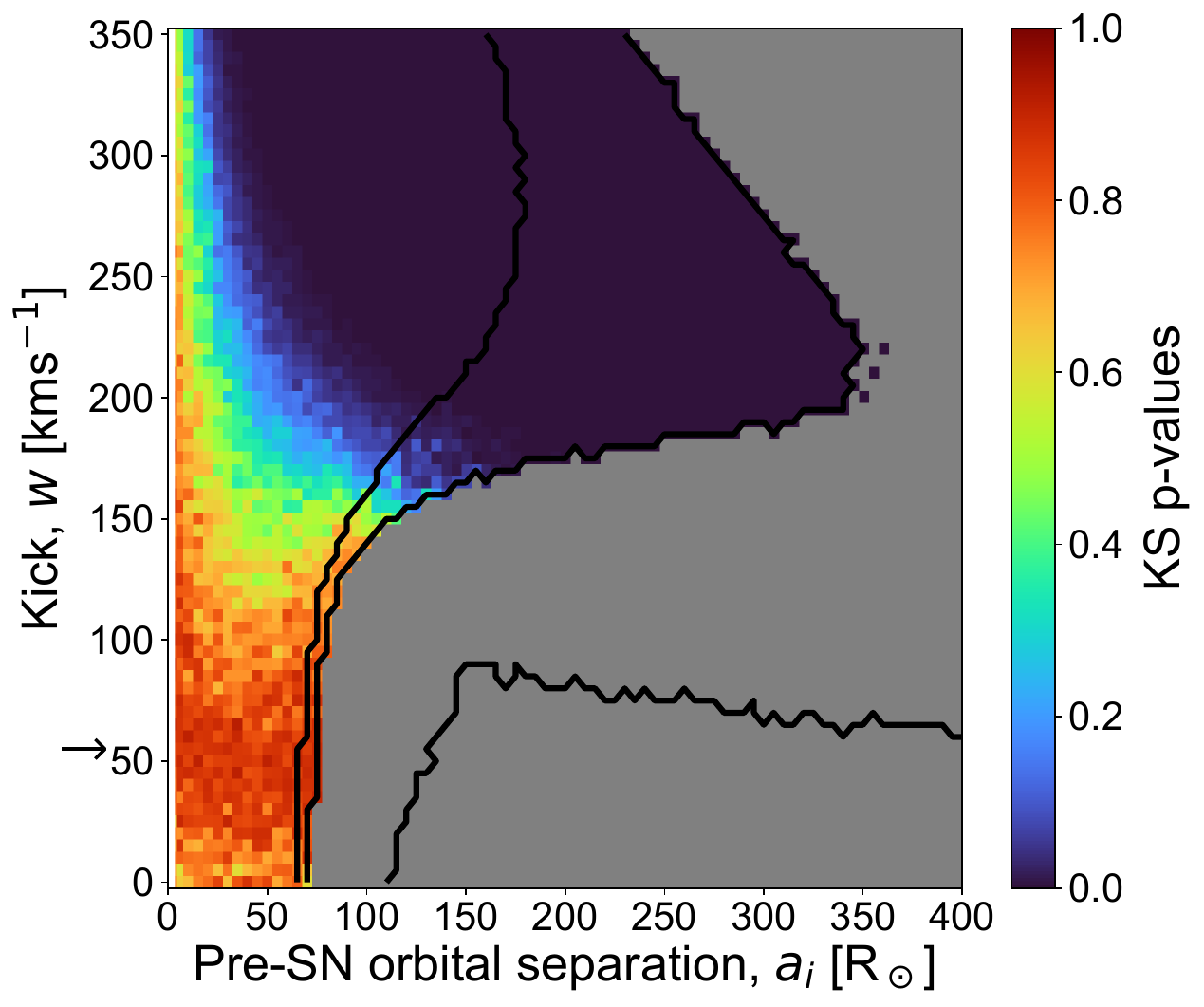}
\caption{Heatmap of p-values from statistical KS-tests comparing the distribution of $\chi_{\rm eff}$ from our simulations including spin-axis tossing with the empirical LVK data, as a function of pre-SN orbital separation and kick velocity. The black lines indicate (left to right) limits where 10\%, 5\% and 0\% of the trial systems become detectable BH+BH mergers. The arrow indicates the median value of $w$ ($55\;{\rm km\,s}^{-1}$) that yields the highest p-values across the $a_i$ range.}
\label{fig:heatmap-p-values}
\end{figure}

Most notable from Fig.~\ref{fig:rangeTester} (which includes spin-axis tossing in all cases) is that the simulations in panels~A, B and D all yield high p-values of the statistical tests and median $\chi_{\rm eff}$ values of $0.05-0.06$, similar to the median value of $\chi_{\rm eff}=0.06$ in the empirical LVK data (Fig.~\ref{fig:GWTC3_BH+BH}).  
A heatmap of the resulting p-values from the KS-test, comparing the distribution of $\chi_{\rm eff}$ from our simulations including spin-axis tossing with the empirical LVK data, as a function of pre-SN orbital separation and kick velocity, is shown in Fig.~\ref{fig:heatmap-p-values}. The SN kick value that yields the highest the p-values has a median value of $w\simeq 55\;{\rm km\,s}^{-1}$.
As no major changes occur due to the varied kicks and pre-SN orbital separations with reasonable limits, a standard kick of $50\;{\rm km\,s}^{-1}$ and a pre-SN orbital separation range of $4-40\;R_\odot$ were adopted in our further analysis.

\subsection{Spin-axis tossing}\label{subsec:spin-axis-tossing}
\begin{figure}
\vspace{-0.7cm}
\hspace*{1.2cm}
 \includegraphics[width=0.66\columnwidth]{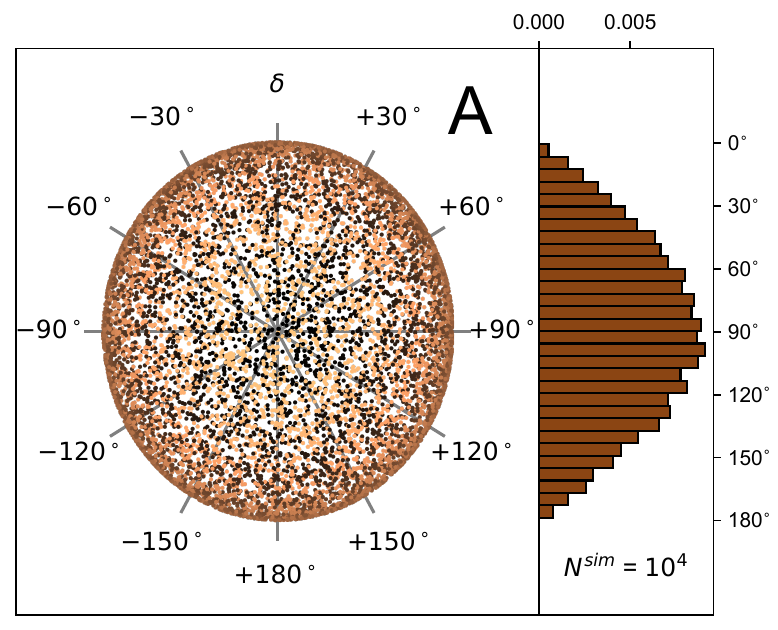}

 \hspace*{1.2cm}
 \includegraphics[width=0.66\columnwidth]{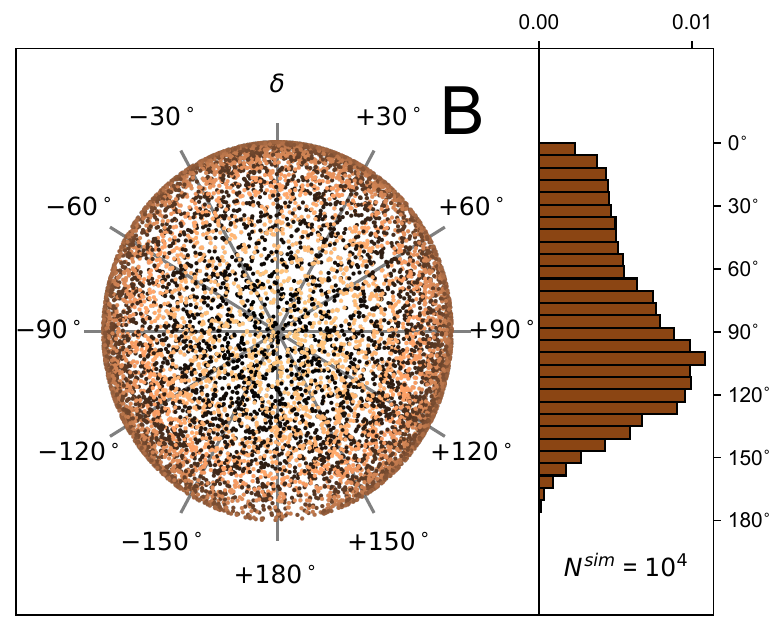}
\caption{Normalized distributions of the spin-axis tossing angle, $\Phi_2$ (here plotted relative to the misalignment angle, $\delta$). Panel~A shows a uniform isotropic distribution, i.e. full tossing: $P(\Phi_2) = \frac{1}{2} \sin(\Phi_2)$. Panel~B shows the distribution of $\Phi_2$ that best fits the empirical LVK data. Both plots were made by randomly selecting $10^4$ values of $\Phi_2$ from the respective distribution.}
\label{fig:tossingSphere}
\end{figure}

In the previous sections, we included tossing of the second-born BH spin axis in a random (isotropic) direction.
In this section, we explore the consequence of varying the distribution of the direction of spin-axis tossing. (We also performed tests {\em without} spin-axis tossing for a comparison, see Section~\ref{subsubsec:no-tossing}).
The assumptions used in this section are as follows:
\begin{itemize}
    \item Pre-SN He-star mass of $M_{\rm He}=M_{\rm BH,2}/0.8$.
    \item Mass reversal is disregarded. 
    \item Uniform pre-SN orbital separation of $a_i \in [4,40]\;R_\odot$.
    \item Constant SN kick of $w=50\;{\rm km\,s}^{-1}$ (isotropic direction).
    \item Different distributions of spin-axis tossing ($\Phi_2$).
    \item $\chi_1$ distributed as $\beta(1.01,9.69)$.
    \item $\chi_2$ distributed as $\beta(2.72,2.34)$.
\end{itemize}
A total of four distributions were tested, two of which are shown in Fig.~\ref{fig:tossingSphere}. We repeat that {\em no tossing} refers to the case where $\Phi_2=\Phi_1=\delta$, and $\delta$ is the post-SN misalignment angle. 
Panel~A in Fig.~\ref{fig:tossingSphere} illustrates a uniform isotropic distribution where the PDF of the spin tilt angle, $\Phi_2$ is given by: $P(\Phi_2)= \frac{1}{2} \sin(\Phi_2)$, which we refer to as {\em full tossing}. 
Panel~B shows a fitted distribution that gives the smallest RMSE with respect to the LSCV curve. This fit was made assuming a simple two-mode distribution for $\Phi_2$ (in units of $\pi$): 
\begin{equation}
    P(\Phi_2) = \frac{1}{2}\Big(\beta(a_1,b_1)+\beta(a_2,b_2)\Big)\,,
\end{equation}
and the best resulting fit yielded the following parameters:
\begin{align}
    P(\Phi_2) = \frac{1}{2}\Big(\beta(1.36,2.51)+\beta(7.9,5.3)\Big)\,.
    \label{eq:fittedphi}
\end{align}
It is interesting that panel~B indicates a distribution of tossing angles that slightly deviates from isotropy (panel~A). 
This could be a signature of the origin of LVK BH+BH mergers being a mixture between isolated binaries and dynamical formation channels (Section~\ref{subsec:dynamical}), or that the degree of tossing depends on e.g. BH mass and/or pre-SN orbital separation.

\begin{figure}[h]
\includegraphics[width=0.45\textwidth]{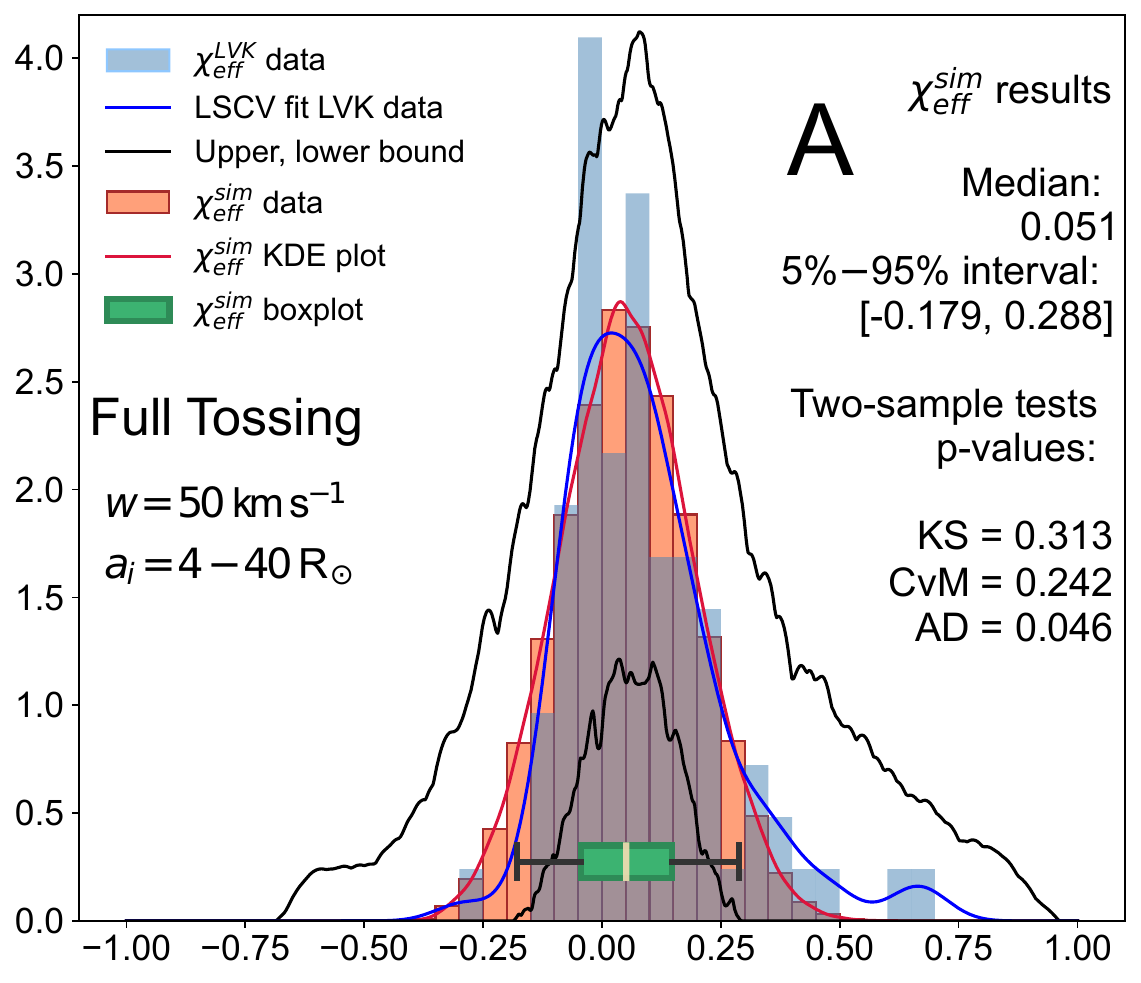}
\includegraphics[width=0.45\textwidth]{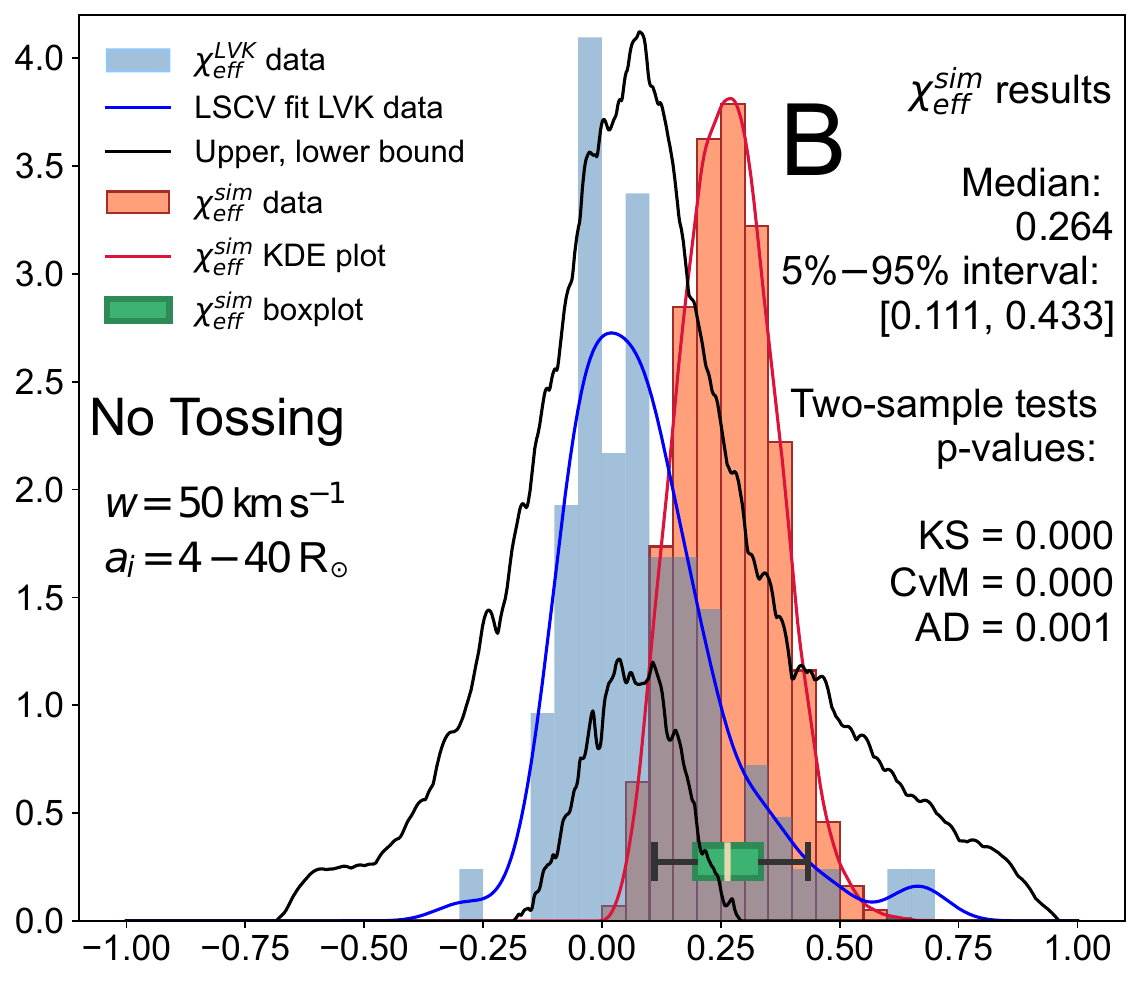}
\includegraphics[width=0.45\textwidth]{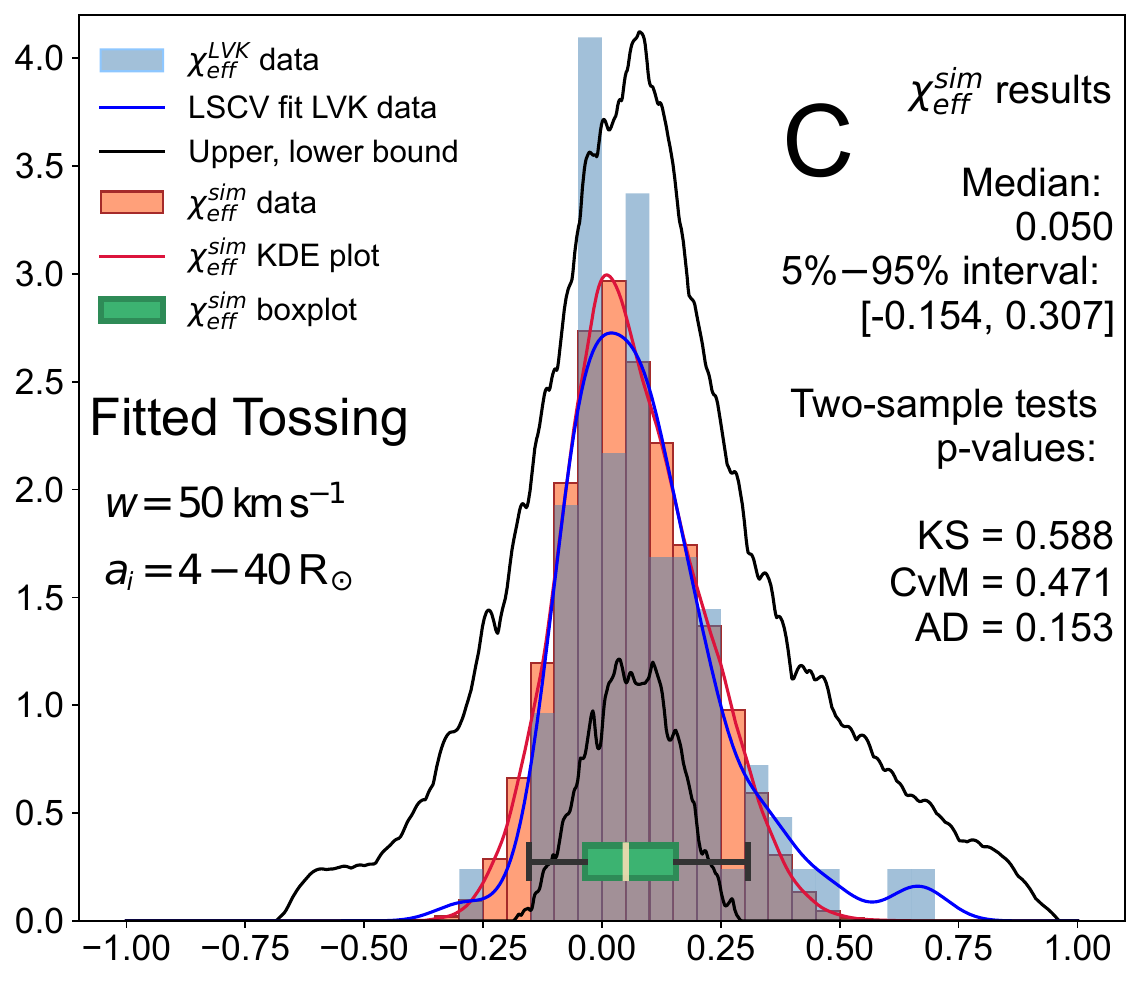}
\caption{Histograms showing how $\chi_{\rm eff}$ is distributed when using different tossing angle ($\Phi_2$) distributions. Panel~A shows {\em full tossing} as previously presented (Fig.~\ref{fig:rangeTester}, panel~B). Panel~B shows the solution for {\em no tossing}, meaning that $\Phi_2 = \Phi_1=\delta$. Panel~C shows a solution for a {\em fitted tossing} distribution using a sum of beta distributions. Along the y-axis is the normalized density.}
\label{fig:tossingQ}
\end{figure}

Figure~\ref{fig:tossingQ} shows how $\chi_{\rm eff}$ is distributed when using the different distributions mentioned above for the tossing angle, $\Phi_2$.
The panels~A, B, and C, refer to the cases of {\em full tossing}, {\em no tossing}, and {\em fitted tossing} (Eq.~\ref{eq:fittedphi}), respectively.
In the no tossing case, $\Phi_2=\delta$, the maximum misalignment angle gained from the Monte Carlo simulation is a mere $\delta_{\rm max}=6.42^\circ$. Thus, in this standard scenario from the literature (i.e. without any spin-axis tossing), we only obtain $\chi_{\rm eff}>0$, which is clearly in contradiction with observations (the statistical p-values are $<0.001$). Therefore, we conclude, similar to \citet{tau22}, that {\em if} all detected BH+BH mergers had formed via isolated binary star evolution, tossing of the spin axis during core collapse {\em must} have taken place.
We find that applying different values of SN kicks and pre-SN orbital separation will not provide any reasonable fits to the LVK data without spin-axis tossing --- see also Fig.~\ref{fig:rangeTester_no-tossing}, and see \citet{tau22} for a large range of qualitative tests.

The fitted tossing distribution in panel~C of Fig.~\ref{fig:tossingQ} results in a slightly better fit to the LSCV curve, as evident from the higher p-values from the two-sided tests when compared to panel~A of Fig.~\ref{fig:tossingQ} with full tossing. 
More data from LVK science runs O4--O5, or the 3G detectors (Einstein Telescope and Cosmic Explorer), are needed to probe this anisotropy aspect further.

One facet not explored here is {\em conditional tossing} based upon initial conditions of the pre-SN system. \citet{Janka_2022} explored how natal BH spins are created from a SN kick perspective, where kick direction and mass ejection asymmetries are able to change the direction of the spin axis relative to that of the progenitor star (i.e. the degree of tossing). Because the progenitor star and its BH companion are tidally locked, the pre-SN spin period of the progenitor star is equal to the pre-SN orbital period. Thus, considering conditions such as amount of mass loss and pre-SN orbital period, one may be able to find different distributions of $\chi_2$ and $\Phi_2$ accordingly.

We conclude from the current data (O1--O3) that while there {\em is} statistical support for spin-axis tossing in BHs (if isolated binaries constitute the prime formation channel of detected BH+BH mergers, see Section~\ref{subsec:dynamical}), the evidence is not yet significant enough to determine whether the {\em direction} of the tossing is fully isotropic or anisotropic --- where the latter case (conditional tossing) depends on the pre-SN structure of the collapsing star and the treatment of BH formation and SN explosion physics. We therefore recommend all future population synthesis studies --- simulating the population of BH+BH mergers from isolated systems --- to {\em include} spin-axis tossing of the second-formed BH.\footnote{The first-formed BH is also expected to experience spin-axis tossing, but its spin axis will later align due to tidal interactions and mass accretion from the progenitor of the second-formed BH \citep[but see][]{bk24}.} For simplicity, we suggest for now applying a fully isotropic spin-axis tossing distribution.

\subsubsection{Simulations without spin-axis tossing}\label{subsubsec:no-tossing}
\begin{figure*}[b]
\vspace{-0.2cm}
\hspace*{0.4cm}
\includegraphics[width=0.43\textwidth]{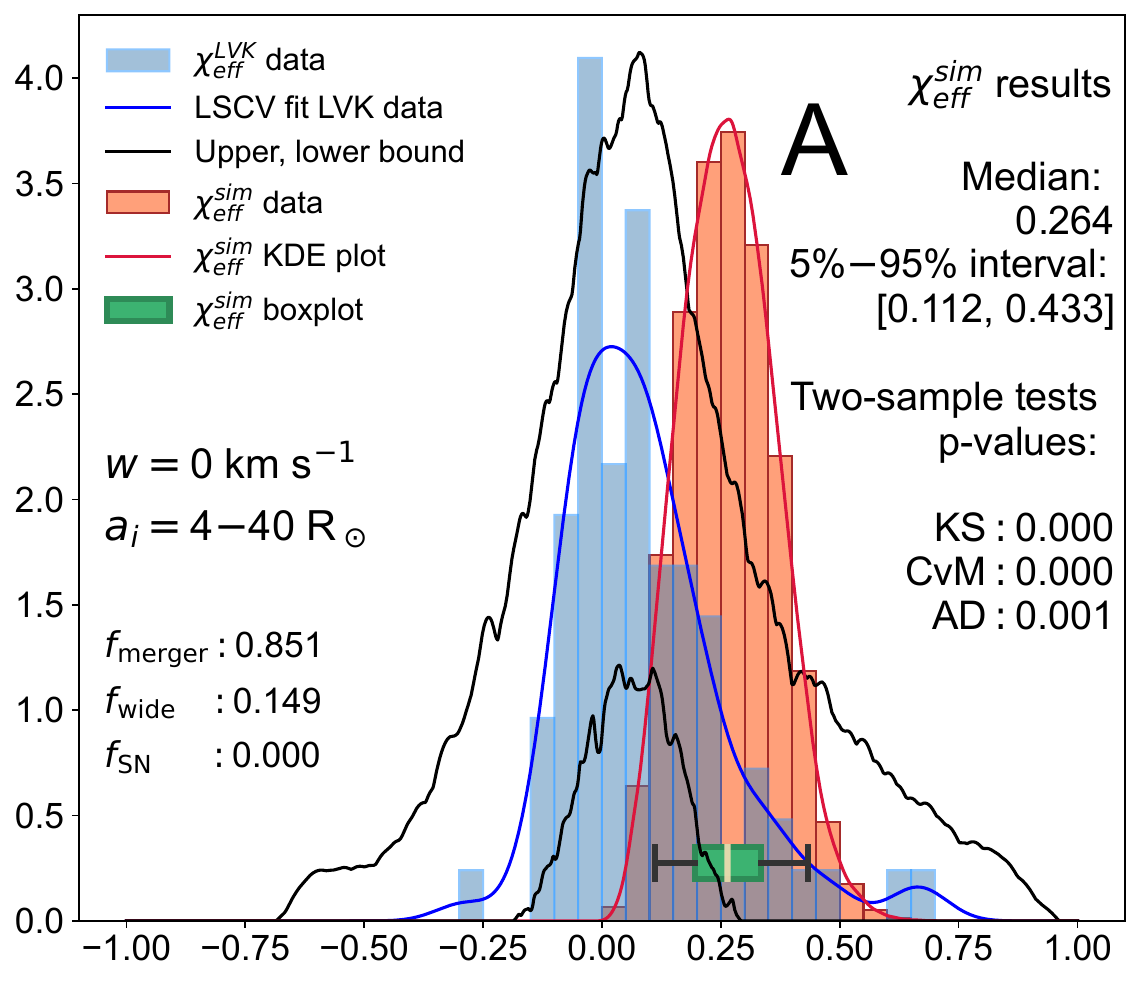}
\hspace*{1.0cm}
\includegraphics[width=0.43\textwidth]{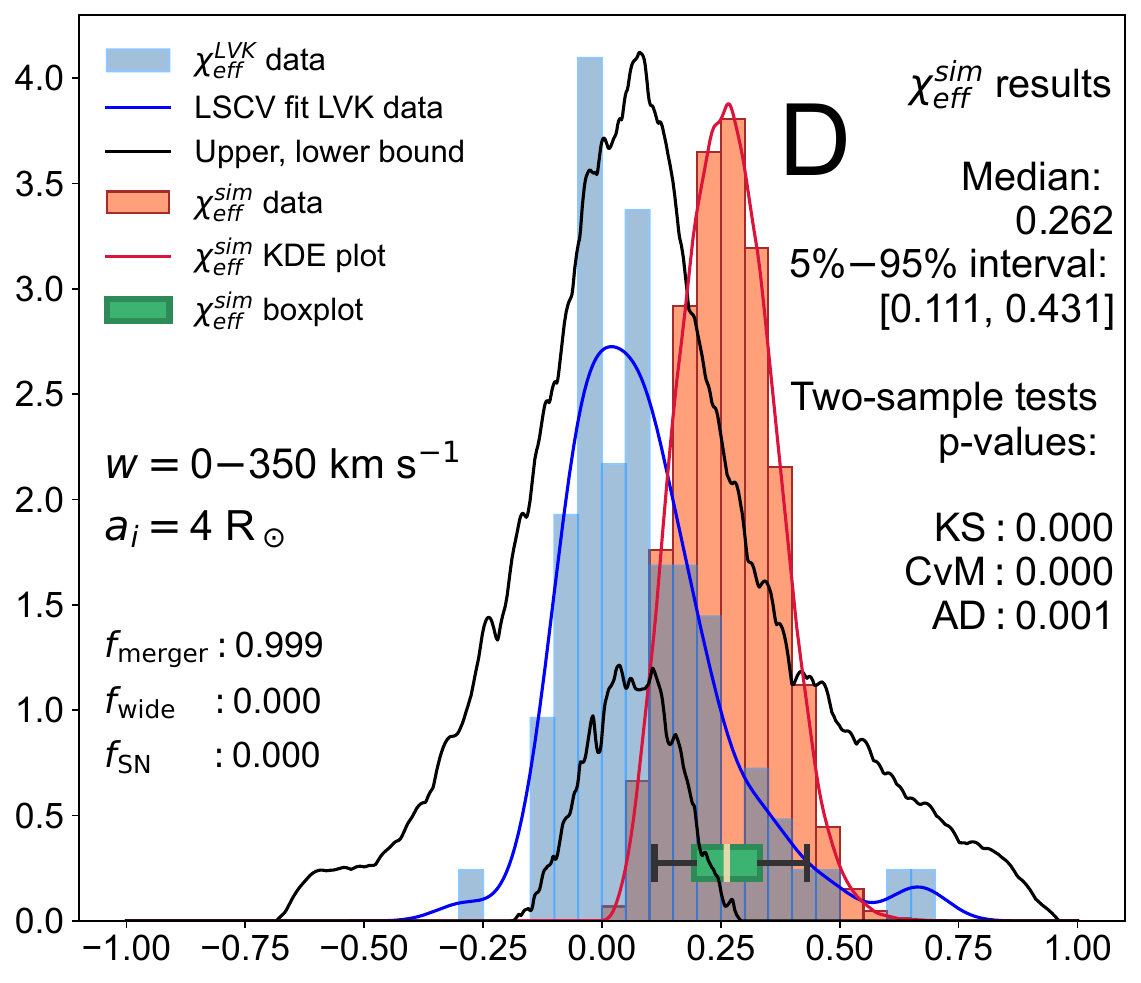}

\vspace*{0.3cm}
\hspace*{0.4cm}
\includegraphics[width=0.43\textwidth]{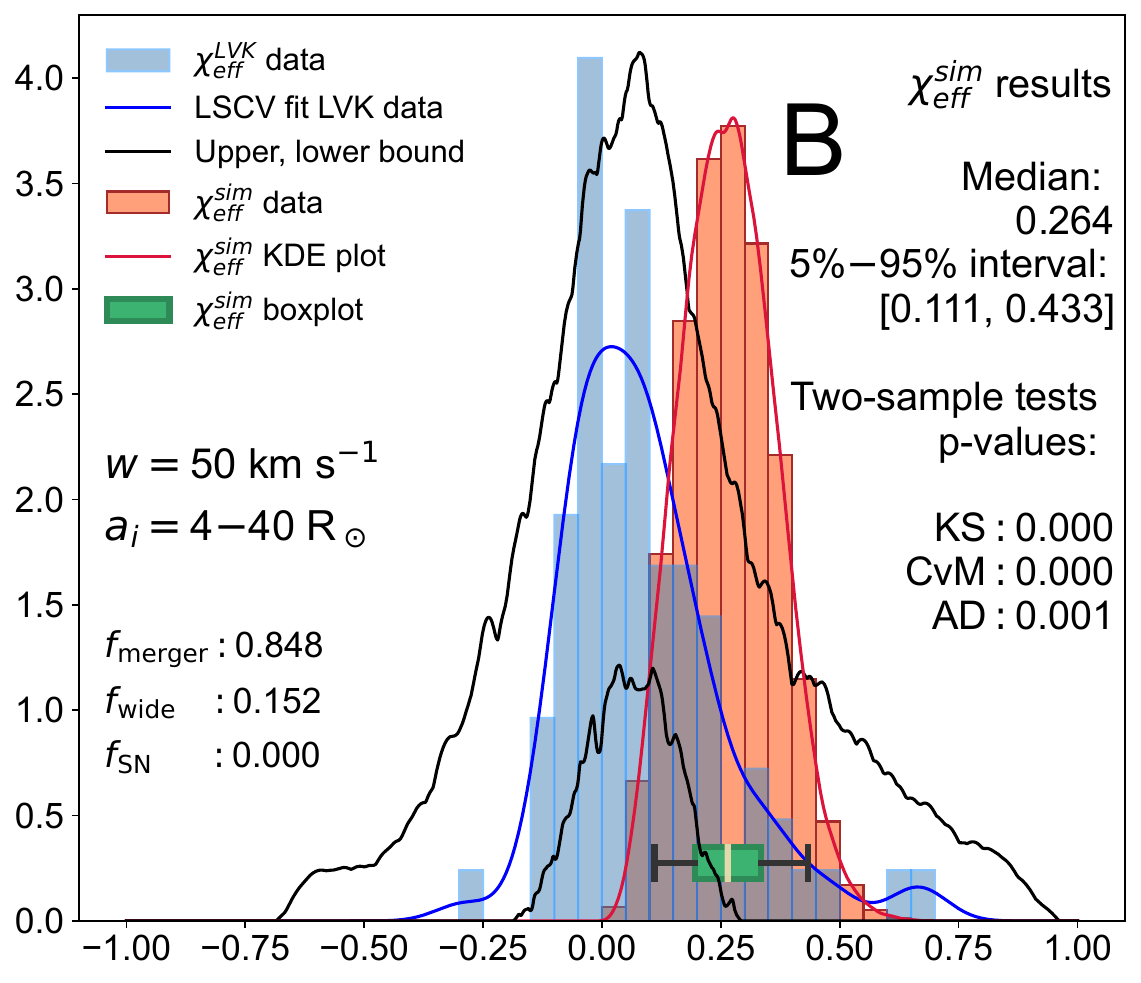}
\hspace*{1.0cm}
\includegraphics[width=0.43\textwidth]{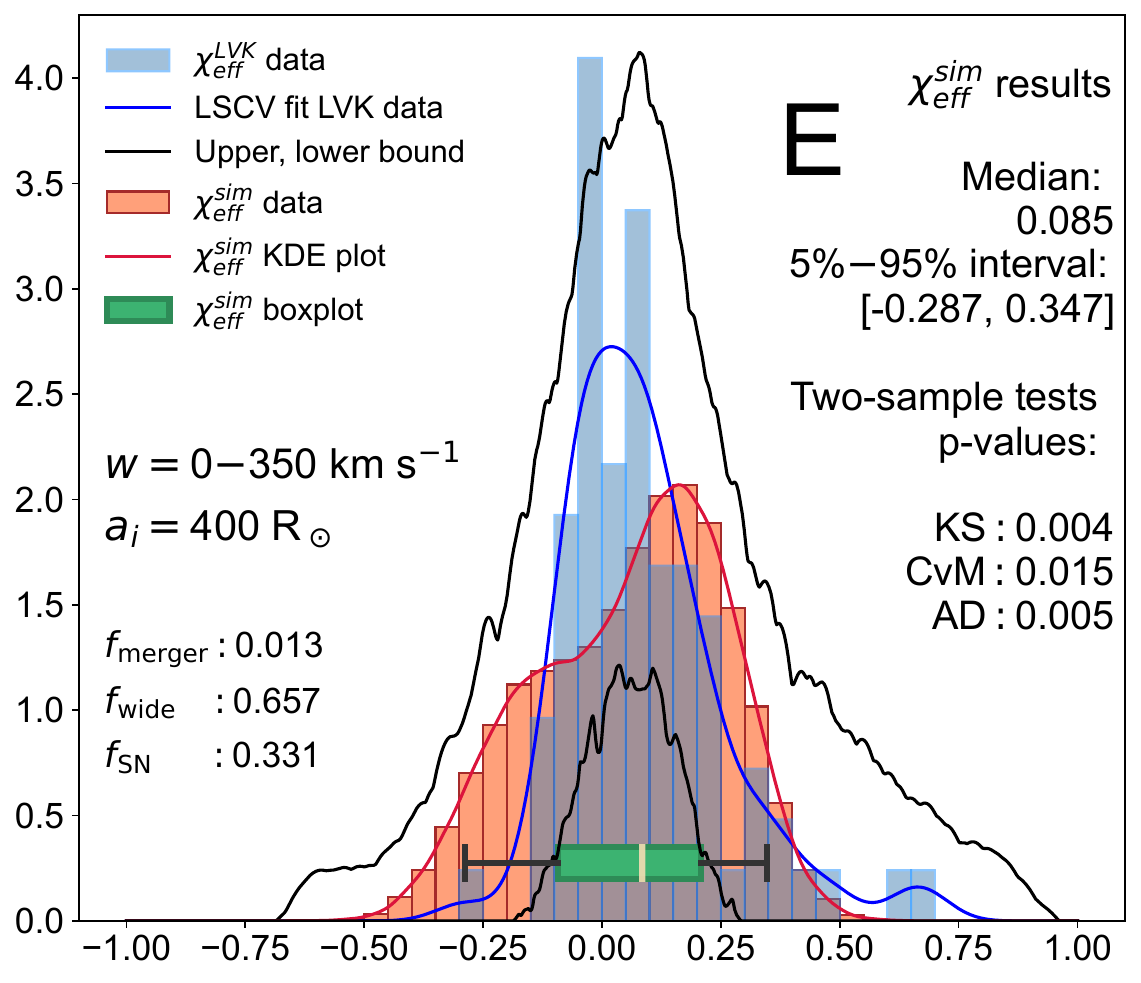}

\vspace*{0.3cm}
\hspace*{0.4cm}
\includegraphics[width=0.43\textwidth]{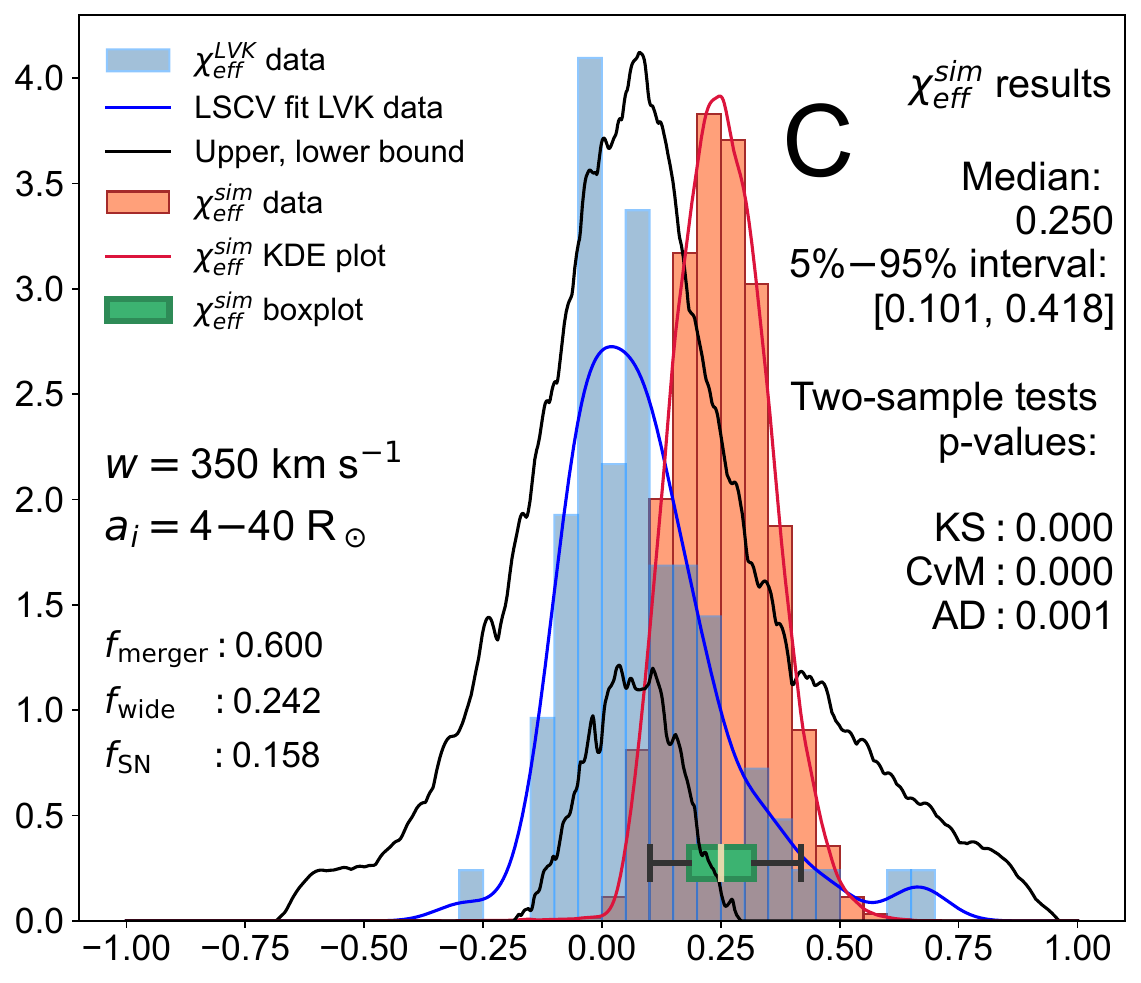}
\hspace*{1.0cm}
\includegraphics[width=0.43\textwidth]{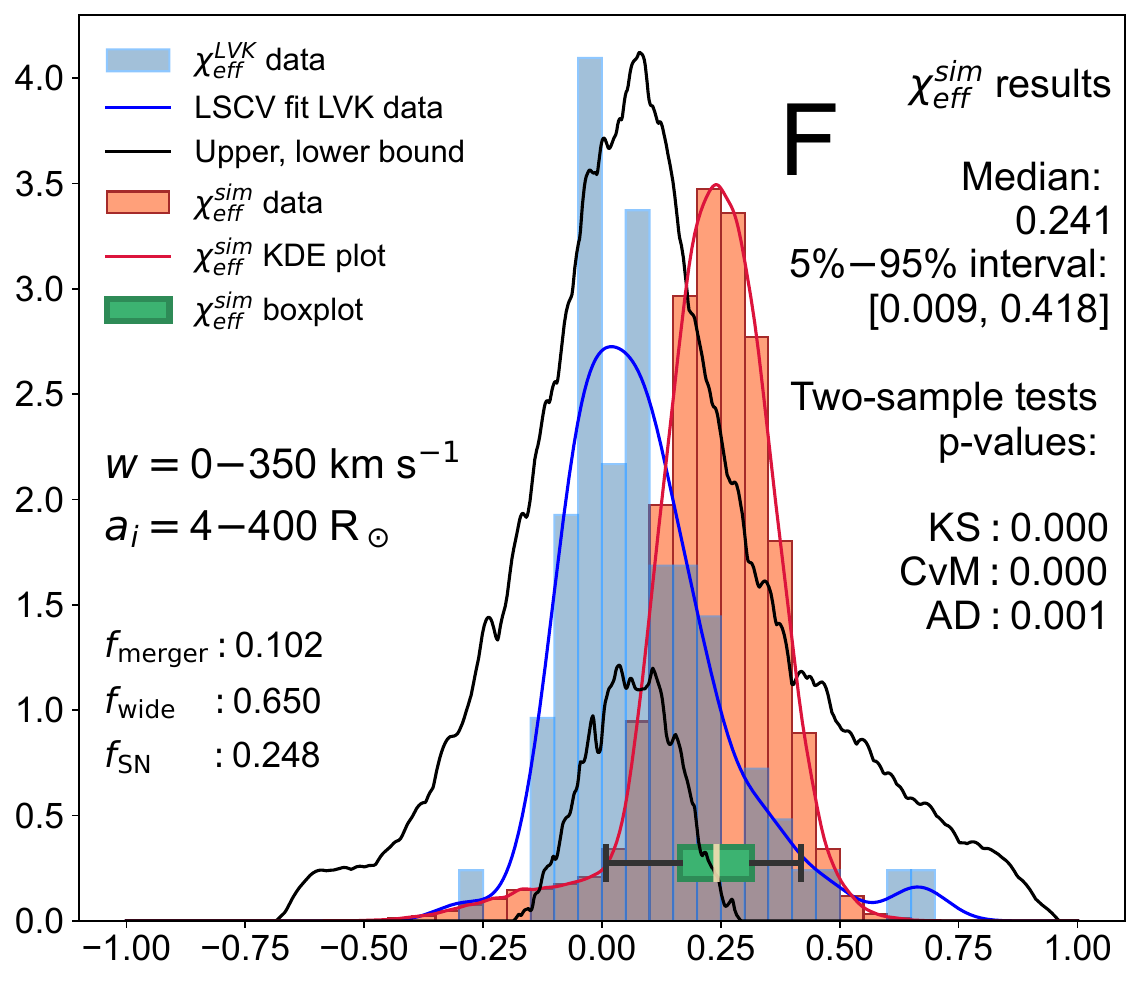}
\caption{Distribution of $\chi_{\rm eff}$ (x-axis) for $N=10^5$ BH+BH systems that merge within 13.7~Gyr, based on simulations testing the effects of natal kicks and orbital separations. All simulations shown here are {\bf {\em without} spin-axis tossing}. The y-axis shows the normalized density of systems. Panels A--F correspond to the same simulation setups as in Fig.~\ref{fig:rangeTester}, except for the omission of spin-axis tossing. Simulations in panel~E yield the ``highest'' p-values, though only at the $\sim 0.01$ level --- allowing us to reject the null hypothesis that the simulated and empirical $\chi_{\rm eff}$ distributions are drawn from the same parent distribution. In addition to this poor fit, we note that for the extreme values of $a_i$ and $w$ used in panel~E, nearly 99\% of trial systems failed to produce detectable BH+BH mergers.
For contrast, see simulations in Fig.~\ref{fig:rangeTester}, which {\em includes} spin-axis tossing.}
\label{fig:rangeTester_no-tossing}
\end{figure*}

We show simulations in Fig.~\ref{fig:rangeTester_no-tossing} {\em without} spin-axis tossing. In all cases, these show poor agreement with the empirical LVK data, typically yielding p-values on the order of $\sim 0.001$.

\clearpage
\subsection{Mass reversal}\label{subsec:mass-reversal}
To account for systems in which the second-born BH is more massive than the first-born, mass reversal must be included as part of isolated binary evolution scenarios. Mass reversal occurs when the initially less massive star (i.e. the secondary at ZAMS) accretes sufficient mass to ultimately form the more massive --- and typically second-born --- BH. Conversely, the initially more massive primary forms the first-born BH, which ends up being the less massive of the two. For a general discussion of mass reversal and SN order in close binaries, see e.g. \citet{wl99,ts00,zdi+17}.

Based on data from LVK science runs O1--O3, \citet{mgbs22} estimate that up to 32\% of the BH+BH mergers were systems in which mass reversal has taken place. This conclusion is based on the number of observed cases in which the more massive BH has a significantly larger spin compared to the least massive BH. This circumstance is thus interpreted as the fraction of mass reversals because the second-born BHs are expected to achieve higher spins due to tidal interactions between the first-born BHs and the helium (Wolf-Rayet) stars that later collapse and produce the second-born BHs \citep{kzkw16,hp17,zkk18}. 

In the previous sections, mass reversal was not considered, although, as a consequence of the overlapping beta distributions for $\chi_1$ and $\chi_2$ (Fig.~\ref{fig:chi12-dist}), in 2.2\% of cases the spin of the larger mass BH ($M_{{\rm BH,1}}$) was greater than the spin of the less massive BH ($M_{\rm BH,2}$). Thus, in order to include mass reversal in the Monte Carlo simulations, a certain fraction of the masses, after they are initially chosen from the $KDE(\text{LVK data})$ distribution, are interchanged to allow for $M_{\rm BH,2}>M_{\rm BH,1}$. The spin distributions for the first- and second-born BHs ($\chi_1$ and $\chi_2$), however, are fixed such that $\chi_2>\chi_1$ (because of the tidal interactions mentioned above between the first-born BH and its helium star companion). 
The assumptions for the simulations presented in this section are thus as follows:
\begin{itemize}
    \item Pre-SN He-star mass of $M_{\rm He}=M_{\rm BH,2}/0.8$.
    \item Mass reversal in a fixed fraction of all systems.    
    \item Uniform pre-SN orbital separation of $a_i \in [4,40]\;R_\odot$.
    \item Constant SN kick of $w=50\;{\rm km\,s}^{-1}$ (isotropic direction).
    \item Spin-axis tossing of second-born BH (isotropic direction).
    \item $\chi_1$ distributed as $\beta(1.01,9.69)$.
    \item $\chi_2$ distributed as $\beta(2.72,2.34)$.
\end{itemize}
i.e. similar to Fig.~\ref{fig:rangeTester} panel~B and Fig.~\ref{fig:tossingQ} panel~A, but now including mass reversal.

\begin{figure}[h]
\vspace{-0.9cm}
\hspace{-0.3cm}
 \includegraphics[width=1.15\columnwidth]{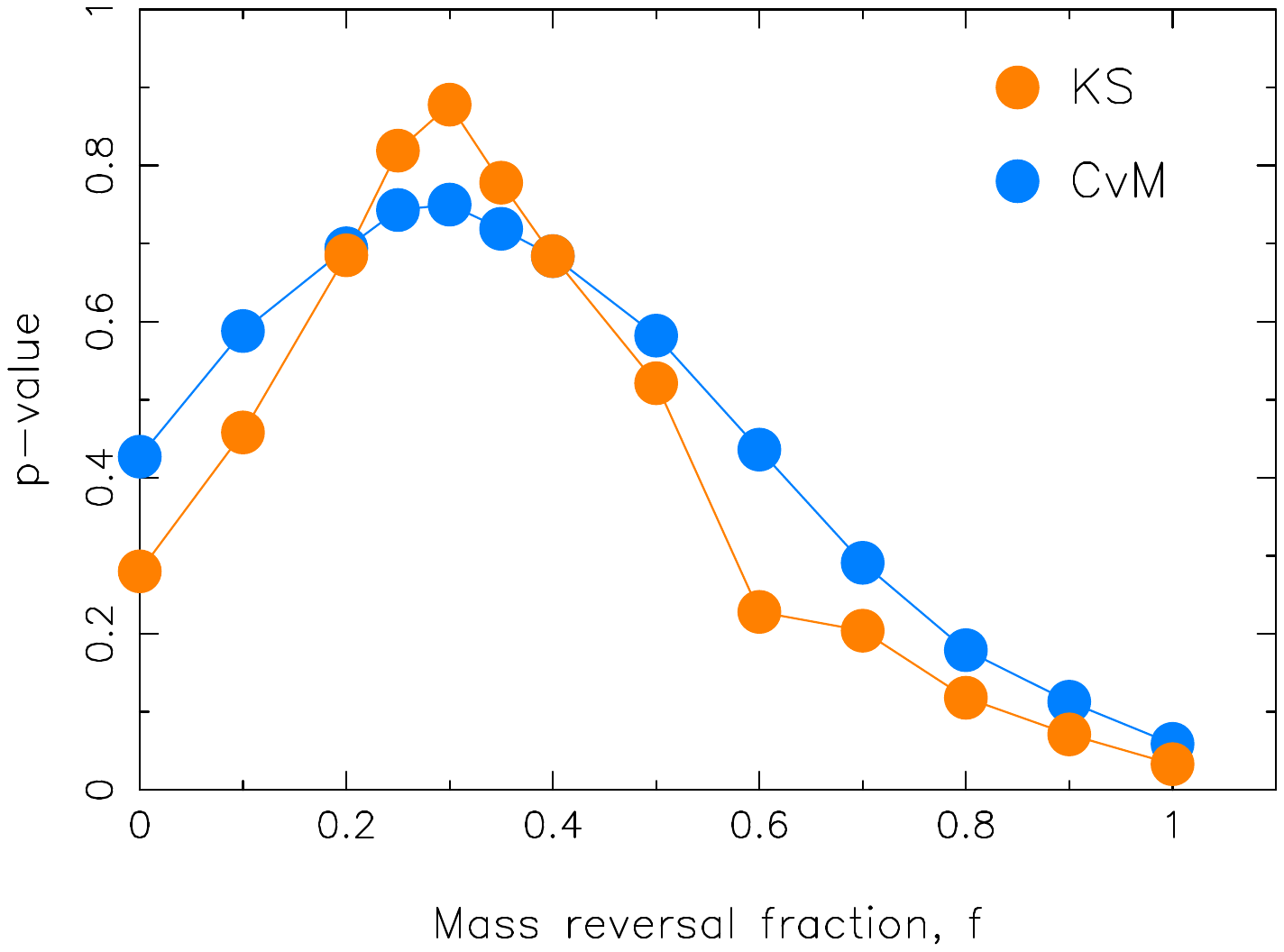}
 \caption{Resulting p-values, from comparison between simulated BH+BH systems and the empirical LVK data using the KS- and CvM-tests, as function of mass reversal fraction, $f$. The peak is at $f\simeq 0.30$.}
 \label{fig:mass-reversal-KS}
\end{figure}

Figure~\ref{fig:mass-reversal-KS} and Table~\ref{Table:mass-reversal} show that choosing the fraction of binaries with mass reversal to about 30\% results in significantly better agreement with the empirical data. Our result is thus very much in line with \citet{mgbs22} who argued for a mass reversal fraction up to 32\% for the recorded BH+BH mergers.  
We also notice that including mass reversal yields a final $\chi_{\rm eff}$ distribution with slightly smaller peaks and a wider range of values, because more BH merger systems obtain values in the tails of the distribution. 
(The change in the 5\% percentile value is $-0.045$, and the change in the 95\% percentile value is +0.029). Thus, as a consequence, mass reversal acts to relax the requirement to the width of the distributions of $\chi_1$ and $\chi_2$, because when having $q>1$ in Eq.(\ref{eq:chi_eff2}), and tossing is included, large spin-parameter values are to a lesser extent necessary to obtain large negative value of $\chi_{\rm eff}$. 

   \begin{table}
   \centering
         \begin{tabular}{llll}
            \hline
            \noalign{\smallskip}
            Mass reversal      &  KS-test & CvM-test & AD-test\\
            \noalign{\smallskip}
            \hline
            \noalign{\smallskip}
            0\%                & 0.280 & 0.427 & 0.197 \\
            10\%               & 0.458 & 0.588 & 0.250$^\ast$  \\
            20\%               & 0.685 & 0.694 & 0.250$^\ast$  \\
            25\%               & 0.819 & 0.743 & 0.250$^\ast$  \\
            30\%               & 0.878 & 0.750 & 0.250$^\ast$  \\
            35\%               & 0.778 & 0.719 & 0.250$^\ast$  \\
            40\%               & 0.684 & 0.684 & 0.250$^\ast$  \\
            50\%               & 0.521 & 0.582 & 0.250$^\ast$  \\
            60\%               & 0.228 & 0.436 & 0.192  \\
            70\%               & 0.204 & 0.291 & 0.136  \\
            80\%               & 0.118 & 0.179 & 0.084  \\
            90\%               & 0.071 & 0.113 & 0.055  \\
            100\%              & 0.033 & 0.059 & 0.028  \\        
            \noalign{\smallskip}
            \hline
         \end{tabular}
         \caption{Mass reversal and resulting p-values. See Fig.~\ref{fig:mass-reversal-KS}.\\$^\ast$ Note: AD p-values were capped at 0.250 \citep{ss87}.}
         \label{Table:mass-reversal}
   \end{table}
%

\subsection{Final simulations using LVK spin components}\label{subsec:final-sim}
\citet{aaa+23b} investigated the spin component distributions of BH+BH mergers (see their fig.~17 for 90\% credibility intervals of the fastest ($\chi_A$) and slowest ($\chi_B$) component spins among BH+BH mergers in GWTC-3).
Their credibility interval enables the possibility of finding the distributions of both BH spins, assuming that the second-born BH in each binary has the fastest spin. We found these spin component distributions by first choosing beta distributions for $\chi_1$ and $\chi_2$, and then comparing the KDE of $\chi_B = min(\chi_1,\chi_2)\,C_B$ and $\chi_A = max(\chi_1,\chi_2)\,C_A$ with fig.~17 in \cite{aaa+23b} ensuring that the distributions are within their 90\% credibility intervals. 
Here $C_A$ and $C_B$ are normalization constants. 

\begin{figure*}
\vspace{-1.0cm}
\centering
\includegraphics[width=0.30\textwidth]{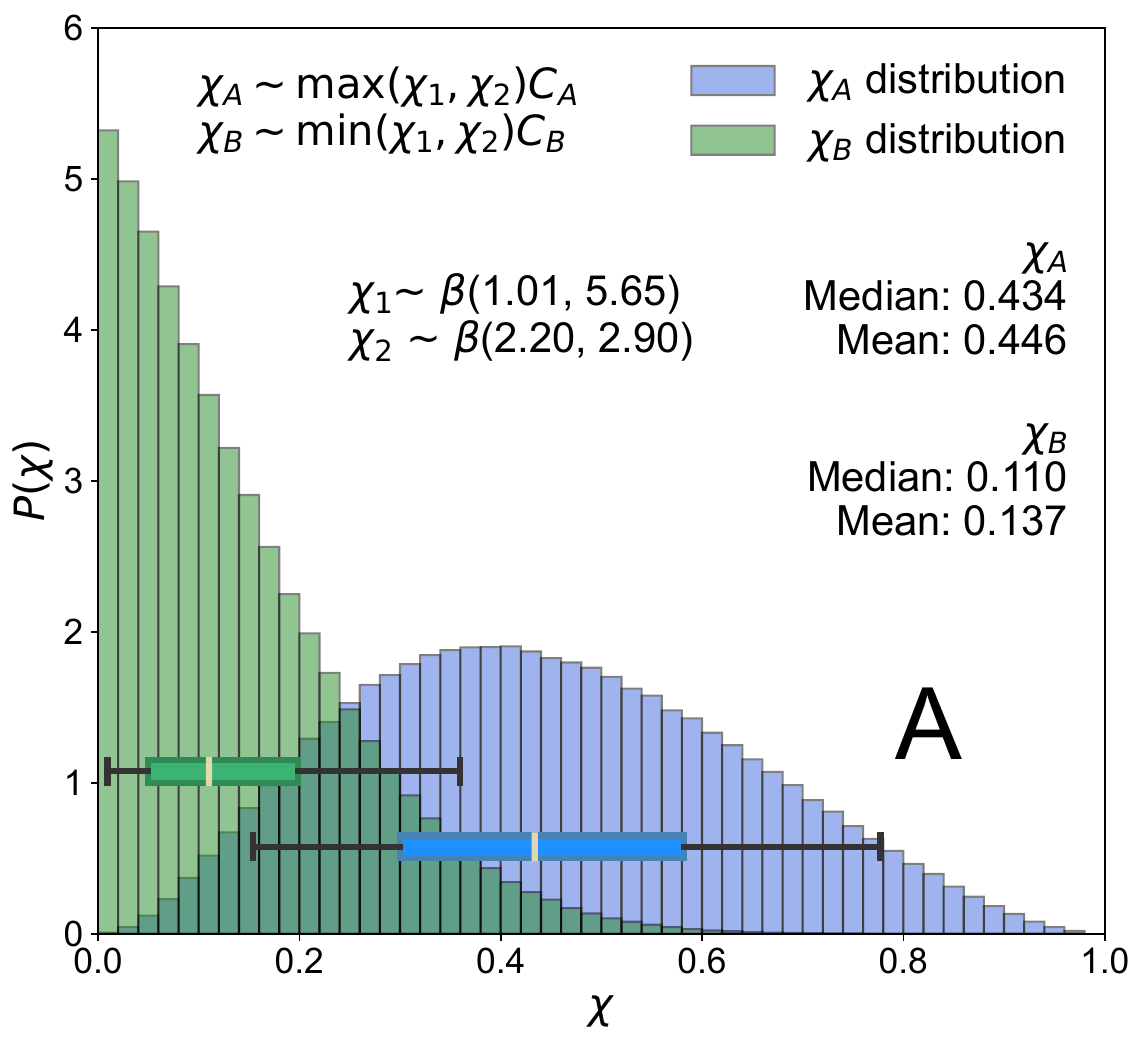}
\includegraphics[width=0.30\textwidth]{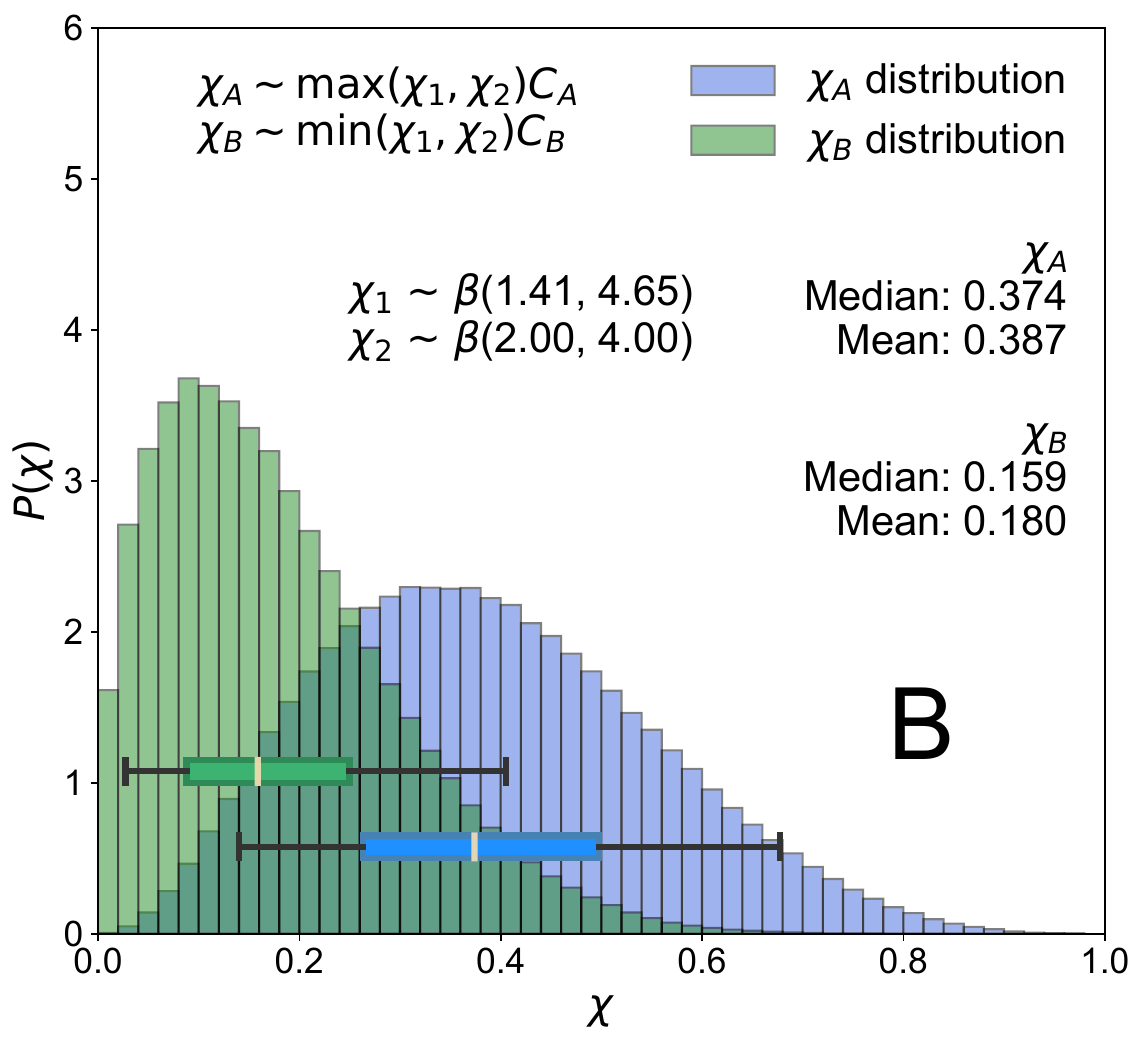}

\includegraphics[width=0.30\textwidth]{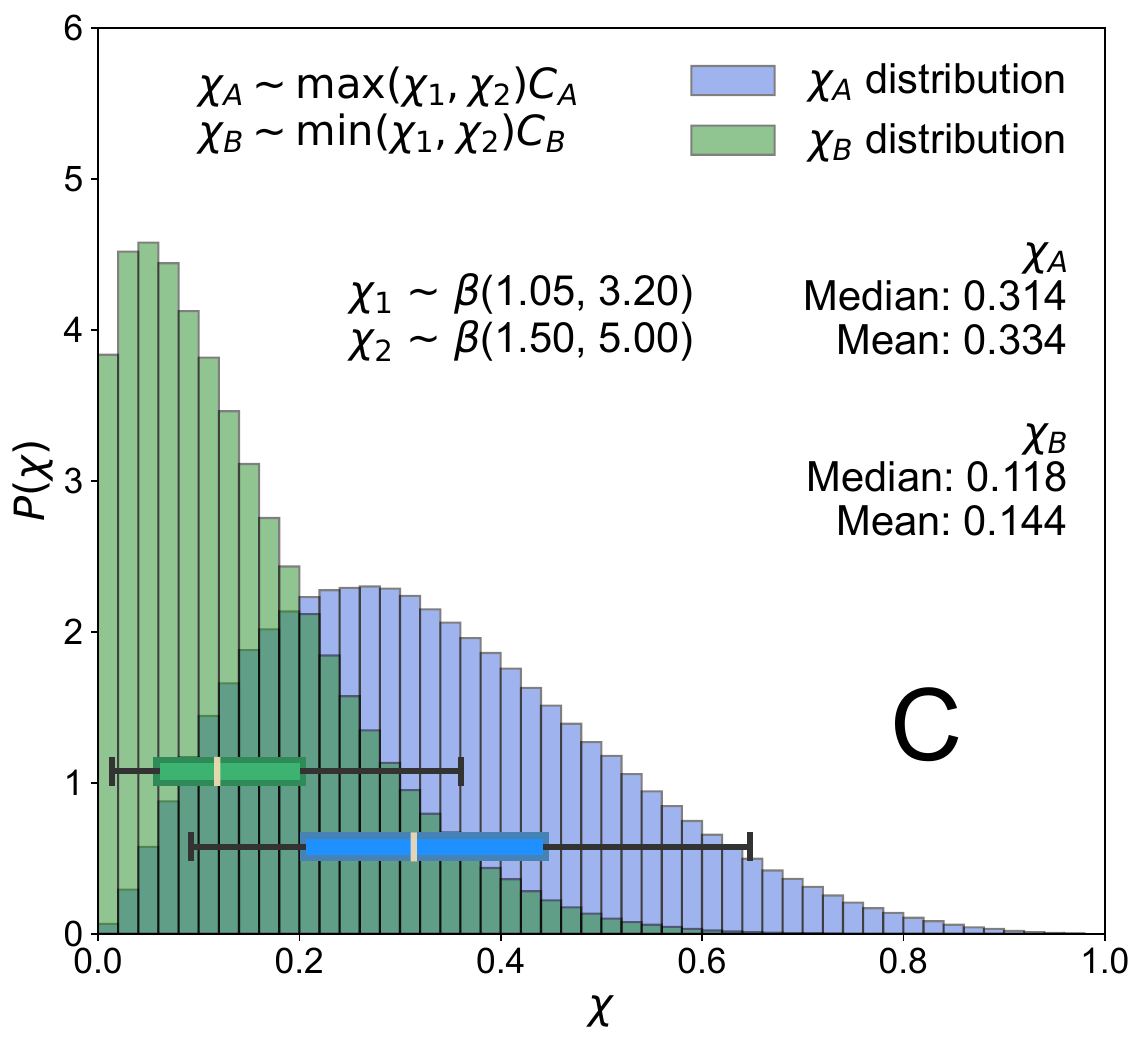}
\includegraphics[width=0.30\textwidth]{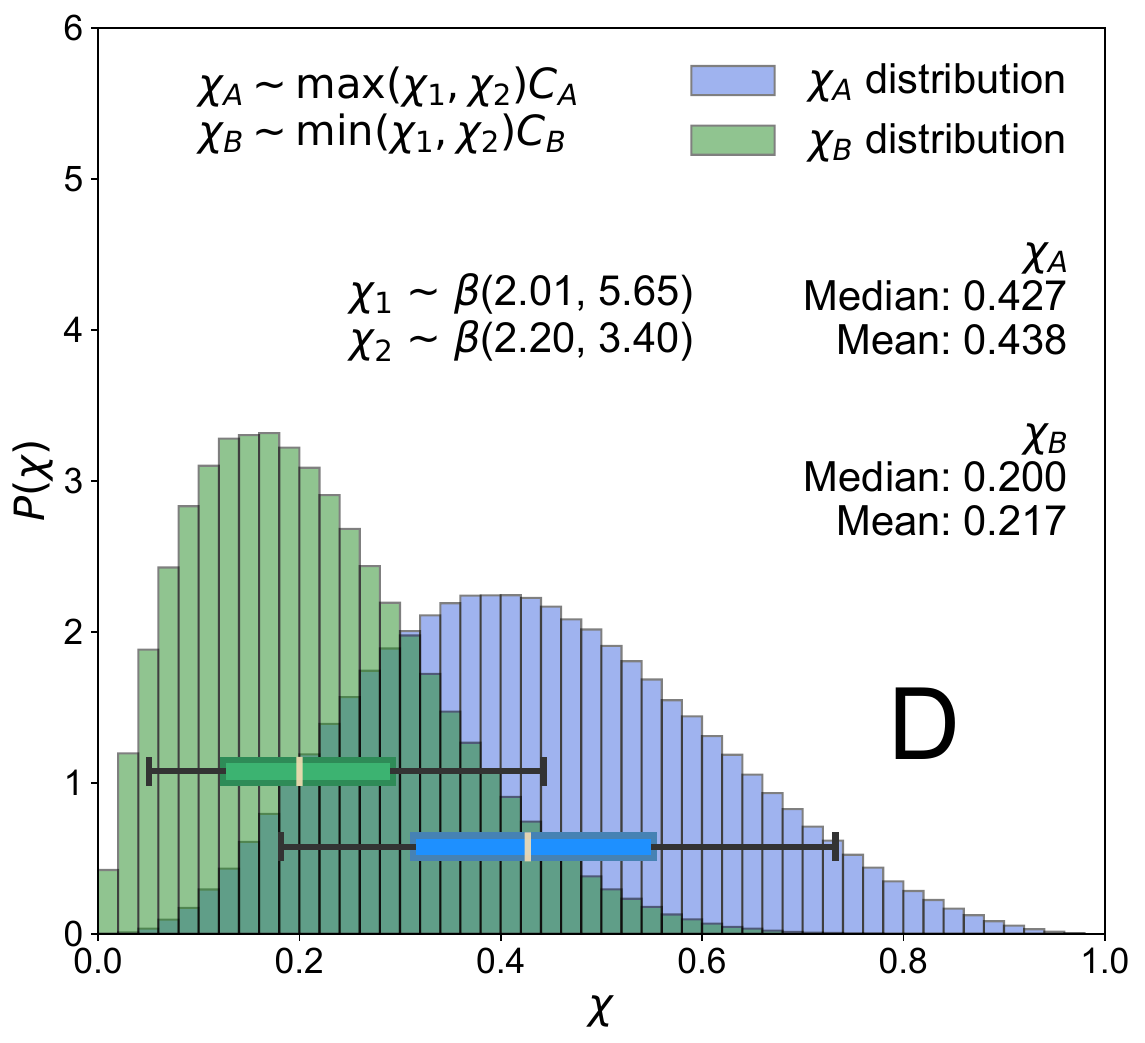}
\caption{Fast-component ($\chi_A$) and slow-component ($\chi_B$) BH spin distributions generated by fitting to the minimum and maximum 90\% credibility intervals from Fig.17 in \citet{aaa+23b}. Panel~A shows a fit with fast-spinning second-born BHs and slow-spinning first-born BHs. Panel~B presents a fit that lies in the middle of the 90\% credibility intervals, referred to as the median fit. Panel~C shows a fit with predominantly slow-spinning BHs, obtained by shifting the distributions as far left as allowed by the credibility intervals. Finally, Panel~D shows a fit with predominantly fast-spinning BHs, created by shifting the distributions as far right as allowed.
The underlying $\chi_{1,2}$ distributions are given in each plot, along with the median and mean values of $\chi_A$ and $\chi_B$.}
\label{fig:LVKchidits}
\end{figure*}

Figure~\ref{fig:LVKchidits} shows four examples of such distributions. Panel~A shows an extreme fit where $\chi_A$ has been maximized and $\chi_B$ minimized, to simulate systems where, in most cases, the spin component $\chi_A$ is much larger than $\chi_B$. Panel~B shows a median fit made to approximately fit the center of the 90\% credibility intervals. Panel~C shows the best fit for slow spinning BHs, which means that both spin components were shifted to the smallest values that the 90\% credibility intervals allow. Panel~D is the opposite case where both spin components were shifted to the highest values that the 90\% credibility intervals allow, thus resulting in pairs of relatively fast spinning BHs.
In our previous simulations, the pairs of $(\chi_1,\chi_2)$ were treated independently, which allowed for cases where $\chi_2 < \chi_1$ due to overlap in the beta distributions. To address this, the new spin distributions have been incorporated into our final simulations to better reflect the expectation that the second-born BH has a greater spin than the first-born BH, while also aligning with the expected BH spin distributions outlined in \cite{aaa+23b}. 

In our next simulations, we applied:
\begin{itemize}
    \item Pre-SN He-star mass of $M_{\rm He}=M_{\rm BH,2}/0.8$.
    \item Mass reversal in variable fractions of all systems. 
    \item Uniform pre-SN orbital separation of $a_i \in [4,40]\;R_\odot$.
    \item Constant SN kick of $w=50\;{\rm km\,s}^{-1}$ (isotropic direction).
    \item Spin-axis tossing of second-born BH (isotropic direction).
    \item First-born BH spin follows $\chi_B =  min(\chi_1,\chi_2)\,C_B$.
    \item Second-born BH spin follows $\chi_A = max(\chi_1,\chi_2)\,C_A$.   
    \item $\chi_{1,2}$ chosen according to the distribution in fig.~17 of\\ \cite{aaa+23b}, as explained in the main text.
\end{itemize}

\begin{figure*}
\hspace*{0.2cm}
\includegraphics[width=0.45\textwidth]{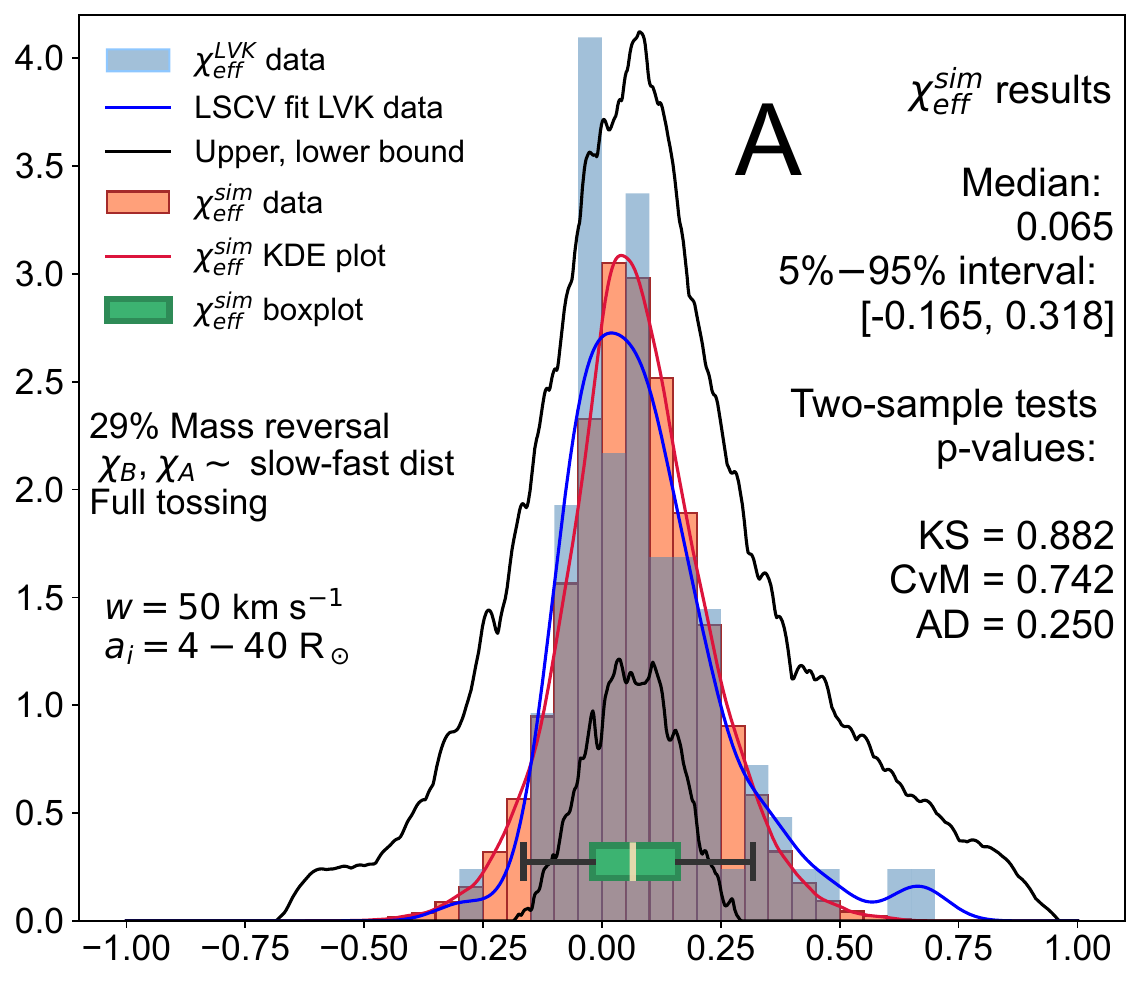}
\hspace*{1.0cm}
\includegraphics[width=0.45\textwidth]{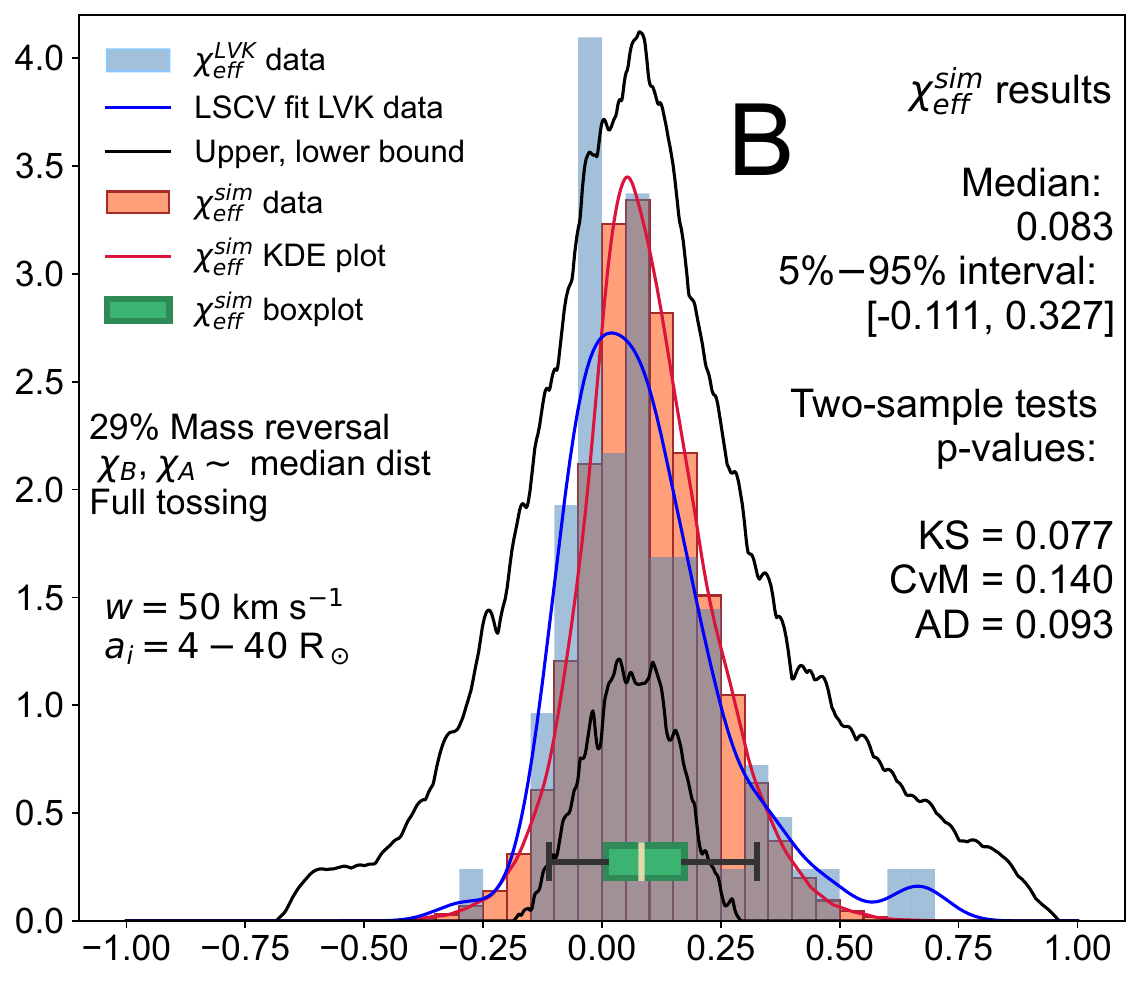}
\caption{Histograms showing the simulated $\chi_{\rm eff}$ distribution when assigning individual BH component spins according to different $\chi_A$ and $\chi_B$ distributions, with 29\% mass reversal included.
Panel~A shows the $\chi_{\rm eff}$ distribution based on the fast–slow $\chi_{A,B}$ distributions from panel~A of Fig.~\ref{fig:LVKchidits}, while panel~B uses the median distributions from panel~B of Fig.~\ref{fig:LVKchidits}. The y-axis shows the normalized density.
\label{fig:final_chi_eff}}
\end{figure*}

Figure~\ref{fig:final_chi_eff} shows our final simulations of $\chi_ {\rm eff}$ using the A and B distributions of the BH component spins from panels~A and B of Fig.~\ref{fig:LVKchidits}, respectively. The resulting p-values from the two-sided tests (KS, CvM, AD), see Section~\ref{sec:stat}, are (0.882, 0.742, 0.25) and (0.077, 0.14, 0.093) for distributions~A and B, respectively. The resulting p-values from distributions~C and D of Fig.~\ref{fig:LVKchidits}, not plotted here, are (0.104, 0.175, 0.058) and (0.034, 0.014, 0.013), respectively.

Both simulations in Fig.~\ref{fig:final_chi_eff} were performed using 29\% mass reversal, as this value provided the highest p-values from our two-sample tests against the empirical LVK data (in agreement with Section~\ref{subsec:mass-reversal} and Table~\ref{Table:mass-reversal}). This value is in agreement with \citet{mgbs22}, and also with \citet{aglt24} who argued that, under a simple ``single spin'' assumption, at least 28\% (90\% credibility) of the LVK binaries contain a primary BH with significant spin, indicative of mass-ratio reversal \citep[see also][who argued for an even higher mass-reversal fraction exceeding $\sim 0.70$]{bst22}.

We notice that the simulation of $\chi_{\rm eff}$ that produced by far the highest p-values from the two-sided tests is the one shown in panel~A of Fig.~\ref{fig:final_chi_eff} which is based on the individual BH spin components plotted in panel~A of Fig.~\ref{fig:LVKchidits}, where the difference in spins between the second-born BH (predominantly fast spinning) and the first-born BH (predominantly slow spinning) is the largest. This is expected as second-born fast-spinning BHs are necessary to obtain sufficiently large negative $\chi_{\rm eff}$ values.
The simulation shown in panel~B of Fig.~\ref{fig:final_chi_eff} shows a similar distribution as in panel~A, but shifted more towards the positive $\chi_{\rm eff}$ values, evident from both the median and the 90\% interval. 
The difference in p-values between panels~A and B of Fig.~\ref{fig:final_chi_eff} demonstrates the strong sensitivity to the underlying BH spin distributions, particularly regarding $\chi_B$. The B model suggests excessively high $\chi_B$ ($\chi_1$) values and/or excessively low $\chi_A$ ($\chi_2$) values, as both would result in higher $\chi_{\rm eff}$ values, as evident from Eq~(\ref{eq:chi_eff2}).

We reemphasize the p-values from the two-sample tests in panel~A of Fig.~\ref{fig:final_chi_eff}: 0.882, 0.742 and 0.250 (capped) for the KS-test, CvM-test and AD-test, respectively --- all of which are very high values compared to the traditionally chosen significance of 0.05, meaning that we cannot reject the $H_0$ hypothesis, namely that the LVK data and the simulated data, including BH spin-axis tossing, come from similar distributions. 
As none of our simulations without BH spin-axis tossing resulted in p-values $>0.001$ (no matter how all other parameters were optimized within reasonable limits, but see panel~E of Fig.~\ref{fig:rangeTester_no-tossing}), we conclude that our investigation provides further support for the spin-axis tossing hypothesis \citep{tau22} {\em if} isolated binaries dominate the formation channel of BH+BH mergers --- a topic which we now discuss in the next section.

\clearpage
\clearpage
\subsection{Isolated versus dynamical binaries}\label{subsec:dynamical}
So far, all our simulations have assumed that the BH+BH mergers detected by LVK were formed in isolated binaries. However, there are compelling reasons to believe that a significant fraction of BH+BH mergers may originate from dynamical formation channels \citep{pm00}. Such environments include e.g. globular clusters \citep{mo10,rzp+16} and AGN disks \citep{smh17,sbd+22}. A study by \citet{zbb+21} concluded that a mixture of formation channels is strongly preferred over any single channel dominating the detected population, and that no individual channel contributes more than $\sim\,$70\% of the observed BH+BH merger events.

Here we test this hypothesis by injecting a fraction $F_{\rm dyn}$ of BH+BH mergers with component spins randomly oriented in space. A purely dynamical population ($F_{\rm dyn}=1$) would produce a symmetric $\chi_{\rm eff}$ distribution centered around zero, which is inconsistent with observations \citep{Abbott_2021,Roulet+2021}. However, based on our statistical analysis, we cannot completely rule out this scenario (see Section~\ref{subsec:ErrorChiEff}).

A mixed origin involving both isolated binary and dynamical formation channels appears likely. For example, the Milky Way hosts 11 double neutron star systems expected to merge within a Hubble time due to GW radiation; two of these ($\sim\,$18\%) reside in a globular cluster, while the rest are located in the Galactic disk. This clear environmental diversity likely extends to BH+BH mergers as well.

For some proposed dynamical formation channels, such as those involving AGN disks, the expected spin-orbit alignment is less well understood (i.e. how BH component spins are oriented with respect to the orbital angular momentum vector). Given this uncertainty, we simply inject a population of BH+BH mergers with random (isotropic) BH component spin orientations into the simulated sample of isolated binaries. We perform this test both with and without BH spin-axis tossing for the isolated binaries, and compare the results to the empirical LVK sample from observing runs O1–-O3.

\subsubsection{Modelling procedure}\label{subsubsec:dynamical-modelling}
We constructed mixed populations of $N=10^5$ BH+BH systems of which a fraction, $F_{\rm dyn}$ represents a dynamical origin (with random final BH component spin directions), and a fraction $1-F_{\rm dyn}$ of the systems stem from our simulated sample of BH+BH mergers produced by isolated binary star evolution. For both populations, we applied similar distributions of BH masses, spin magnitudes, and pre-SN orbital separations as
done in Section~\ref{subsec:final-sim}.
We adopted fixed kicks of $w=50\;{\rm km\,s}^{-1}$ and a mass reversal fraction of 29\%. We performed two lines of simulations: with and without BH spin-axis tossing for the isolated binaries part of the systems.
The assumptions for the simulations of mixed systems presented here are therefore as follows:
\begin{itemize}
    \item BH masses from ePDFs (Fig.~\ref{fig:BimodalMBH1MBH2})
     using $KDE(\text{LVK data})$.
    \item Pre-SN He-star mass of $M_{\rm He}=M_{\rm BH,2}/0.8$.
    \item Mass reversal in 29\% of all systems.    
    \item Uniform pre-SN orbital separation of $a_i \in [4,40]\;R_\odot$.
    \item Constant SN kick of $w=50\;{\rm km\,s}^{-1}$ (isotropic direction).
    \item First-born BH spin follows $\chi_B =  min(\chi_1,\chi_2)\,C_B$.
    \item Second-born BH spin follows $\chi_A = max(\chi_1,\chi_2)\,C_A$.   
    \item $\chi_{1,2}$ chosen in accordance with panel~A of Fig.~\ref{fig:LVKchidits}.
    \item Fraction of dynamical origin binaries, $0\le F_{\rm dyn}\le 1$.
    \item Dynamical: random distribution of BH component spins.  
    \item Isolated: spin-axis tossing/no tossing of second-born BH.
\end{itemize}

\begin{figure}[t]
\vspace{-0.9cm}
\hspace{-0.3cm}
 \includegraphics[width=1.15\columnwidth]{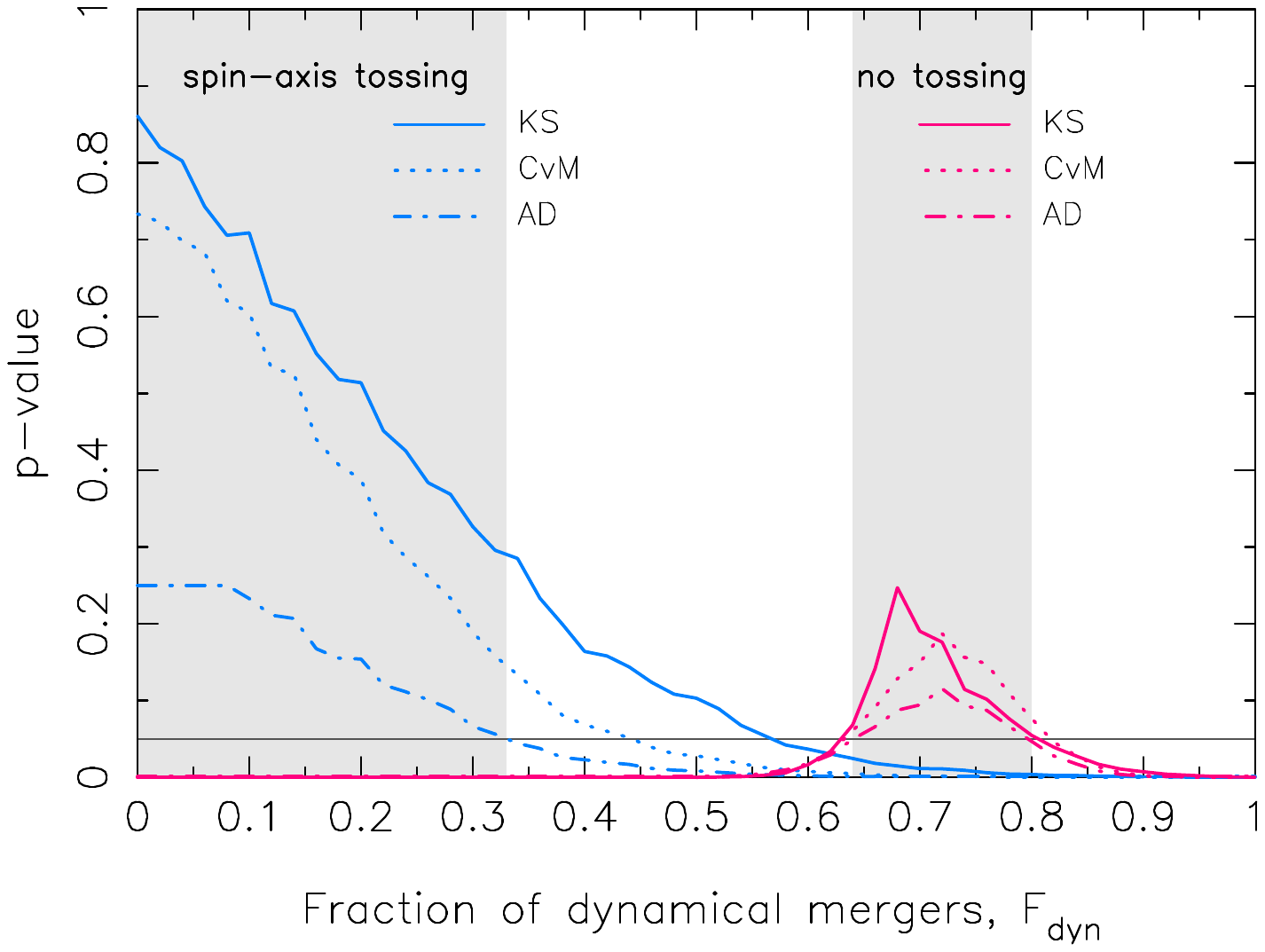}
 \caption{Resulting p-values from comparison between simulated mixed populations of BH+BH systems and the empirical LVK data, using the KS-, CvM-, and AD-tests, as function of the fraction of all systems with a dynamical origin, $F_{\rm dyn}$. (The fraction of systems produced in isolated binaries is thus $1-F_{\rm dyn}$). The three blue lines represent simulations that {\em include} BH spin-axis tossing of the second-born BHs in isolated systems. The three red lines represent simulations {\em without} BH spin-axis tossing.
 Two clear regions of solutions are shaded in gray color. The horizontal dark-gray line represents a critical p-value of 0.05.}
 \label{fig:mixed-p-values}
\end{figure}

\subsubsection{Resulting mixing fractions}\label{subsubsec:dynamical-results}
Figure~\ref{fig:mixed-p-values} shows a comparison between our simulations of a mixed population of BH+BH mergers and the empirical LVK sample, using the methodology and KS-, CvM-, and AD-tests described earlier. Two key results emerge from this figure:
\begin{enumerate}[i)]
\item if BHs {\em do} have their spin-axis tossed in a random direction at birth, the fraction of detected BH+BH mergers that come from a dynamical environment could be up to 33\% (possibly slightly higher), i.e. the fraction of BH+BH mergers produced by isolated binaries is most likely higher than 67\% (certainly higher than $\sim\,$50\%).
\item if there is {\em no} BH spin-axis tossing at the formation of BHs, we find that about $72\pm 8\,\%$ of all detected BH+BH mergers come from a dynamical environment if  their resulting BH spin component directions are random.
\end{enumerate}

\begin{figure}[t]
 \includegraphics[width=1.0\columnwidth]{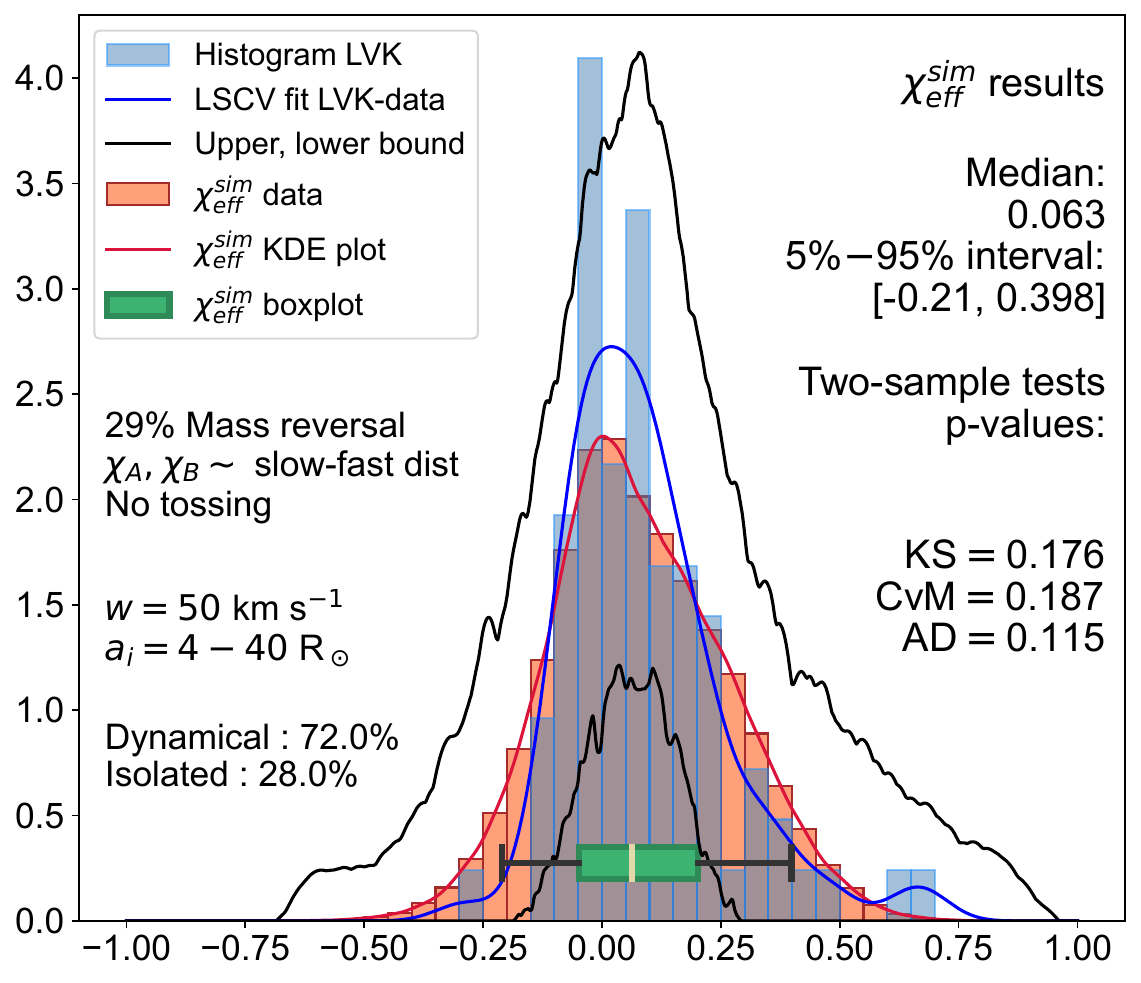}
 \caption{Expected distribution of $\chi_{\rm eff}$ (orange histogram) based on a mixed population of 28\% isolated binaries ({\em without} BH spin-axis tossing) and 72\% dynamical binaries for which we assumed random (isotropic) BH spin component directions. The resulting p-values for all statistical tests are above 0.05 and thus this solution cannot be rejected to explain the empirical LVK data (blue histogram).}
 \label{fig:mixed-no-tossing}
\end{figure}

In Fig.~\ref{fig:mixed-no-tossing} we plot an example for the second solution: a mixed population of 72\% BH+BH mergers that have a dynamical origin (random BH spin component directions) and 28\% BH+BH mergers that come from isolated binaries {\em without} BH spin-axis tossing.
The resulting p-values for all three statistical tests are above the critical limit of 0.05 and thus this solution cannot be rejected to explain the empirical LVK data. 

\begin{figure*}[b]
\hspace*{-0.1cm}
\includegraphics[width=0.48\textwidth]{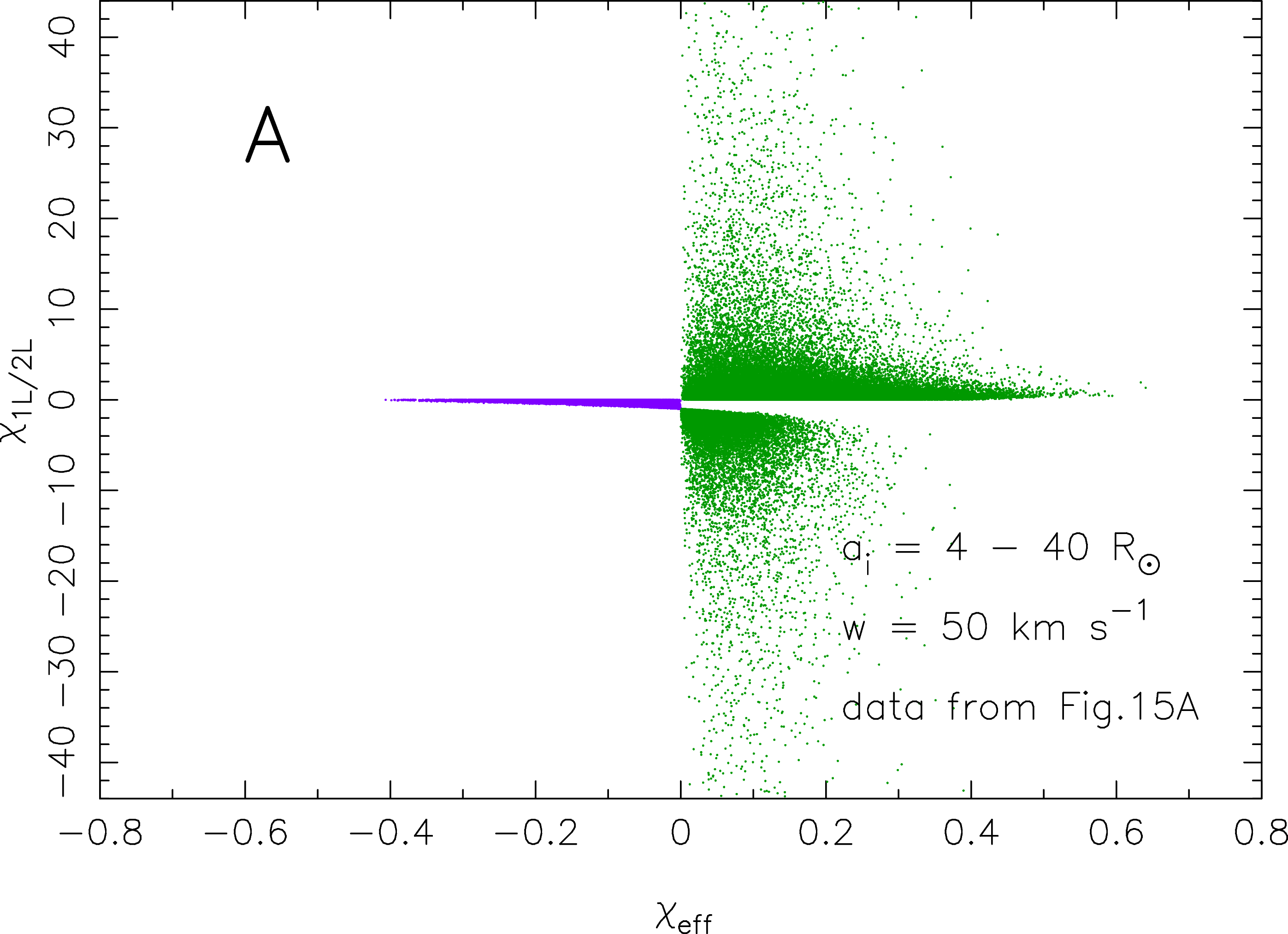}
\hspace*{0.6cm}
\includegraphics[width=0.48\textwidth]{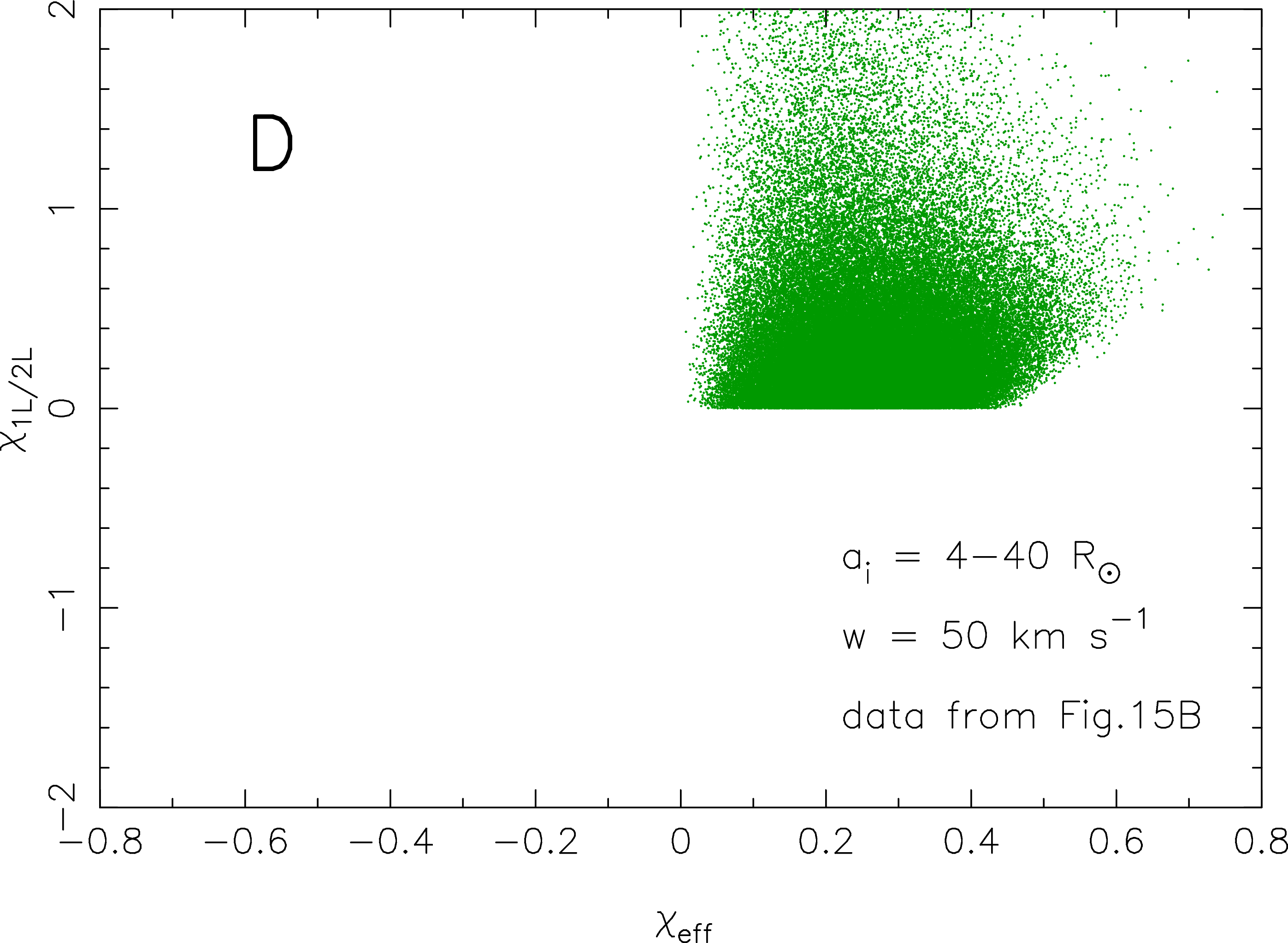}

\vspace*{1.2cm}
\hspace*{-0.1cm}
\includegraphics[width=0.48\textwidth]{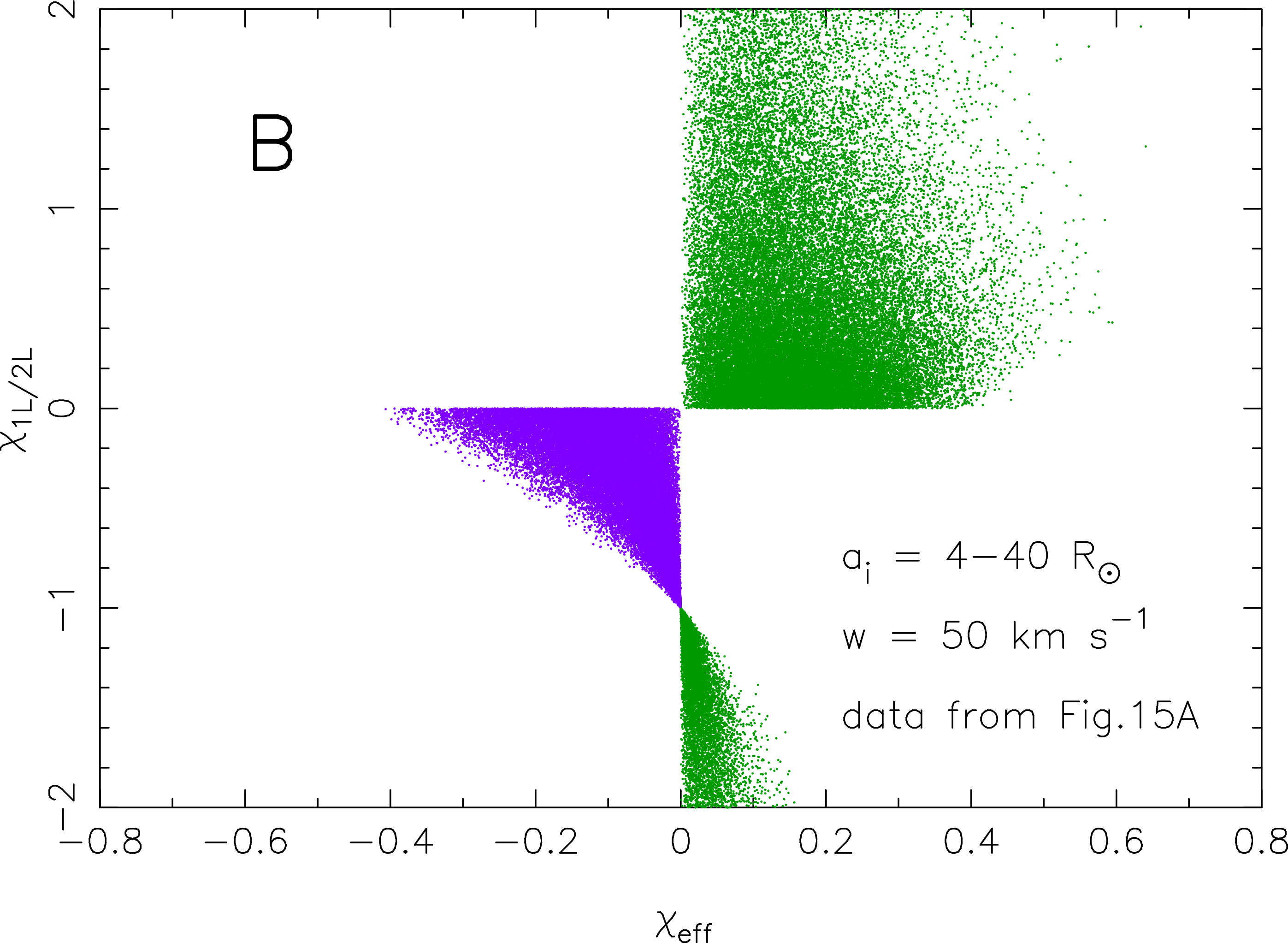}
\hspace*{0.6cm}
\includegraphics[width=0.48\textwidth]{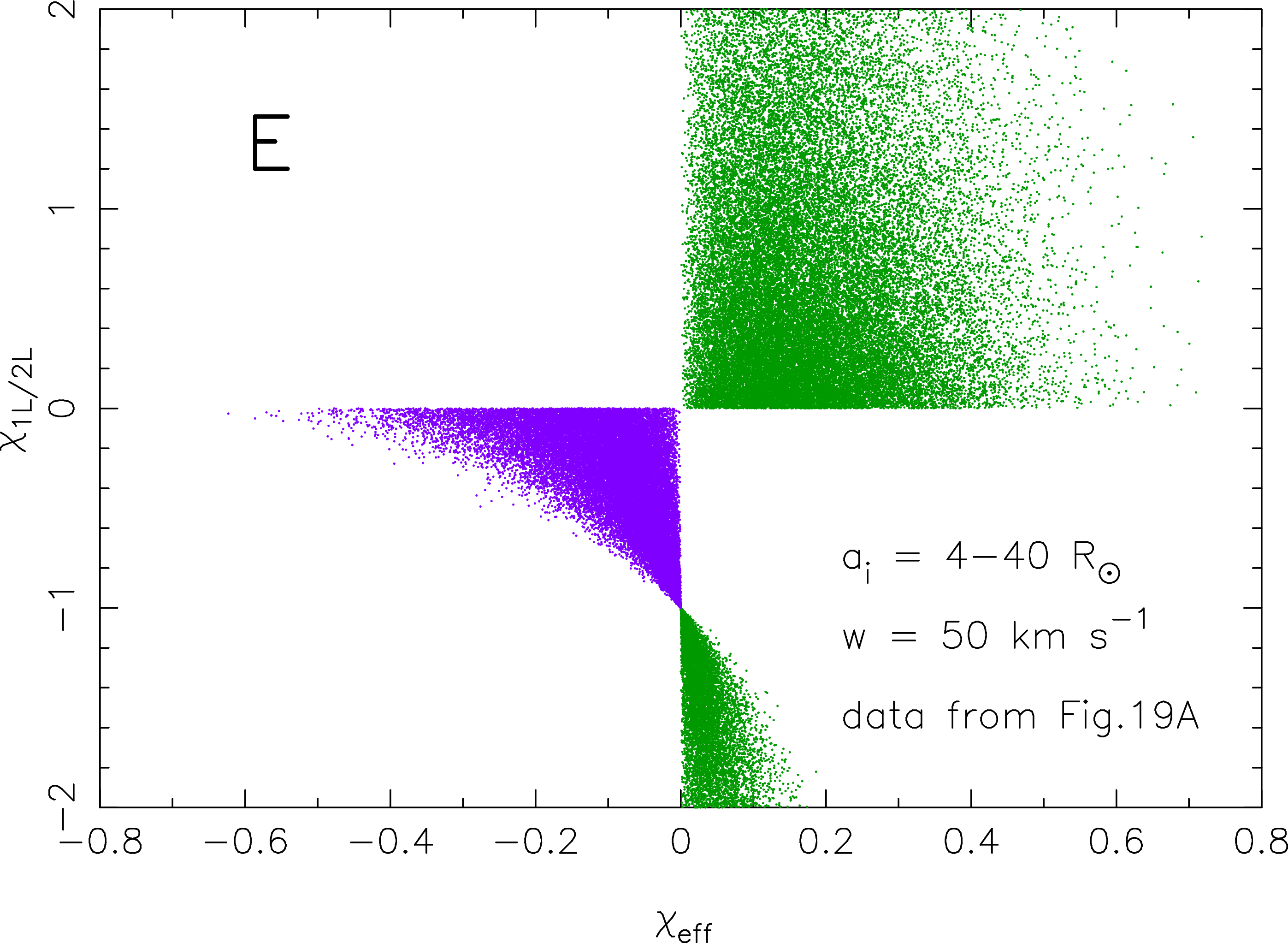}

\vspace*{1.2cm}
\hspace*{-0.1cm}
\includegraphics[width=0.48\textwidth]{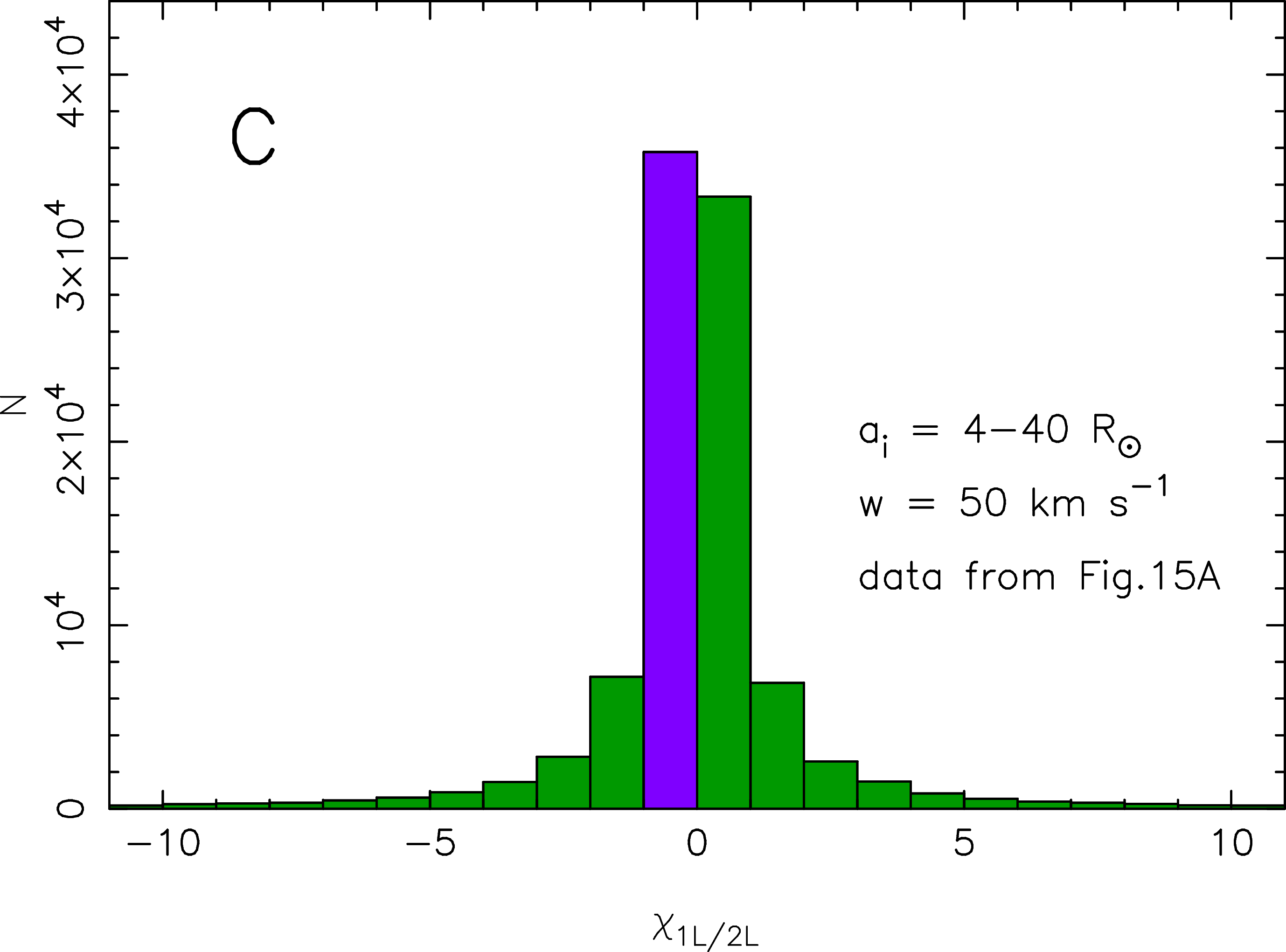}
\hspace*{0.6cm}
\includegraphics[width=0.48\textwidth]{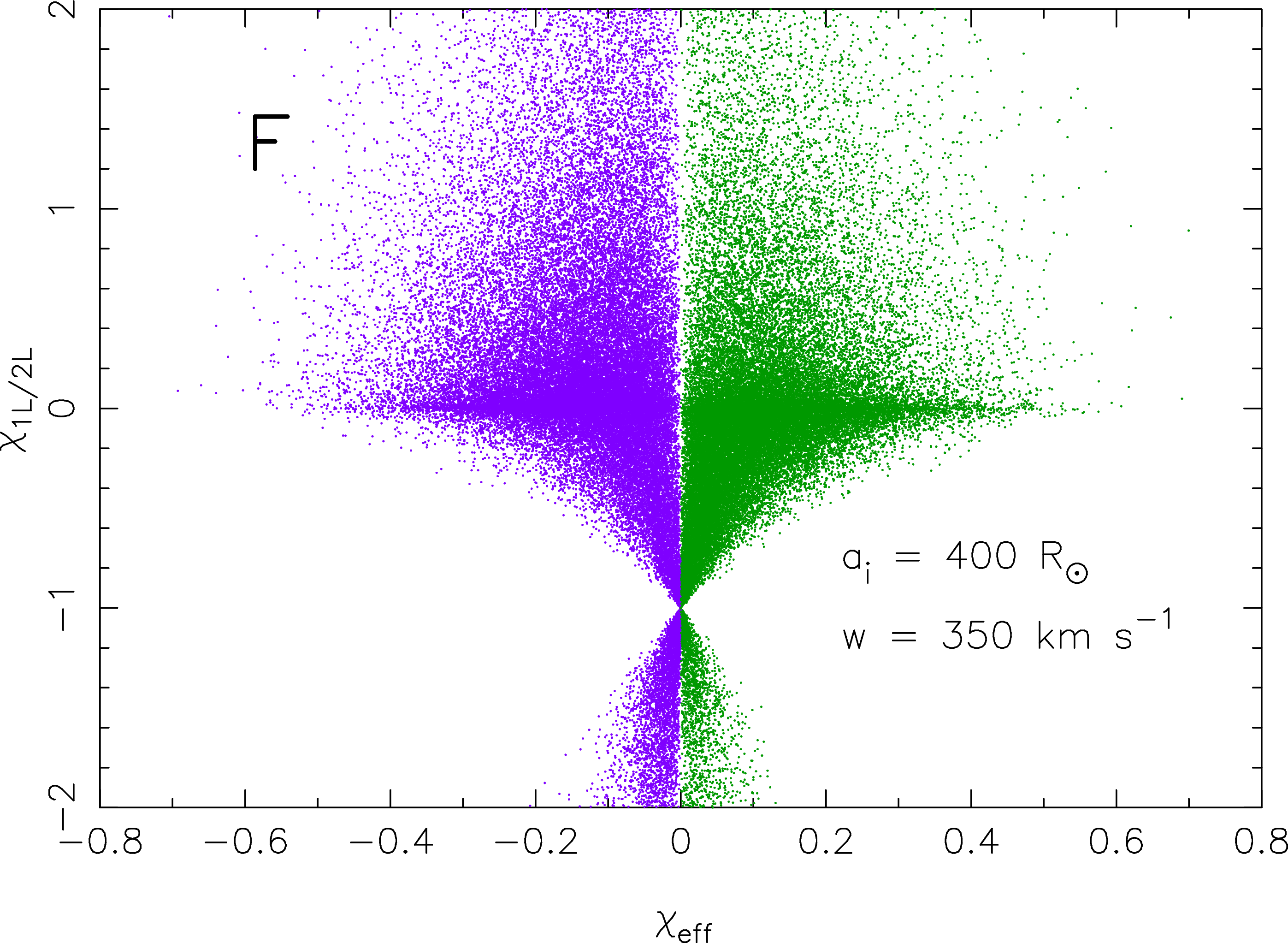}

%
%
\caption{Plots showing the relationship between $\chi_{\rm eff}$ and the projected spin ratio, $\chi_{1L/2L} \equiv (M_1\chi_1\cos\Theta_1) / (M_2\chi_2\cos\Theta_2)$ (i.e. ratio of the two individual mass-weighted BH spin components projected along the orbital angular momentum vector, $\vec{L}$). Data points in green (blue) correspond to cases where $\chi_{\rm eff} > 0$ ($\chi_{\rm eff} < 0$). Panels~A, B, and C are based on data from Fig.~\ref{fig:tossingQ}A (full BH tossing). Panel~D is based on data from Fig.~\ref{fig:tossingQ}B (no BH tossing). Panel~E is based on data from Fig.~\ref{fig:final_chi_eff}A (full BH tossing and mass reversal). Panel~F follows a similar setup as panel~E but uses $a_i=400\;R_\odot$ and $w=350\;{\rm km\,s}^{-1}$. See Section~\ref{subsec:relative-BH-spins} for discussion.}
\label{fig:chi12-relative}
\end{figure*}
\clearpage
\clearpage

With current data from LVK O1--O3, the statistical p-values for our simulations that consist of pure isolated or mixed populations up to a fraction, $F_{\rm dyn}\lesssim 0.33$ and {\em including} spin-axis tossing, however, are considerably higher (up to $p\simeq 0.88$) than those that result from a mixture of dynamical binaries and isolated binaries {\em without} spin-axis tossing ($p<0.25$). 
Redoing this investigation once the empirical BH+BH mergers from O4+O5 are published may enable us to put tighter constrains on the fraction of mergers with an isolated vs. a dynamical origin. And more importantly, future GW data might provide the final evidence for spin-axis tossing to occur in nature --- in addition to evidence from measurements of BH spin geometry in (near)Galactic X-ray binaries \citep[e.g.][]{pvb+22}.

\subsubsection{Hierarchical second- or third-generation BHs}
Our simulated BH+BH populations throughout this paper (with and without spin-axis tossing, isolated and dynamical binaries) have a deficit of mergers with $\chi_{\rm eff}>0.60$. However, the density of the LVK histogram in this region is based on just two data points. This limited representation introduces potential ambiguity when attempting to reproduce these specific $\chi_{\rm eff}$ values, especially given that the total current dataset consists of only 83 data points. However, these two data points may indicate that some of the detected BH+BH mergers could include hierarchical second- or even third-generation BHs \citep{rza+19,rkg+20,ktb+21} that are not modeled in our work.
Hence, based on this preliminary argument alone, we might at least expect some (minor) fraction of detected BH+BH mergers to have a dynamical origin. 
Similarly, our simulations seem to produce an excess of mergers with $\chi_{\rm eff}<-0.15$. But again, we still deal with small number statistics. 

\subsection{BH component contributions to $\chi_{\rm eff}$}\label{subsec:relative-BH-spins}

As argued in Section~\ref{sec:stat}, the BH spin tilts ($\Theta_1$ and $\Theta_2$) cannot be measured directly with current technology. Our study, therefore, focuses on measurable quantities that can be directly compared with simulations. Nevertheless, it is insightful to examine the contributions of each BH spin to $\chi_{\rm eff}$.  
To facilitate this, we introduce the {\em projected spin ratio}:
\begin{equation}
    \chi_{1L/2L} \equiv \frac{M_1\chi_1\cos\Theta_1}{M_2\chi_2\cos\Theta_2}\;,
\end{equation}
which represents the mass-weighted spin of the first-formed BH projected onto the orbital angular momentum vector, $\vec{L}$, divided by the corresponding contribution from the second-formed BH.

Figure~\ref{fig:chi12-relative} presents various plots illustrating the relationship between $\chi_{1L/2L}$ and $\chi_{\rm eff}$. Based on expectations from binary stellar evolution \cite[see e.g. summary of arguments in][]{tau22}, the first-formed BH is typically the more massive of the two but also the slowest spinning (see Sections~\ref{subsec:BH-spins} and \ref{subsec:mass-reversal}). If BH spin-axis tossing does not occur, the tilt angles remain small and equal ($\Theta_1=\Theta_2=\delta$, see introduction to Section~\ref{sec:MCMC}) because it typically requires a (unrealistic) kick exceeding $500\;{\rm km\,s}^{-1}$ to produce a post-SN retrograde orbit. In such a scenario, both BH spins contribute positively to $\chi_{\rm eff}$, meaning that only mergers with $\chi_{\rm eff}>0$ are expected. This prediction is confirmed by panel~D, which is based on simulated data from Fig.~\ref{fig:tossingQ}B. Additionally, the plot reveals that in most cases, $M_1\chi_1\cos\Theta_1 < M_2\chi_2\cos\Theta_2$, reflecting the fact that although $M_1 > M_2$, the spin magnitude of the first-formed BH is typically smaller ($\chi_1 < \chi_2$ by a factor of 2–4; see Fig.~\ref{fig:LVKchidits}).

If BH spin-axis tossing is included, however, a different scenario emerges where $\Theta_2 \gg \Theta_1 = \delta$, with $\delta \lesssim 6^\circ$ for BH kicks of order $w \sim 50\;{\rm km\,s}^{-1}$. 
Since $\Theta_2 > 90^\circ$ causes $\vec{\chi}_2$ to be anti-aligned with $\vec{L}$, spin-axis tossing can result in mergers with $\chi_{\rm eff} < 0$. This behavior is evident in panels~A and B (where panel~B provides a zoom-in of panel~A), using data from Fig.~\ref{fig:tossingQ}A. The histogram in panel~C shows that approximately 70\% of these systems have $-1 \leq \chi_{1L/2L} \leq 1$, indicating that the second-formed BH contributes most significantly to $\chi_{\rm eff}$.

Panel~E, based on data from Fig.~\ref{fig:final_chi_eff}A, includes mass reversal in $\sim\!30\%$ of the simulated BH+BH systems. As a result, cases with even more negative $\chi_{\rm eff}$ values emerge when $M_2 > M_1$. 

Notably, in all cases discussed above in which $\chi_{\rm eff} < 0$ (blue points) it follows that $-1 \leq \chi_{1L/2L} < 0$. For extreme cases with very large BH kicks and/or wide pre-SN orbits (panel~F, where $w = 350\;{\rm km\,s}^{-1}$ and $a_i = 400\;R_\odot$), the post-SN orbit can become retrograde ($\Theta_1 = \delta > 90^\circ$). Consequently, some systems with $\chi_{\rm eff} < 0$ may still have $\chi_{1L/2L} > 0$ if both BH spin components have anti-aligned projections onto $\vec{L}$. Alternatively, when $0 < \Theta_2 < 90^\circ$, systems may have $\chi_{1L/2L} < -1$. In other words, if the post-SN orbit is retrograde, $\chi_{1L/2L}$ can take a wide range of values within reasonable limits.

Looking ahead, the metric $\chi_{1L/2L}$ may serve as a key diagnostic for understanding the formation of BH+BH mergers and SN dynamics. If future third-generation detectors such as the Einstein Telescope and Cosmic Explorer are capable of measuring individual BH spins, they will enable direct observational constraints on $\chi_{1L/2L}$.


\section{Summary and Conclusions}\label{sec:conclusions}
We investigated the final formation stage of double BH systems in isolated binaries, using Monte Carlo simulations that included kicks and spin-axis tossing of the second-born BH, to reproduce the $\chi_{\rm eff}$ distribution of the empirical LVK data of BH+BH mergers.
For statistical analysis of our numerical experiments, we applied kernel density estimations (KDEs) and functional analysis (boxplots). 
Kolmogorov-Smirnov, Cramer von Mises, and Anderson-Darling tests were applied to compare our theoretical simulations with LVK data.

We began by parameterizing the empirical $\chi_{\rm eff}$ distribution with a Gaussian fit and found that the LVK effective spin data have a mean, median, and standard deviation of $0.09$, $0.06$, and $0.17$, respectively. The distribution was also found to be right-skewed. 
Two related questions were investigated: i) can negative effective spin parameters be excluded, and ii) whether the distribution of the effective spins is symmetric around $\chi_{\rm eff}=0$. 
We found no solutions for the data set where all BH+BH mergers have $\chi_{\rm eff}>0$ (the probability of this was found to be $3.01 \times 10^{-19}$, applying a simple uniform PDF over the credibility intervals). 
However, while the present data favor a slightly asymmetric distribution of $\chi_{\rm eff}$, a fully symmetric distribution cannot be entirely ruled out with the current limited data sample. Analysis of LVK science runs O4+05 might change this.
 
The spin component parameters of the two BHs ($\chi_1$ and $\chi_2$) were investigated by assuming underlying beta distributions. The results showed that a larger spin is required for the second-born BH, compared to the first-born BH, in agreement with findings of several other studies \citep[e.g.][]{bivv21,aaa+23b}.

Our analysis provides clear evidence supporting the spin-axis tossing hypothesis put forward by \citet{tau22}, but only if isolated binaries dominate the formation channel of BH+BH mergers (Fig.~\ref{fig:final_chi_eff}A).
Given reasonable initial conditions of pre-SN orbital separation, masses and applied kick magnitudes ($w\le 350\;{\rm km\,s}^{-1}$), there is no possibility for obtaining $\chi_{\rm eff}<0$ without spin-axis tossing.
Whether the direction of the spin-axis tossing is random (isotropic) or not, is still an open question and more data is needed to resolve it.
Our simulations produce a best fit to the empirical LVK data for relatively small kicks ($w\sim 55\;{\rm km\,s}^{-1}$, Fig.~\ref{fig:heatmap-p-values}) and tight pre-SN orbits ($a_i \in [4,40]\;R_\odot$, Fig.~\ref{fig:f_detect}).

Mass reversal in isolated progenitor binaries of BH+BH mergers was tested too. 
The possibility of mass reversal flattens the $\chi_{\rm eff}$ distributions in a way that the central peak becomes slightly less defined and also results in more negative effective spins.
Our best results (i.e. highest p-values when comparing simulations to empirical data) were obtained using spin-axis tossing and a mass-reversal fraction of 29\% (Fig.~\ref{fig:mass-reversal-KS}). The Kolmogorov-Smirnov, Cramer-von Mises, and Anderson-Darling tests yielded in this case p-values of 0.882, 0.742, and 0.25 (capped), respectively, meaning that our simulations including BH spin-axis tossing and empirical data agree to a high degree. In stark contrast, all our statistical tests with reasonable input parameters produced p-values $\lesssim 0.001$ without spin-axis tossing (Fig.~\ref{fig:rangeTester_no-tossing}), thereby firmly rejecting the possibility of reproducing the empirical data using simulations of BH+BH mergers from isolated systems without BH spin-axis tossing.

Finally, we investigated simulations using a mixed population of BH+BH mergers that originated from isolated binaries and environments with dynamical interactions (assuming random spin directions of both BH components in the latter case). A comparison with the empirical LVK data resulted in two possible solutions (Fig.~\ref{fig:mixed-p-values}) that links the question of BH spin-axis tossing to the mixing ratio of formation channels of BH+BH mergers: 
\begin{enumerate}[i)]
\item  if BHs have their spin-axis tossed in a random direction at birth, the fraction of detected BH+BH mergers that come from a dynamical environment is less than about 33\%, i.e. the fraction of BH+BH mergers produced by isolated binaries is most likely higher than 67\%. 
\item if BHs are formed without spin-axis tossing, we find that a confined range of $72\pm 8\,\%$ of all detected BH+BH mergers {\em must} come from a dynamical environment. (This latter solution is less statistical significant compared to the former.)
\end{enumerate}

Because BH spin tilts ($\Theta_1$ and $\Theta_2$) cannot be measured directly with current GW detector technology and sensitivity, we introduce the projected spin ratio as a metric to help disentangle the individual BH component contributions to $\chi_{\rm eff}$.

The rich empirical data from LVK O4+O5 (and the 3G detectors, Einstein Telescope and Cosmic Explorer) will enable tighter constrains on both the formation channels of BH+BH mergers and the question of BH spin-axis tossing. 
We strongly recommend future population synthesis studies to include spin-axis tossing and urge further theoretical studies of the origin and degree of spin-axis tossing \citep[e.g.][]{Janka_2022}.

Finally, we recommend that the LVK Collaboration further investigate the question of spin-axis tossing by combining simulations and analyzing the empirical sample using their improved posterior probability distributions --- within the estimated 90\% credibility intervals for each measured $\chi_{\rm eff}$ --- and applying various waveform models for parameter estimation.

\section*{Acknowledgements} 
 This investigation was originally initiated as one semester FYS7 AAU PBL-style cross-disciplinary project at Aalborg University, supervised jointly by T.M. Tauris (astrophysics) and C.A.N. Biscio (math, statistics). We thank the anonymous referee (and the editor Simon Portegies Zwart) for their time and for the constructive, insightful, and encouraging comments.

\section*{Author contributions}
Conceptualization: T.M.T. 
Data curation: H.C.G.L, C.P.P.  
Formal analysis: H.C.G.L., C.P.P., A.S., C.L., C.A.N.B.  
Investigation: H.C.G.L., C.P.P., A.S., C.L.
Methodology: T.M.T, C.A.N.B.
Supervision: T.M.T., C.A.N.B.
Validation: H.C.G.L., C.P.P., T.M.T., A.S., C.L., C.A.N.B.  
Visualization: H.C.G.L., C.P.P., T.M.T.
Writing – original draft: H.C.G.L., C.P.P., T.M.T., A.S., C.L.
Writing – review and editing: H.C.G.L., C.P.P., T.M.T., C.A.N.B.

Declaration of generative AI and AI-assisted technologies in the writing process.
During the preparation of this work, the authors occasionally used ChatGPT to assist with language improvements. After using this tool/service, the authors reviewed and edited the content as needed and take full responsibility for the content of the publication.

\appendix
\section{Applied statistics theory}\label{Appendix:A}
\subsection{Kernel Density Estimation (KDE)}\label{appendix:KDE}
To describe this approach, let us begin by considering the following data sample $X_1, X_2, X_3, X_4, X_5,...., X_n$. The relative frequency of different values can be visualized by using a histogram. KDE can be described by formalizing the definition of a histogram. This can be achieved by aggregating the data points in intervals of the form  $(x-\Tilde{h}, x+\Tilde{h})$, the bin length $\Tilde{h}>0$, and $x \in \mathbb{R}$. The definition of histogram will be: 
\begin{equation}
    \label{eq:histogram2}
    \hat{f}_{H}(x, \Tilde{h}) = \frac{1}{2n\Tilde{h}} \sum_{i=1}^n 1_{\{x-\Tilde{h} < X_i < x+\Tilde{h}\}} \,.
\end{equation}
The above definition implies that for each data point $X_i$, the statement $x-\Tilde{h} < X_i < x+\Tilde{h}$ is considered. If the statement is true we add one. A normalized histogram is obtained by dividing the sum with ${2n \Tilde{h}}$.  
In the KDE, the histogram is smoothed by an appropriate kernel function, in other words, Eq.~(\ref{eq:histogram2}) becomes:
\begin{align}
    \label{eq:histogram3}
    \hat{f}_{{H}}(x, \Tilde{h}) &= \frac{1}{2n \Tilde{h}} \sum_{i=1}^n 1_{\{x-\Tilde{h} < X_i < x+\Tilde{h}\}} \notag \\
    &= \frac{1}{n\Tilde{h}} \sum_{i=1}^n \frac{1}{2} 1_{\{-1 < \frac{x-X_i}{\Tilde{h}} < 1\}} \notag \\
    &\approx \frac{1}{nh} \sum_{i=1}^n K \Big(\frac{x-X_i}{h} \Big).   
\end{align}
Thus the KDE is defined as:
\begin{equation}
    \hat{f}(x;h) = \frac{1}{nh} \sum_{i=1}^n K \Big(\frac{x-X_i}{h} \Big)\,, \label{eq:kde} 
\end{equation}
where $K$ is a suitable function called kernel.
It is important to note that now $h$ is used instead of $\Tilde{h}$, which denotes the bandwidth parameter, that gives the optimal KDE. In most cases $\Tilde{h} \neq h$. 

The generalization is to choose any density function $K$ called the kernel, that is centered around each data point $X_i$ from the sample. The expression $\sum_{i=1}^n K \Big(\frac{x-X_i}{h} \Big)$ is the summation of $n$ kernels. To guarantee good properties of the estimation in terms of biases, the kernel is assumed to be symmetric, centered, and unimodal around 0. 
The choice of kernel is important in getting the curve that represents the data best. 

We use a Gaussian kernel for the LVK data, and an Epanechnikov kernel for the permuted data used for the functional boxplots. The latter is chosen in order to not oversmooth the obtained KDE curves. 

The choice of bandwidth can lead to oversmoothing or overfitting: if the bandwidth is too small, overfitting occurs; and on the other hand if $h$ is too big, oversmoothing will occur. The necessary mathematical background for bandwidth estimation is presented below. 

\subsection{Choice of Bandwidth}
When using KDE, the bandwidth controls the smoothness of the estimator function, and how well the function describes some sample $X_1, X_2, ..., X_n$, drawn from a population $X$.
Therefore, choosing a bandwidth should be done with caution; but no consensus exists in the literature. The method chosen to calculate the bandwidth needs to be suitable for the situation. 
In this investigation two different methods have been chosen as to accommodate two different situations where an estimate for the bandwidth is needed. These will now be described in Appendices~\ref{appendix:LSCV} and \ref{appendix:simple-bandwidth}.

\subsubsection{Least-Squares Cross-Validation}\label{appendix:LSCV}
Firstly, a method for estimating the bandwidth for one specific sample $X_1, X_2, ..., X_n$ is considered. This can be done by minimizing the estimation error for the population density $f(x)$, where an error criterion could be the integrated squared error (ISE),

\begin{align}
    \mathrm{ISE} [\hat{f}(\cdot ; h)] := \int (\hat{f}(x;h) - f(x))^2 \mathrm{d}x,
\end{align}
here $\hat{f}(\cdot ; h)$ is the estimator of $f(x)$ over the entire real line $\mathbb{R}$. It is seen that the ISE is the squared distance between the estimator function and the actual density. 
From this error criterion it is, however, hard to find an optimal bandwidth, as the ISE depends on the sample which is being used to estimate the actual density of the population. It is therefore useful to introduce the mean integrated squared error (MISE),
\begin{align}
    \mathrm{MISE} [\hat{f}(\cdot ; h)] &:= \mathrm{E} \left[\mathrm{ISE} [\hat{f}(\cdot ; h)]\right] =\mathrm{E} \left[\int (\hat{f}(x;h) - f(x))^2 \mathrm{d}x \right],
\end{align}
as it is easier to find solutions for \citep{KernelSmoothing23}. 
The MISE can then be used to estimate a bandwidth using cross validation, which means that the sample is used firstly to compute some KDE for a given bandwidth and then secondly to evaluate it at that bandwidth. To do this, the MISE is written out into the terms:
\begin{align}
    \mathrm{MISE} [\hat{f}(\cdot ; h)] 
     = & \;\mathrm{E} \left[\int \hat{f}(x;h)^2 \mathrm{d}x \right] -2\mathrm{E} \left[\int \hat{f}(x;h)f(x) \mathrm{d}x \right] \nonumber \\ 
    & + \int f(x)^2 \mathrm{d}x,
\end{align}
because the last term does not depend on $h$, only the first two terms need to be minimized in order to minimize the MISE. The last term only shifts the MISE.
These terms are, however, unknown if the prior $f(x)$ is not given, but they can be estimated. To do this an expression, which does not depend on the prior, for the second term is needed. By assuming $f(x) \mathrm{d} x = \mathrm{d} F_n (x)$ and using the Lebesgue–Stieltjes integral \citep{Garcia-Portugues2023} grants the following: 
\begin{equation}
    \int \hat{f}(x;h) \mathrm{d}F_n(x)= \mathrm{E}[\hat{f}(X;h)] \approx \frac{1}{n} \sum^{n}_{i=1} \hat{f}(X_i;h).
\end{equation}
The leave-one-out estimator is now defined such that it can replace $\hat{f}(X_i;h)$ in the sum above in order to get rid of the dependency of the sample $X$: 
\begin{equation}
    \hat{f}_{-i}(x;h) = \frac{1}{n-1} \sum^{n}_{\substack{j=1 \\ j \neq i}}  K\left(\frac{x - X_j}{h} \right).
\end{equation}
The two terms can now be estimated to obtain the least-square cross-validation (LSCV) function:
\begin{equation}
    \mathrm{LSCV} (h)  =\int \hat{f}(x;h)^2 \mathrm{d}x  -\frac{2}{n} \sum^{n}_{i=1} \hat{f}_{-i}(X_i;h).
\end{equation}
This function can be used to minimize the error of an estimator with a given kernel considering different bandwidths. The optimal bandwidth is defined as: 
\begin{equation}
     \hat{h}_{LSCV} = \mathrm{arg \ min}\ \mathrm{LSCV}(h),
\end{equation}
which means that the argument $h$, where the LSCV is minimal, is the optimal bandwidth \citep{KernelSmoothing_3_3}. 

\begin{figure}
\hspace{1.0cm}
 \includegraphics[width=0.75\columnwidth]{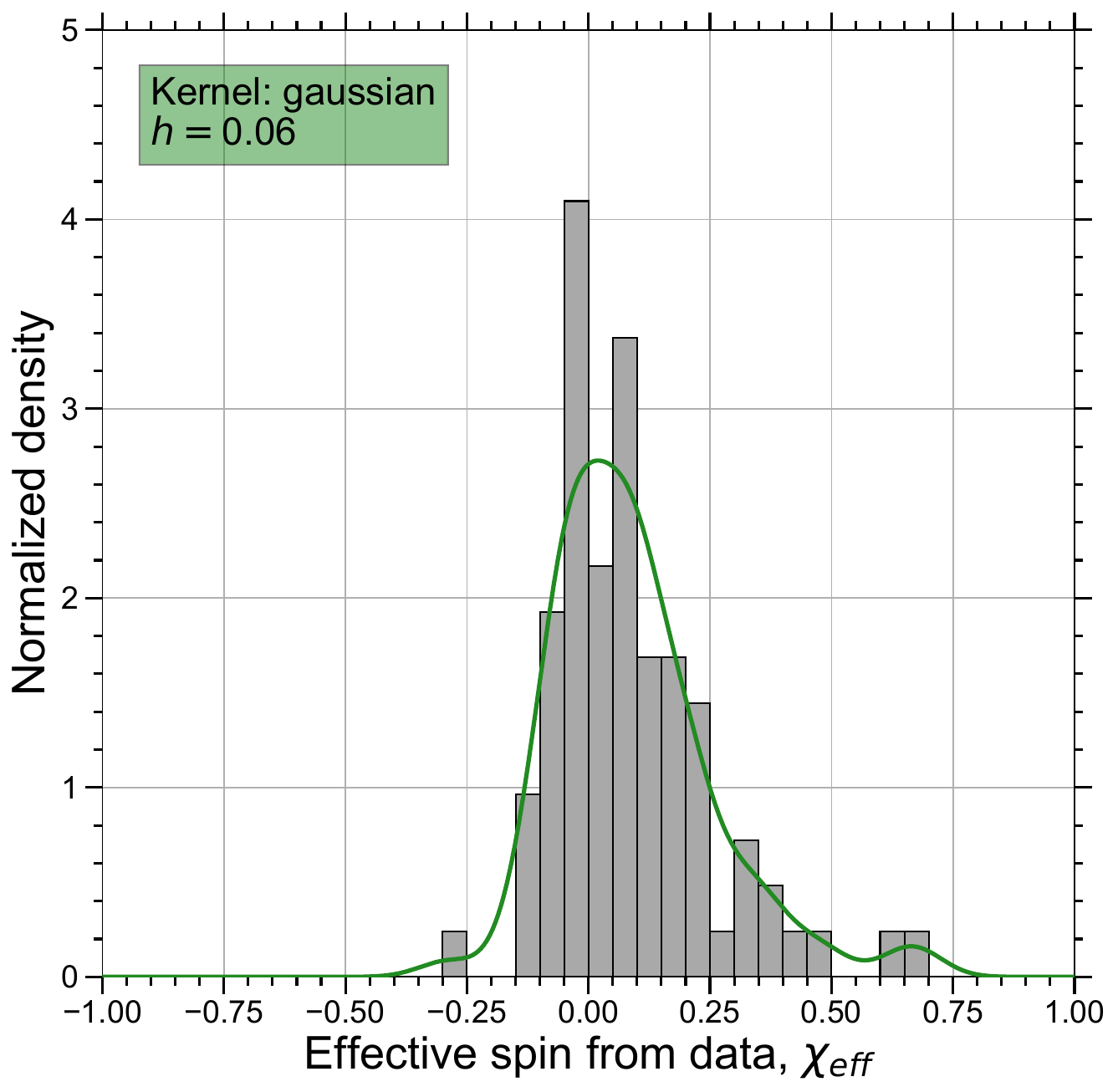}
 \caption{KDE (green line) of the effective spin distribution from LVK data using a Gaussian kernel. The bandwidth which minimized the LSCV function has been applied. The LVK data is also shown as a histogram.}
 \label{fig:LSCV_gaussian}
\end{figure}

An example of using LSCV is seen in Fig.~\ref{fig:LSCV_gaussian}, where a Gaussian kernel has been used and the bandwidth was estimated to be $h=0.06$. The KDE of the LVK data is seen to be more smooth, however, this is a consequence of the Gaussian kernel. 

\subsubsection{Quick and Simple Bandwidth Selector}\label{appendix:simple-bandwidth}
The second case, where another type of bandwidth selector is needed, is when computational limitations require a quick and simple method. For example, when bandwidths, for a lot of different samples, need to be calculated. In this investigation a bandwidth estimator from the \texttt{scikit-learn} library in \texttt{Python} has been used \citep{scikit-learn}.
This bandwidth selector uses a nearest neighbor approach to compute the bandwidth of some sample. If a data set $X_k$ of size $k$ is given, then it is the distances to the $n$'th nearest neighbors, where $n = 1,2,\dots, k-1$, that is used to calculate the bandwidth for a data set. The set of distances between the neighbors to the $i$'th point, is defined to be: 
\begin{align*}
    d^i := \left\{|  x_j - x_i  | , \ j \neq i   \right\},  
\end{align*}
where $j, i=1,2,...,k$ and $x_j, x_i$ are data points in $X_k$. The distance to the $n$'th nearest neighbor is denoted by $d^i_{(n)}$, where the subscript $(n)$ is notation for order statistics and means that the $n$'th smallest distance between $x_j$ and $x_i$ is used. 
From this the bandwidth is defined as:
\begin{align}
    h_{n} = \sum_{i=1}^k \frac{d^i_{(n)}}{k},
\end{align}
where a mean between the $k$ distances is used.

This method to find a bandwidth proves to be less computationally heavy than LSCV. The $n$'th neighbor must be chosen such that the KDE is somewhat reasonable when considering the sample.
As a rule of thumb, $n$ is often chosen to equal the $30^{\rm th}$ percentile of the number of data points, but no consensus exists.
However, we have chosen $n$ to equal the $15^{\rm th}$ percentile (for our data this means that $n=12$), which gives approximately the same bandwidth for the LVK median data, as the LSCV method.

\subsection{Functional Boxplot}\label{subsec:functionalboxplot}
In this investigation, simulations will be performed based on the BH+BH effective spins. A total of 83 different BH+BH events were observed with effective spins, each with a corresponding confidence interval. Many different fits can be obtained if the entire, or part of the confidence intervals for the effective spins are considered. In one histogram set one effective spin parameter from each confidence interval of the 83 different merger events will be used.
In this sense, many different histograms will lead to a lot of possible KDEs, and a method of finding the curve, that is the optimal KDE based on all the variations, is needed. 
In some sense, functional data analysis, which is applied to find the most central curve(s), can be considered to be a further abstraction of finding the average of data points. In this section, the fundamental theoretical framework for functional data analysis will be presented with an emphasis on practical applications to interpret the results. The notion of band depth (BD) and modified band depth (MBD) that is introduced in this section is quite robust and feasible computationally because the calculations do not require complicated mathematical functions.

Let us begin with a finite set of real functions such as $x_1(t), x_2(t), x_3(t),...,x_n(t)$, where each $x_i$ is a function depending on variable $t$. Only continuous functions which are defined on an interval $U$ are considered. The graph of a function $x_i$ will formally be defined as \citep{lopez2009concept}:
\begin{equation}
    G(x_i) = \{(t,x_i(t)): t \in U \}. \label{eq:funboxplot}
\end{equation}
The band delimited by the subset $\{x_{i_1},x_{i_2},x_{i_3},....,x_{i_k}\}$ is defined as:
\begin{align}
  & B(x_{i_1},x_{i_2},x_{i_3},....,x_{i_k}) = \nonumber \\ & \qquad\{(t,y):t\in U,\, \min_{r=1,2,..,k}\, x_{i_r}(t) \leq y \leq \, \max_{r=1,2,..,k} x_{i_r}(t) \} \label{eq:bdelimited}. 
\end{align}
The definition above means that when considering a subset of curves, the entire area between any pair of curves is covered. For instance, if a specific point $t$ is considered, then the inequality $\min_{r=1,2,..,k}\, x_{i_r}(t) \leq y \leq \text{max}_{r=1,2,..,k}\, x_{i_r}(t)$ implies that $y$ runs vertically from the $x_{i_r}$ which has the smallest value at $t$ to the $x_{i_{R}}$ with the largest value, at $t$. 
It is important to note that the delimited band $B$ will become an area no matter which subset of functions we consider. From the bands, a measure called band depth can be introduced in order to determine the centrality af each curve with respect to all the others. The formulae for the band depth is \citep{lopez2009concept}:
\begin{gather}
    BD^{(j)}_n(x) =  \binom{n}{j}^{-1} \sum_{1 \leq i_1 < i_2 <..<i_j \leq n } I\{G(x) \subseteq B(x_{i_1},x_{i_2},x_{i_3},....,x_{i_j})\}, \label{eq:BDj}
\end{gather}
where $x \in \{x_1, x_2,...., x_n\}$. This definition means that we are considering the proportion of bands $B(x_{i_1},x_{i_2},x_{i_3},....,x_{i_j})$ that cover the entire graph $G(x)$. $I$ is the indicator function which acts similar to a delta function i.e. if the statement $A$ is true then $I(A)$ is 1, otherwise $I(A) = 0$. This means that if the function $G(x)$ is entirely contained in one of the bands of $B(x_{i_1},x_{i_2},x_{i_3},....,x_{i_j})$ then $I = 1$. This is then repeated for each possible band of $B(x_{i_1},x_{i_2},x_{i_3},....,x_{i_j})$. The normalization factor $\binom{n}{j}^{-1}$ is multiplied when considering $j$ out of $n$ curves. It is paramount to remember that $j \geq 2$ since we are testing whether $G(x)$ lies between at least two other curves. Although in theory, $j$ can go from 2 to $n$, it is extremely computationally demanding to consider $j \geq 3$, especially for more than $10^6$ curves, which are used in this investigation. In numerical implementation, it is essential to limit $j=2$. The calculations would otherwise be too extensive. The band depth is given by: 
\begin{gather}
    BD^{(2)}_n(x) =  \binom{n}{2}^{-1} \sum_{1 \leq i_1 < i_2 \leq n } I\{G(x) \subseteq B(x_{i_1},x_{i_2})\}. 
\end{gather}
The sum $\sum_{1 \leq i_1 < i_2  \leq n }$ implies that one must consider all possible pairs of functions without regard to order. The total number of such combinations is $\binom{n}{2}$, by dividing with this factor it is ensured that $BD^{(2)}_n(x)$ is at most one. 
The curve with the largest band depth is defined as the median curve,
\begin{gather}
    \Hat{M}_n = \max_{x \in \{x_1,x_2,x_3,...,x_n\}} \,
 BD_n(x). \label{eq:centralcurve}
\end{gather}
The indicator function, $I$, was used as an important component in the definition of band depth; however, from a numerical implementation point of view, this function is tedious and unflexible. It is therefore practical to have a more flexible definition of band depth without indicator functions. This leads to the concept of MBD. One can begin by defining \citep{lopez2009concept}:
\begin{align}
    A_j(x) & = A(x;x_{i_1}, x_{i_2},x_{i_2},.....,x_{i_j}) \nonumber \\ & =\; \{ t \in U: \min_{r=i_1,i_2,..i_j} x_r(t) \leq x(t) \leq \max_{r=i_1,i_2,..i_j} x_r(t)\}.
    \label{eq:Ajx}
\end{align}
The function $x \in \{x_1,x_2,..,x_n\}$ and $2\leq j\leq n$. The definition Eq.(\ref{eq:Ajx}) can intuitively be understood as the amount of "time" that the curve of function $x$ lies between the functions $x_{i_1}, x_{i_2},x_{i_2},.....,x_{i_j}$. The proportion of time rather than the amount of time can be expressed as \citep{lopez2009concept}:
\begin{gather}
    \lambda_r(A_j(x)) = \lambda(A_j(x))/\lambda(U). \label{eq:timeprop}
\end{gather}
The length of an interval $U$ for instance is defined as $\lambda(U)$. A more flexible version of Eq.~(\ref{eq:BDj}) becomes:
\begin{gather}
    MBD^{(2)}_n = \binom{n}{2}^{-1} \sum_{1 \leq i_1 \leq i_2 <..<i_j \leq n } \lambda_r(A(x;x_{i_1}, x_{i_2},x_{i_2},.....,x_{i_j})). \label{eq:MBDJ}
\end{gather}

These tools are relevant when discussing the physics and result in the main text. 
They have proven useful in handling the LVK data.

\subsection{Test statistics}
\subsubsection{Two-sided Kolmogorov–Smirnov test}
Firstly, we consider the two-sided Kolmogorov–Smirnov (KS) test and briefly summarize its mathematical framework here. The KS-test is completely distribution-free as it does not make any prior assumptions about the types of distributions that are used to obtain the stochastic observations.
We assume that the samples $\{X_1, X_2,....,X_n\}$ and $\{Y_1, Y_2,....,Y_m\}$ come from unknown distributions, $F$ and $G$, respectively. In general, the two-sided KS-test is defined as  \citep{massey1951kolmogorov}:
\begin{equation}
    H_0:\, F =G\,,\quad \text{versus} \quad H_1:\, F \neq G\,.
\end{equation}
The null hypothesis, $H_0$ states that the two observation sets come from the same underlying distribution. The alternative hypothesis, $H_1$ states that the two observation sets come from different distributions. One should bear in mind that the alternative hypothesis does not provide information on how the distribution $F$ differs from the distribution $G$. 
The central parameter in assessing whether $H_0$ can be rejected or not is the p-value, which can at most be 1. In the case of $p=1$, there is no difference between the two samples and $F=G$ \citep{scott2015multivariate}. A value of $p=0.70$, for instance, implies that if one repeats the process of taking 10 samples from each of the underlying distributions $F$ and $G$ and compares all of $100$ different sample pairs each time, on average the null hypothesis can be rejected in 30\% of the comparisons. 
For the currently available LVK data, a significance value of 0.05 will be considered significant. Hence, we will comply to the general consensus that only if $p<0.05$ then the null hypothesis is rejected \citep{scott2015multivariate}.

The KS-test can be described by the supremum distance between $F_n$ and $G_m$, which are the empirical cumulative distribution functions (eCDF) of $F$ and $G$, respectively. For example, $F_n(x)$ can be written as \citep{massey1951kolmogorov}:
\begin{equation}
    F_n(x) = \frac{1}{n}\sum^n_{i=1} 1_{X_i \leq x} \,. \label{eq:ecdff}
\end{equation}
The notation above means that we take a value $X_i$ from the observation set $\{X_1, X_2,....,X_n\}$ and assess whether $X_i \leq x$. If this statement is true, we add one and else zero. This process will continue until all observations have been evaluated. 
One can define a parameter, $D_{n,m}$ which will be useful in understanding the underlying principle of a KS-test: 
\begin{equation}
    D_{n,m} = \sqrt{\frac{nm}{n+m}} \sup_{x \in \mathbb{R}}\big|F_n(x)-G_m(x)\big|\,. \label{eq:supdistance} 
\end{equation}
Although the definition is completely valid for any $x\in \mathbb{R}$, it is essential to remember that the only values of $x$ relevant for a KS-test are $x = X_i$ or $x = Y_j$, where $1\leq i \leq n$ and $1\leq j \leq m$. Since $x$ is a discrete variable in this case, it is computationally less demanding to provide an alternative definition. Firstly, the following definitions are required:
\begin{gather}
      D_{n,m,1} = \sqrt{\frac{nm}{n+m}} \max_{1\leq i \leq n} \Big|\frac{i}{n} - G_m(X_i)\Big|\,, \label{eq:supdistance1} \\
      D_{n,m,2}  =  \sqrt{\frac{nm}{n+m}} \max_{1\leq j \leq m} \Big|F_n(Y_j) - \frac{j}{m}\Big|\,. \label{eq:supdistance2} 
\end{gather}
In Eq.~(\ref{eq:supdistance1}), one starts by considering $ x \in \{X_1, X_2,....,X_n\}$ and find the value of $i$ that maximizes $\Big|\frac{i}{n} - G_m(X_i)\Big|$. Since $\{X_1, X_2,....,X_n\}$ is a well-ordered set of numbers, i.e. $X_i \leq X_{i+1}$, it is known that $F_n(X_i) = \frac{i}{n}$ by using Eq.~(\ref{eq:ecdff}). Similarly, in the definition in Eq.~(\ref{eq:supdistance2}) one can do the exact same thing with $ \{Y_1, Y_2,....,Y_m\}$ instead of $\{X_1, X_2,....,X_n\}$. It is clear that $D_{n,m} = \max\{D_{n,m,1}, D_{n,m,2}\}$.
In general, large values of $D_{n,m}$ suggest that the null hypothesis must be rejected, while small values imply the opposite \citep{massey1951kolmogorov}. 

There are well-established mathematical formulas for calculating the p-values and confidence intervals. These methods are beyond the scope of this paper. The KS-test is relatively simple and easy to implement, but it is more sensitive to central deviations (i.e. at small $|\chi_{\rm eff}|$ values) rather than deviations in the tails.
This should be taken into account when considering the $\chi_{\rm eff}$ distribution of BH+BH mergers (see Fig.~\ref{fig:GWTC3_BH+BH}), which is a main scope of this study.

\subsubsection{Cramér–von Mises test}
Another test, that is generally considered to be more robust than the KS-test, is the Cramér–von Mises (CvM) test \citep{quessy2012cramer}. This test uses the quadratic distance between the eCDFs $F_n$ and $G_m$, which is defined\footnote{Note, here: $dH_{n,m}(z)=h_{n,m}(z)\,dz$} by the parameter $W_{n,m}$:
\begin{equation}
    W_{n,m} = \frac{nm}{n+m} \int \left(F_n(z)-G_m(z)\right)^2\, dH_{n,m}(z)\,. \label{eq:pooledw}
\end{equation} 
This function can be considered to be the eCDF of the pooled sample $Z_1,Z_2,...,Z_n,Z_{n+1},....,Z_{n+m}$, where $Z_i = X_i, \; i\in \{1,2,..,n\}$ and $Z_{n+j} = Y_{n+j}, \, j \in \{1,2,3,...,m\}$ \citep{quessy2012cramer}. For instance, we can write $H_{n,m}(z_0)$ as:
\begin{equation}
    H_{n,m}(z_0) = \frac{1}{n+m}\sum^n_{i=1} 1_{X_i \leq z_0} + \frac{1}{n+m} \sum^m_{j=1} 1_{Y_j \leq z_0}\,. \label{eq:Hnmz0} 
\end{equation}
The argument $z_0 \in I$, where $I$ is the interval that spans all of the possible observations. 
Numerically, the integral can be calculated by choosing a fixed step size $\Delta H_{n,m}$ and estimating $(F_n(z)-G_m(z))^2$ for all relevant values of $z \in I$. Since $H_{n,m}$ is the pooled eCDF, the integral $\int dH_{n,m} =1$.
Since there is a total of $n+m$ observations, the relevant step size will be $\Delta H_{n,m} = \frac{1}{n+m}$. This step size can further be understood from the discretization of the integral, which will become:
\begin{equation}
    \sum^{n+m}_{i=1} \Delta H_{n,m} = 1 \quad \Longleftrightarrow \quad \Delta H_{n,m} = \frac{1}{n+m}. \label{eq:discHnm} 
\end{equation}
By using Eq.~(\ref{eq:discHnm}), and the fact that $z \in \{Z_1, Z_2,...,Z_{n+m} \}$, we can write the discrete version of Eq.~(\ref{eq:pooledw}), namely:
\begin{equation}
    W_{n,m} = \frac{nm}{(n+m)^2} \sum^{n+m}_{i=1} \Big(F_n(Z_i)-G_m(Z_i)\Big)^2. 
\end{equation}
If the null hypothesis, $H_0$ is true, then $F=G$ and the value of $W_{n,m}$ will be very small because the observations drawn from $F$ and $G$ will from the same distribution. On the other hand, if $F \neq G$, then $W_{n,m}$ can be large. 

The quadratic distance in the CvM-test makes it more versatile, and in general more powerful than the KS-test \citep{razali2011power}, since high variations in a distribution function are better handled in the integral that involves the quadratic distance $W^2_{n,m}$ rather than the supremum distance. 

\subsubsection{Anderson-Darling test}
If there is a large difference in the central region of two distributions (i.e. for small values of $\chi_{\rm eff}$), it will have a very dominating effect on $W^2_{n,m}$. However, the distribution tails are less accounted for in the KS and CvM-tests. The Anderson-Darling (AD) test, on the other hand, places a stronger emphasis on tails by weighting the quadratic distance in the following way:
\begin{equation}
    A^2_{n,m} = \frac{nm}{n+m} \int \frac{\big(F_n(z)-G_m(z)\big)^2}{H_{n+m}(z)(1-H_{n+m}(z))}\, dH_{n,m}(z)\,. \label{eq:AD2}
\end{equation}
Similar to the discrete version of $W^2_{n,m}$, if $z \in \{Z_1,Z_2,....,Z_{n+m}\}$, then the discrete version of $A^2_{n,m}$ will become:
\begin{equation}
    A^2_{n,m} = \frac{nm}{(n+m)^2} \sum^{n+m}_{i=1}  \frac{\big(F_n(Z_i)-G_m(Z_i)\big)^2}{H_{n+m}(Z_i)(1-H_{n+m}(Z_i))}\,. \label{eq:AD2dis}
\end{equation}
The weighing in the AD-test thus makes it more sensitive to the differences in the tails of two distributions (i.e. for effective spin magnitudes $|\chi_{\rm eff}|\rightarrow 1$). Some empirical evidence suggests that the AD-test is more versatile than the KS and CvM-tests \citep{razali2011power}. To summerize, each test must be chosen with caution, and the strengths and weaknesses of each test must be considered in two-sample tests. 

With the mathematical background established, we can now apply the relevant formulas to the statistical analysis of LVK BH+BH merger data and compare the results with our simulations of BH+BH binaries.

\clearpage
\section{Dependence of $\chi_{\rm eff}$ on mass ratio, $q$}\label{Appendix:B}

\begin{figure}
\hspace{1.0cm}
 \includegraphics[width=0.75\columnwidth]{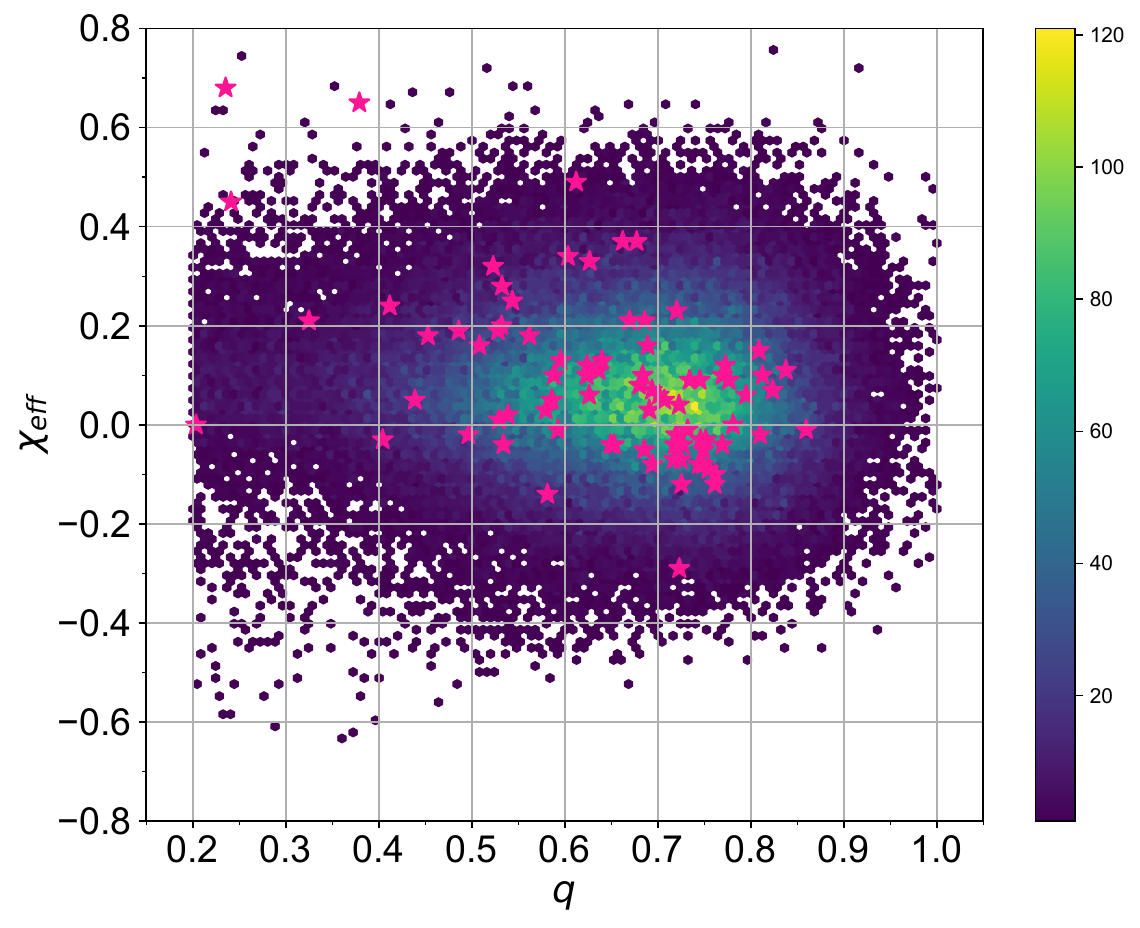}
 \caption{Effective spin ($\chi_{\rm eff}$) as a function of BH mass ratio ($q$). Observations are marked with magenta stars and simulations (in which $q$ and $\chi_{\rm eff}$ are chosen independently, see Section~\ref{sec:MCMC}) are plotted as points (color indicates density of points). By definition $q\in [0;1]$, but the simulations were restricted to $q\in [0.2;1]$ based on observations. If any correlation exists among the observed BH+BH events, it appears to be weak.}
 \label{fig:chi_eff-q}
\end{figure}

To check for any correlation between $\chi_{\rm eff}$ and $q$, we plotted $\chi_{\rm eff}$ as a function of $q$ in Fig.~\ref{fig:chi_eff-q}. If a correlation exists among the observed BH+BH events, it appears to be weak. A quick Pearson correlation test using median values of the 83 BH+BH mergers yields a correlation coefficient of $-0.48$, indicating a moderate negative correlation between the two variables; however, this is not necessarily a strong relationship --- see also \citet{hvb+24,hmv25,rvf25}.
For this reason, in our simulations presented in Section~\ref{sec:MCMC}, we assume $q$ and $\chi_{\rm eff}$ to be independent. Nonetheless, we cannot exclude the possibility of a bias and emphasize the need for further investigation into the potential correlation between $\chi_{\rm eff}$ and $q$.

\clearpage

\bibliographystyle{elsarticle-harv} 
\bibliography{references}






\end{document}